\documentclass{aa}
\usepackage[varg]{txfonts}
\usepackage{graphicx}
\usepackage{t1enc}
\newcommand{\phn}   {\phantom{0}}

\newcommand{\sii}   {[\ion{S}{ii}]}

\newcommand{\fe}    {[\ion{Fe}{ii}]}
\newcommand{\feP}   {\ion{Fe}{ii}}
\newcommand{\hmol}  {H$_2\ \upsilon=1$--0 S(1)}
\newcommand{\hii}   {\ion{H}{ii}}
\newcommand{\oi}    {[\ion{O}{i}]}

\newcommand{\kms}   {km~s$^{-1}$}

\begin{document}

\title{ 
A survey of IRAS young stellar object candidates: \\
searching for large-scale Herbig-Haro objects
\thanks{
Based on observations obtained at the Centro Astron\'omico Hispano-Alem\'an de Calar Alto (CAHA), Spain.} 
}

\titlerunning{A survey of IRAS YSOs candidates}

\author{
Rosario L\'opez\inst{1}
\and
Angels Riera\inst{2,\thanks{Deceased on 2017 September 27}}
\and
Robert Estalella\inst{1}
\and
Gabriel G\'omez \inst{3,4,5}, 
}

\institute{
Departament de F\'{\i}sica Qu\`antica i Astrof\'{\i}sica,
Institut de Ci\`encies del Cosmos, Universitat de Barcelona, IEEC-UB,
Mart\'{\i} i Franqu\`es, 1, E-08028 Barcelona, Spain.
\and
Departament de F\'{\i}sica i Enginyeria Nuclear, Universitat Polit\`ecnica de Catalunya, Av.\ Eduard Maristany, 16, E-08019 Barcelona, Spain.
\and
Grantecan S.A., Centro de Astrof\'isica de La Palma, Cuesta de S. Jos\'e, E-38712 Bre\~na Baja, La Palma, Spain.
\and
Instituto de Astrof\'isica de Canarias, Via L\'actea s/n,  E-38200 La Laguna, Tenerife, Spain.
\and
Departamento de Astrof\'isica, Universidad de La Laguna, E-38205 La Laguna, Tenerife, Spain.
}

\authorrunning{L\'opez et al.}

\date{Version 2021  /
Received /
Accepted
}

\abstract{
Jets and outflows are associated with young stellar objects across the stellar mass spectrum, from brown dwarf protostars to massive, Ae/Be stars. 
Frequently, the jet morphology is spatially discontinuous because of the temporal variability of the ejection from the driving source. 
Images covering a wide field of view around the jet driving-source are useful to map the large-scale jet emission and to explore the mass ejection history.
}{
The aim of this work was to search for large-scale optical Herbig-Haro (HH) objects lying in a wide field around a sample of IRAS sources, candidates to trace young stellar objects. 
}{
Deep, narrow-band images through the H$\alpha$ and \sii{} emission lines, and through an off-line continuum filter, covering a wide ($\sim15'$) field around the IRAS targets were acquired. 
The images in the three filters were analyzed to identify shock-excited line emission (i.e., HH) in contrast to scattered line emission. 	
}{
New images of a sample of fifteen IRAS sources, obtained in an homogeneous way are presented. 
HH emission was detected in six fields, and the astrometry of the knot features is given.
The nature of the extended emission as scattered emission around nine of the IRAS targets is confirmed. 
For seven IRAS sources, with unclear optical counterpart, a more plausible counterpart is proposed.
A refined value of the source distance is reported for seven targets. 
An update of the main data available for each  of the sampled fields, including images from public data archives, is also presented.
}{}


\maketitle
\section{Introduction}

Protostellar jets and outflows are found everywhere in star forming regions as  fundamental events occurring during the star formation process. 
They are believed to regulate processes such as the removal of angular momentum excess from the star-disk system, and the dispersion of the parent cloud. 
They are observed in all evolutionary stages of young stellar objects (YSOs) where accretion is occurring, from Class 0 to Class III, and across the stellar mass spectrum, from  brown dwarf protostars \citep{Whe05,Ria17} to massive, high luminosity young  stars \citep{Guz12}, and  are detected over a wide wavelength range through continuum emission and line emission from molecular, neutral and ionized atomic transitions. 

Herbig-Haro objects (HHs) are the optical manifestation of outflow events. 
They are small shock-excited nebulae visible in low excitation lines (eg. \oi{}, H$\alpha$, \sii{}) produced by the radiative cooling in post-shock zones.  
Most HHs appear as a string of knots aligned in a highly collimated jet (HH-jet) ending in a bright bow-shock where the jet rams into the  surrounding medium.
The spatial scales covered by the HH-jets range from a few au, in the case of microjets of T Tauri stars \citep{Aga11} to several parsecs \citep[giant Herbig-Haro jets,][]{Dev99, Reip19}. 
The knots trace internal shocks driven by velocity ejection variability. 
Frequently the jet morphology shows discontinuities in its spatial distribution, which is usually attributed to the temporal variability of the ejection material from the source \citep{Rag90}. 
Because of this, images covering a wide field of view (FOV) around the suspected driving source are essential to explore the mass ejection history.
HH-jets interact with the natal molecular cloud as they travel outwards entraining ambient material and giving rise to large-scale molecular outflows (see e.g., reviews from \citet{Fra14} and \citet{Bal16} for a more complete picture of protostellar outflow theory and observations).

We carried out a project aimed at obtaining new deep optical, narrow-band (\sii{} and H$\alpha$) images of a sample of IRAS sources, candidates to be tracing YSOs. 
A sample of fifteen targets was selected to be imaged in a homogeneous way, covering a wide field of view ($\sim15'$) around the IRAS counterpart, looking for shocked emission that could be associated with the target. 
In addition to the narrow-band images, an image through an off-H$\alpha$ filter was obtained to get the continuum emission, with the aim of distinguishing between reflected and shocked emission in the neighborhood of each of the sources. 
Because of the properties of the sources selected, we expected to find HH jets with different evolutionary ages and spatial scales.
Most of the fields were not previously imaged through narrow-band filters, nor with a several arcmin wide FOV. 
Such a spatial coverage is useful to explore whether the HH jet shows discontinuities in its spatial emission, indicative of episodic mass ejections. 
It was expected that the sample likely included  TTauri and Herbig Ae/Be stars being still actively accreting matter from circumstellar disks and driving small scale jets, 
so that a product of the project would be increasing the sample of known low-mass (TTauri) and intermediate-mass (Herbig Ae/Be) sources driving HH jets.
In this work, in addition to the characterization of the optical emission associated with the YSOs traced by the IRAS sources, we updated the observational data available from optical and near-IR data archives for the targets observed, and we performed new accurate astrometry of the jet knots and sources mapped in the observed fields. 

The work is structured as follow: 
in \S 2 we present the criteria followed to select the sample to be observed; 
in \S 3 we describe the observations and data reduction; 
in \S 4 we present an updated description of the data available and the results obtained for each of the observed fields, and 
in \S 5 we summarize the global results derived from our survey. 

\section{Sample selection criteria}

\begin{figure}[htb]
\centering
\resizebox{\hsize}{!}{\includegraphics{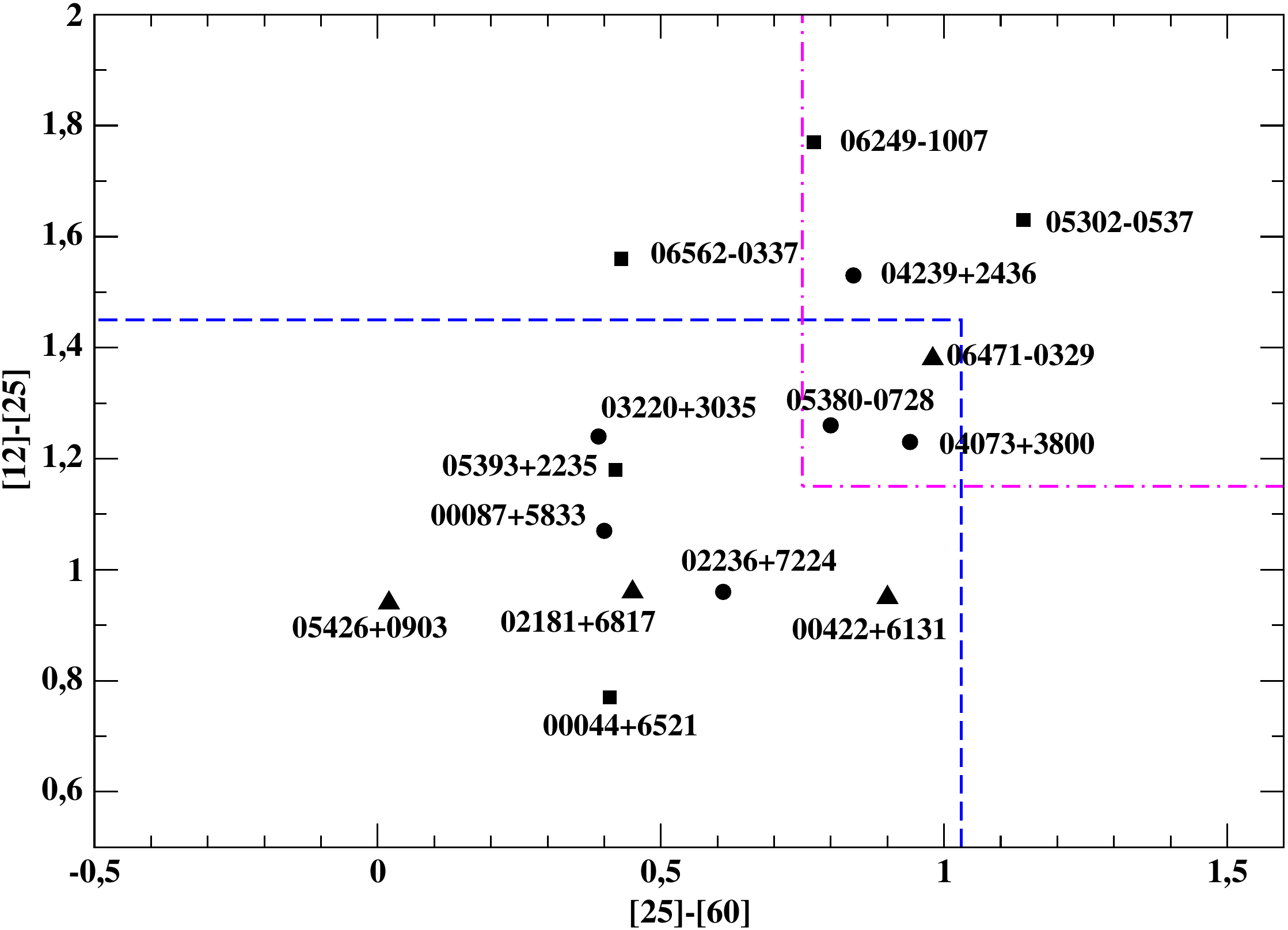}}
\caption{
Location of the IRAS sources observed in the ($[12]-[25],[25]-[60]$) color--color diagram. 
The blue lines mark the region of young stellar objects, and the magenta lines that of ultra-compact \hii{} regions.
Circles correspond to the sources driving jets, squares correspond to the sources with extended nebular emission and triangles correspond to the point-like sources (see Table \ref{sources}). 
\label{i_c}}
\end{figure}

\begin{figure}[htb]
\centering
\resizebox{\hsize}{!}{\includegraphics{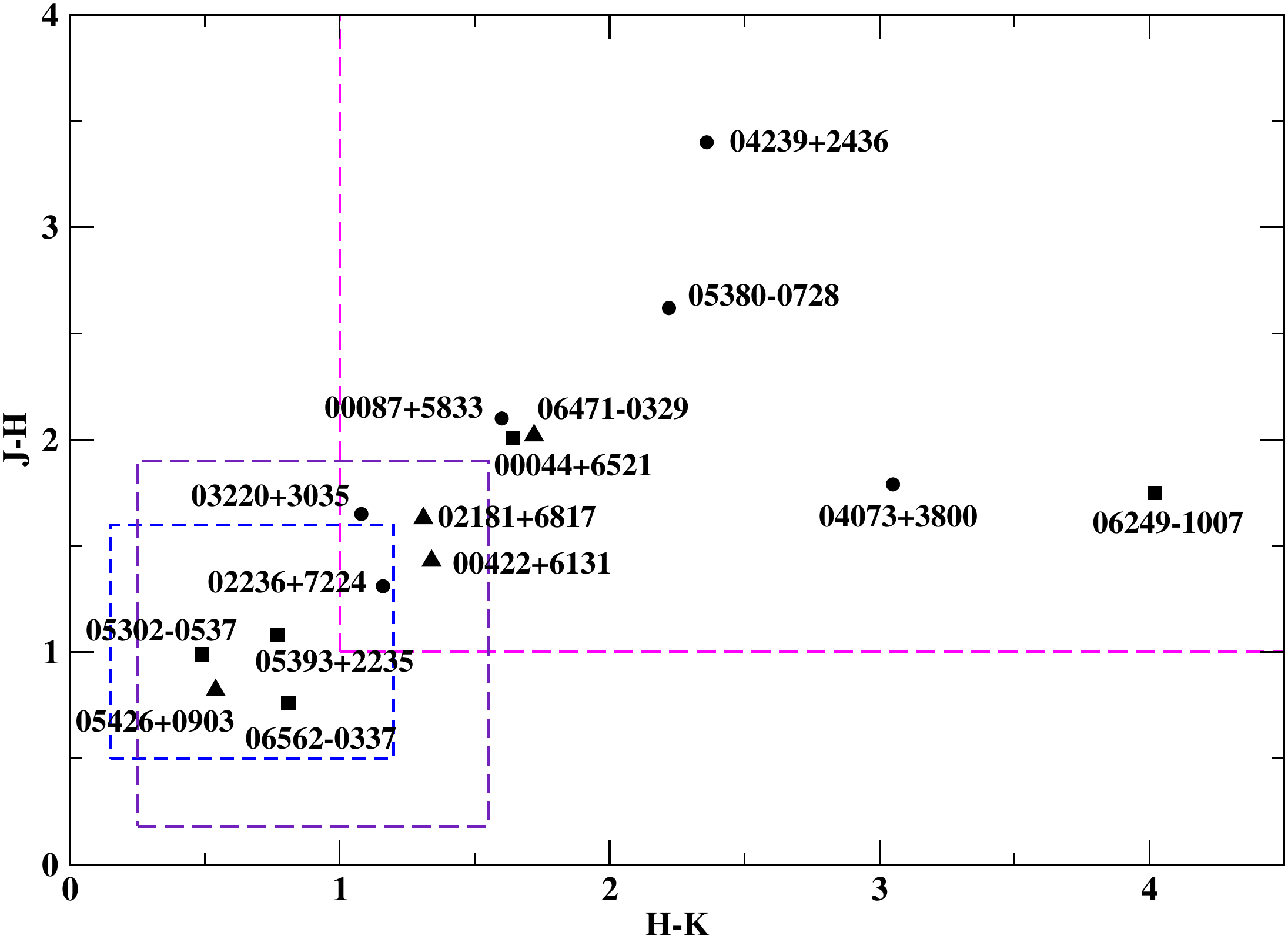}}
\caption{Location of the IRAS sources counterparts in the ($J-H,H-K$) color--color diagram.
The blue lines mark the TTauri region,  the purple lines the region of the Herbig Ae/Be stars, and the magenta lines that of Class I sources.
The symbols are the same as in Fig.~\ref{i_c}.
\label{n_c}}
\end{figure}

The sample observed  was extracted from the GLMP catalogues \citep{Gar91, Gar97} and the optical survey of \citet{Sua06} of nearby IRAS sources. 
GLMP uses the $[12]-[25]$ versus $[25]-[60]$ color--color (CC) diagram to identify YSO candidates.
Later on \citet{Sua06} confirm the YSO nature of several YSO candidates of the GLMP sample, based on the optical spectrum of their IRAS counterpart, their luminosity class, and their location in a star-forming region.
We selected targets from the GLMP catalogues and the \citet{Sua06} survey that, in addition, showed optical emission in the DSS plates.
As can be seen in Figures~\ref{i_c} and \ref{n_c}, all of the selected targets have infrared (IR) color indices characteristic of YSOs. 
Some of the targets are likely to be TTauri or Herbig Ae/Be, which would be visible because they had already emerged from their native environment. 
Thus, they would still be actively accreting matter from circumstellar disks and could be driving small-scale jets (microjets). 
The resulting sample consists of fifteen targets that were observable from the Calar Alto Observatory (CAHA) in an observing run during the winter period.
Table \ref{sources} lists the sample of fields mapped.

\begin{sidewaystable*}
\centering
\caption{Observed sample}
\label{sources}
\begin{tabular}{lcclllllcc}  
\hline\hline
& $\alpha_{2000}$ & $\delta_{2000}$  & & & & & & \multicolumn{2}{c}{Distance}  \\
\cline{9-10}            
IRAS source & {($^\mathrm{h~m~s}$)} & {($^\circ$ $\arcmin$ $\arcsec$)} & Counterpart & Classif.\tablefootmark{a} & Association\tablefootmark{b} & HH in FOV & Location & (pc) & Ref.\\
\hline
\multicolumn{6}{l}{I: Sources associated with jet emission} \\
\hline
00087+5833   & 00 11 26.5 & $+58$ 49 50 & LkH$\alpha$ 198   & HAeBe     &$\cdots$    & 161--164, 462 & L1265  & $\phn344\pm17\phn$ & 1 \\
02236+7224   & 02 28 16.4 & $+72$ 37 36 & J02281661+7237328 & TT        & RNO 7 (c)  & 488           & L1340  & $\phn861\pm22\phn$ & 2 \\
03220+3035   & 03 25 09.2 & $+30$ 46 21 & L1448-IRS1        & TT, b     & RNO 13 (n) & 194--196      & L1448  & $\phn240\pm12\phn$ & 1 \\
04073+3800   & 04 10 41.2 & $+38$ 07 54 & PP 13             & TT/FUO, b & PP 13 (n)  & 463--465      & L1473  & $350$              & 3 \\
04239+2436   & 04 26 55.3 & $+24$ 43 34 & $\cdots$          & CI, b     & $\cdots$   & 300           & Taurus & $129.0\pm0.8\phn$  & 4 \\
05380$-$0728 & 05 40 27.7 & $-07$ 27 28 & Re 50N            & CI        & $\cdots$   & 1121--1122    & L1641  & $460$              & 5 \\
\hline
\multicolumn{6}{l}{II: Sources with extended nebular emission} \\
\hline
00044+6521   & 00 07 03.5 & $+65$ 38 41 & PP 1, McC H12       & HAeBe & $\cdots$         & $\cdots$ & Cepheus IV & $\phn845\pm110$    & 6 \\
05302$-$0537 & 05 32 41.7 & $-05$ 35 48 & J05324165$-$0535461 & CI, b & $\cdots$         & $\cdots$ & Orion A    & $\phn319\pm12\phn$ & 2 \\
05393+2235   & 05 42 21.3 & $+22$ 36 47 & J05422123+2236471   & FUO   & RNO 54 (n)       & $\cdots$ & $\cdots$   &    $1540\pm106$    & 2 \\
06249$-$1007 & 06 27 18.1 & $-10$ 09 41 & J06271812$-$1009387 & TT    & HHL 43 (n)       & $\cdots$ & $\cdots$   & $\phn860\pm70\phn$ & 7 \\
06562$-$0337 & 06 58 44.4 & $-03$ 41 12 & J06584435$-$0341099 & HAeBe & Iron-clad Nebula & $\cdots$ & $\cdots$   &    $5650\pm430$    & 7 \\
\hline
\multicolumn{6}{l}{III: Point-like sources} \\
\hline
00422+6131   & 00 45 09.9 & $+61$ 47 57 & J00450982+61147574  & TT       & $\cdots$ & $\cdots$ & $\cdots$   & $2400^{+920}_{-520}\phn\phn$ & 2 \\
02181+6817   & 02 22 22.6 & $+68$ 30 43 & CO Cas              & HAeBe, v & $\cdots$ & $\cdots$ & $\cdots$   & $\phn664\pm25\phn$           & 2 \\
05426+0903   & 05 45 22.4 & $+09$ 04 13 & FU Ori              & FUO      & $\cdots$ & $\cdots$ & $\cdots$   & $\phn416\pm9\phn\phn$        & 2 \\
06471$-$0329 & 06 49 40.4 & $-03$ 32 51 & J06494021$-$0332523 & CI       & $\cdots$ & $\cdots$ & G216$-$2.5 &    $2110\pm21\phn$           & 1 \\
\hline 
\end{tabular}
\tablefoot{
\tablefoottext{a}{b: binary; CI: Class I; FUO: FU Ori; HAeBe: Herbig Ae/Be; TT: TTauri; v: variable.}
\tablefoottext{b}{c: star cluster; n: extended nebular object.}
}
\tablebib{
(1) \citet{Zuc20};
(2) \citet{Gai18};
(3) \citet{Coh83};
(4) \citet{Gal19};
(5) \citet{Coh90};
(6) \citet{Mac68};
(7) See text.
}
\end{sidewaystable*}

According to the nature and morphology of the emission detected in this work (see \S \ref{sec_results}),
and the data available in the literature, the sample was separated in three groups, which are indicated in Table~\ref{sources} and used in Figs.\ \ref{i_c} and \ref{n_c}. 
The first group (I) included fields with extended, pure-line emission, suggestive of being produced by shocked (jet) emission. 
The second group (II) included IRAS targets associated with nebular, reflected emission.
Finally the third group (III) included IRAS targets showing a point-like emission and for which no extended emission, shocked or reflected, was detected in the entire field mapped around the IRAS source.
Each observed field is identified with the name of the central IRAS source (listed in column 1) on which the images were centered and its catalogue position is given in columns 2--3. 
The following columns list the most probably optical counterpart assigned to the IRAS source in the available literature. 
The classification of the counterpart (Class I, TTauri, FU Ori, Herbig Ae/Be), and the membership of a cluster or extended nebular nature of the object is also indicated.
The rest of columns list the HH objects catalogued in the FOV mapped,
the star-forming region where the IRAS source is located, when known, and finally, 
the distance to the source. Some of the distances have been updated in this work from Gaia data \citep{Gai18}, and from the improved method based on parallaxes of high-mass star-forming regions of \citet{Rei19}.\footnote{\label{footnote:distance}
http://bessel.vlbi-astrometry.org/node/378, Parallax-Based Distance Calculator V2}

As already mentioned, the observed targets have IR colors that are characteristic of YSOs. 
Figure~\ref{i_c} presents the IRAS ($[12]-[25], [25]-[60]$) CC  diagram of the IRAS sources of the sample.  
The region of the plane where YSOs are located is delimited by the blue lines \citep{Pal90}, and the region delimited by the magenta lines corresponds to the location of ultra-compact \hii{} regions \citep{Woo89}, probably associated with YSOs too. 
Figure~\ref{n_c} presents the ($J-H, H-K$) CC diagram of the assigned near-IR IRAS counterpart, when known. 
The $J, H, K$ magnitudes are taken from the 2MASS catalogue\footnote{
This publication makes use of data products from the Two Micron All Sky Survey, which is a joint project of the University of Massachusetts and the Infrared Processing and Analysis Center/California Institute of Technology, funded by the National Aeronautics and Space Administration and the National Science Foundation}. 
  
The region of the plane delimited by the blue square corresponds to the location of TTauri stars \citep{Mey97}, the purple lines delimits the region of Herbig Ae/Be \citep{Man06}, and the magenta lines delimits the region of the luminous Class~I protostars \citep{Lad92}. 

\section{Observations and Data Reduction}

Observations were carried out on November 2016 with the 2.2~m telescope in the  Calar Alto Observatory (CAHA)  using the Calar Alto Faint Object Spectrograph (CAFOS) in direct imaging mode. 
The instrument was equipped with a 2048 $\times$ 2048 CCD, giving a spatial scale of $0\farcs53$~pixel$^{-1}$ and a field of view of $16'$. 
Three narrow-band filters were used: 
the line filters of 
H$\alpha$ (central wavelength $\lambda=6569$~\AA, bandpass $\Delta\lambda=50$~\AA), and 
\sii{} ($\lambda=6744$~\AA, $\Delta\lambda=97$~\AA), 
which included the emission from the \sii{} $\lambda=6717, 6731$~\AA\ lines, and an off-H$\alpha$ filter ($\lambda=6607$~\AA, $\Delta\lambda=43$~\AA) 
to look for the continuum emission nearby to the H$\alpha$ and \sii{} lines. 
Conditions were not photometric and the seeing values varied from $1\farcs5$ up to $\sim3''$ during the observing run.

Fourteen fields of our survey were mapped homogeneously in three narrow-band filters, the H$\alpha$ and \sii{} lines and an off-line, nearby continuum filter. 
One of the targets (IRAS 05302$-$0535) was only imaged through the \sii{} filter due to problems during the runs.
The images  include a total of eighteen IRAS sources.
Three fields (centered on IRAS 02236+7224, 03220+3035 and 05380$-$0728) included a second IRAS source.

We obtained images in H$\alpha$ and \sii{} of 1~hr of total integration time by combining three frames of 1200~s exposure each, and an additional continuum image of 1200~s integration after combining two frames of 600~s exposure.

All the images were processed with the standard tasks of the IRAF\footnote{
IRAF is distributed by the National Optical Astronomy Observatories, which are operated by the Association of Universities for Research in Astronomy, Inc., under cooperative agreement with the National Science Foundation.}
reduction package, which included bias subtraction and flatfielding corrections, using sky flats. 
In order to correct for the misalignment between individual exposures, the frames were recentered using the reference positions of field stars well distributed around the source. 
Astrometric calibration of the  images was performed in order to compare the optical emission with the positions of the objects reported in the field and, in particular, with the nominal position of the IRAS sources. 
The images were registered using the $(\alpha, \delta)$ coordinates from the  USNO Catalogue\footnote{
The USNOFS Image and Catalogue Archive is operated by the United States Naval Observatory, Flagstaff Station.}
of ten field stars well distributed in the observed field. 
The rms of the transformation was $0\farcs2$ in both coordinates.

\section{Results}
\label{sec_results}
In the following, we present the study carried out for each field, following the classification of the sample in the three groups listed in Table~\ref{sources}.  
The structure of each mapped field is as follows.
First, we include a summary updating  the relevant information reported in the literature. 
Next, we detail the new results found from our survey:
(i) Wide field image in the narrow-band, off-line continuum filter, with
the IRAS source and other YSOs potentially related to the source marked. 
(ii) Close-ups of selected regions obtained from the images acquired through the H$\alpha$ and \sii{} line filters. 
The sub-images allowed us to visualize and characterize  the structure of the extended emission around the IRAS source. 
(iii) For some targets we also show close-ups from images retrieved from public data archives,  with the aim of comparing the emission in other wavelengths with the emission detected in our observations.
(iv) A table with the detailed astrometry of the shocked emission, if detected. (v) In some cases, we propose a new identification of the IRAS counterpart, based on updated catalogue information and in a more accurate astrometry obtained from our images. 

\subsection{Group I: Sources associated with jet emission}

\subsubsection{\object{IRAS 00087+5833}}

\begin{figure}[htb]
\resizebox{\hsize}{!}{\includegraphics{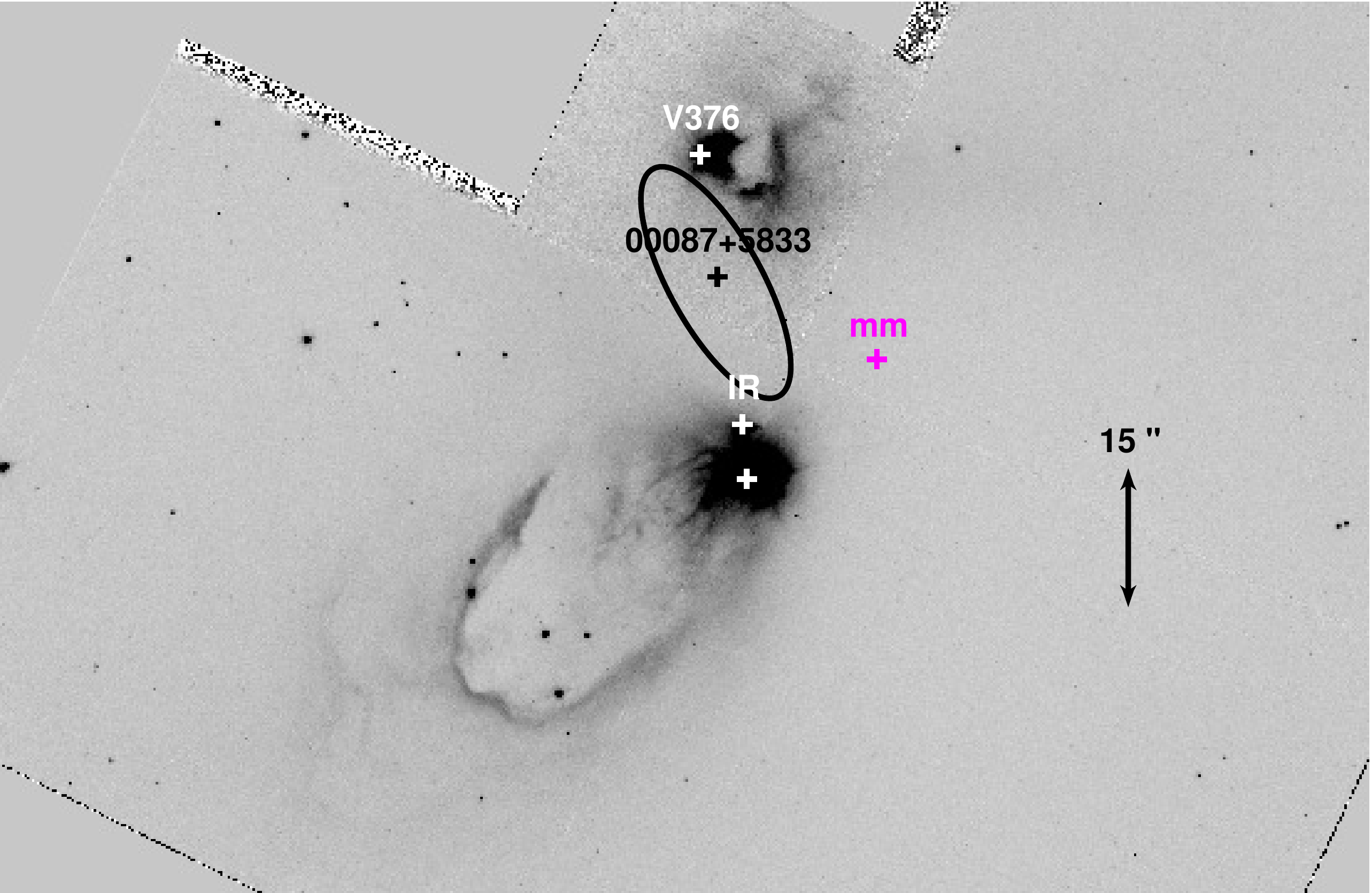}}
\caption{
IRAS 00087+5833: \emph{HST} WFPC2 image in the F814W filter of the field around the IRAS source. 
The optical, near-IR and millimeter sources reported in the field, and the IRAS position with the error ellipse are plotted. 
The coordinates of these objects are given in Table~\ref{objects}. 
All figures are oriented north up and east left.
\label{HST}} 
\end{figure}

IRAS 00087+5833 is located in the dark cloud L1265, in Cassiopeia, at a distance of $344\pm17$~pc \citep{Zuc20}. In the following we present a short description of the field around the source.
\begin{description}
    \item[\textit{Young stellar objects:}]
     They are plotted in Fig.~\ref{HST}, a \emph{Hubble Space Telescope} (\emph{HST}) image (\emph{HST} Legacy Archive. PI K. Stapelfeldt. Program ID: 8216. Cycle 8). Their positions are given in Table~\ref{objects}. Namely:
    \begin{description}
        \item[V376~Cas:] 
         Located north of IRAS.
        Classified as a Herbig Ae/Be star \citep{Lag93}.
        \item[LkH$\alpha$~198]  (also known as V~633 Cas):
        Located south of IRAS.
        Classified as a Herbig Ae/Be star \citep{Asp00}, and
        a spectroscopic binary \citet{Smi05}.
        \item[LkH$\alpha$~198-IR:]
         Located $6''$ north of LkH$\alpha$ 198.
        Classified as a deeply embedded object.
        Has counterparts at 10 $\mu$m \citep{Lag93} and 1.3 mm \citep{Boi11}.
        \item[LkH$\alpha$~198-mm:]
         Located $\sim19''$ northwest of LkH$\alpha$ 198.
        Classified as a deeply embedded object, with neither visible, nor IR emission detected. 
        Has counterparts at 1.3~mm \citep{Hen98} and 3~mm \citep{Boi11}.
    \end{description}

    \item[\textit{Proposed counterpart of IRAS:}]  LkH$\alpha$ 198, but it is offset $\sim20''$ from the IRAS position, outside the IRAS error ellipse, as are all the other YSOs.
   
    \item[\textit{Molecular outflows:}]
     Several CO outflows, observed with low \citep{Can84} and high angular resolution \citep{Mat07}, including:
        \begin{itemize}
        \item  Outflow with lobes east and west of LkH$\alpha$~198, with signs of precession \citep{Smi05} (LkH$\alpha$~198 is a binary system). 
        \item  Outflow in the north--south direction, centered on LkH$\alpha$ 198-mm \citep{Smi05}.
    \end{itemize}

    \item[\textit{Optical outflows/Herbig-Haro objects:}]
     Several knots in \sii{} near YSOs \citep{Str86,Goo93,Cor95,Asp00}.
    \begin{description}
        \item[Identification of knots \citep{Rei00b}:]
        \begin{description}\item[]
            \item[A and B:]   correspond to \object{HH 161}, driven by LkH$\alpha$ 198-IR \citep{Cor95};
            \item[C, D, E, and F:]  correspond to \object{HH 164}, driven by  LkH$\alpha$ 198 \citep{Cor95};
            \item[G, H, I, and J:]  correspond to \object{HH 162}.
            \item[\object{HH 461}:]  a new HH, located $\sim82''$ southeast of LkH$\alpha$ 198, proposed to be a distant bow-shock of HH 164 \citep{Asp00}.
            \item[\object{HHs 800-802}:] proposed to be likely an extension of the LkH$\alpha$ 198 parsec-scale outflow \citep{McG04}. They are outside the FOV of our images.
        \end{description}
    \end{description}

\end{description}

\begin{figure}[htb]
\resizebox{\hsize}{!}{\includegraphics{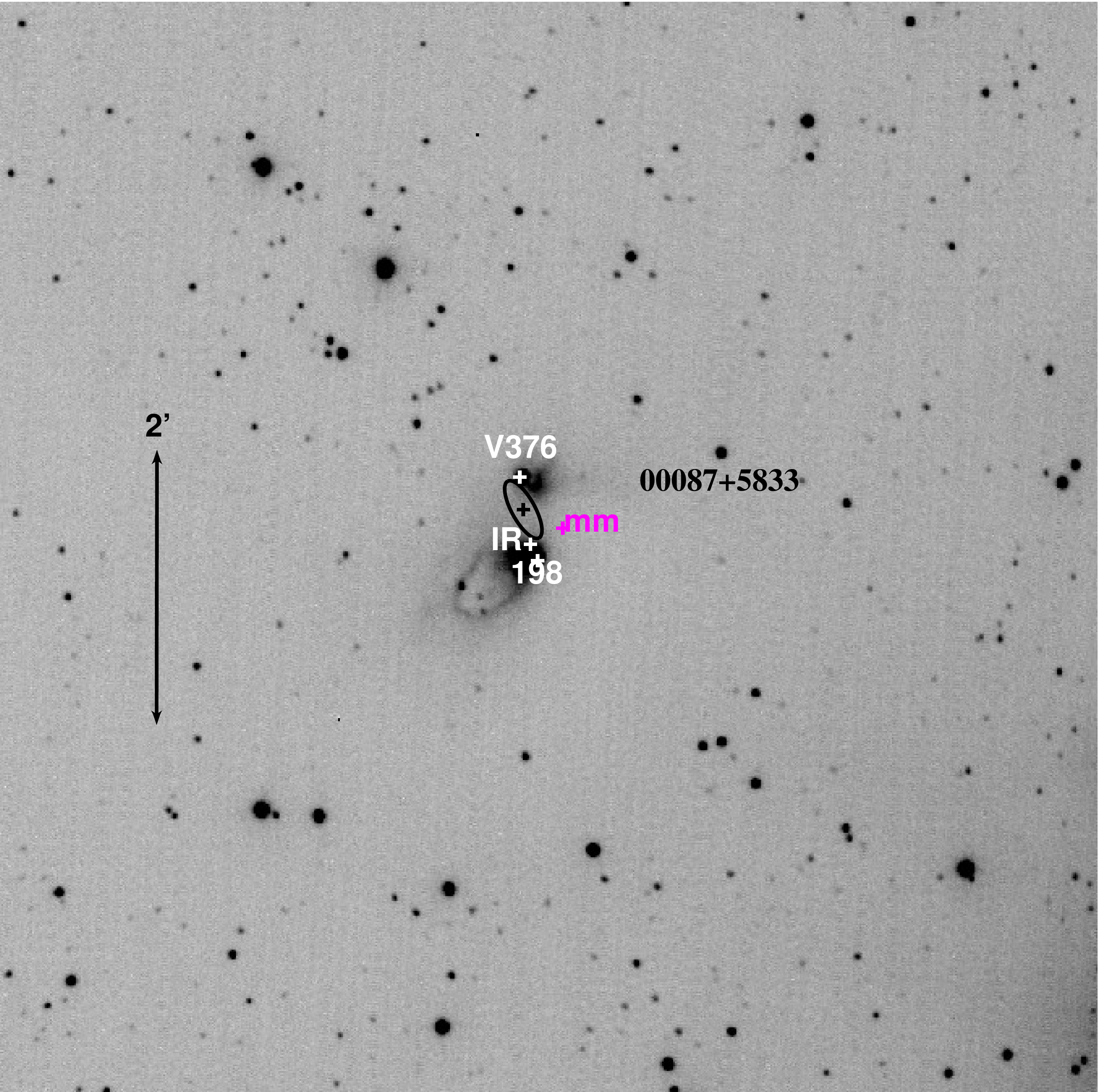}}
\caption{
IRAS 00087+5833: CAHA image of the field in the continuum filter.
\label{2_cont}} 
\end{figure}

\begin{figure}[htb]
\resizebox{\hsize}{!}{\includegraphics{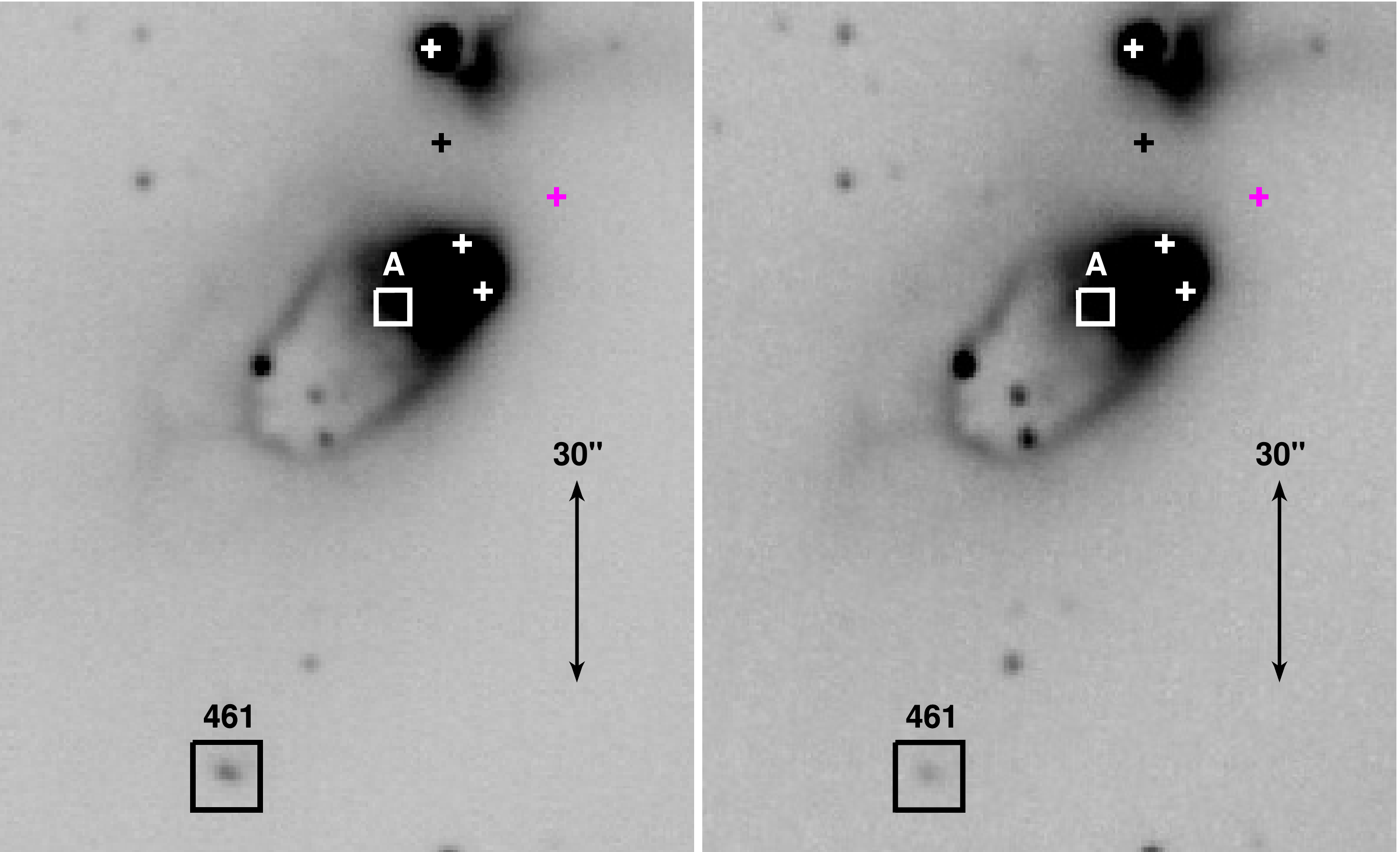}}
\caption{
IRAS 00087+5833: Close-up of the CAHA images in the H$\alpha$ (left) and \sii{}  (right) filters. 
The YSOs are marked as in Fig.~\ref{2_cont}.
The knots HH 161A and HH 461 are enclosed in boxes.
\label{2_line}} 
\end{figure}
 
\begin{table}[htb]
\centering
\caption[ ]{Young stellar objects  around  IRAS 00087+5833}
\label{objects}
\begin{tabular}{lccc}  
\hline\hline
&\multicolumn{2}{c}{Position} \\
\cline{2-3}
&$\alpha_{2000}$ & $\delta_{2000}$ \\
Source & {($^\mathrm{h~m~s})$} & {($^\circ$ $\arcmin$ $\arcsec$)} & Ref. \\
\hline
V376        & 00 11 26.7 &+58 50 04 & 1 \\
00087+5833  & 00 11 26.5 &+58 49 50 & 2 \\
198-mm      & 00 11 24.3 &+58 49 42 & 3 \\
198-IR      & 00 11 26.1 &+58 49 35 & 1 \\
198         & 00 11 25.7 &+58 49 28 & 4 \\
HH 161A     & 00 11 27.4 &+58 49 26 & 5 \\
HH 461      & 00 11 30.6 &+58 48 16 & 4, 5\\
\hline
\end{tabular}
\tablebib{
(1) \citet{Lag93}; 
(2) IRAS PSC; 
(3) \citet{San94}; 
(4) \citet{Asp00};
(5) This work.
}
\end{table}

Figure~\ref{2_cont} displays the field around IRAS 00087+5833 mapped through the continuum filter in our survey.
We did not detect any new extended line emission in H$\alpha$ nor in \sii{} in the field mapped.
The YSOs around the IRAS source, marked in the figure, lie all of them outside the IRAS position error ellipse. \textbf{}
This discards that any of these YSOs could be the counterpart of IRAS 00087+5833.
Figure~\ref{2_line} shows a close-up of the H$\alpha$ and \sii{} images showing the extended emission associated with the YSOs. 
We detected line emission from the HH knots HH 161A and HH 461. We did not find any evidence of being related to the IRAS source.
Regarding HH 461, the emission of the knot is stronger in H$\alpha$ than in \sii{}, which is a characteristic of the emission from bow shocks. 
This confirms the nature of HH 461 as a bow shock previously proposed by \citet{Asp00}, based only on geometrical arguments.

\subsubsection{\object{IRAS 02236+7224}}

IRAS 02236+7224 is located in the dark cloud L1340, at a distance of $861\pm22$~pc \citep{Gai18}. In the following we present a short description of the field around source.

\begin{description}
    
   \item[\textit{Other IRAS sources in the field:}] \object{IRAS 02238+7222}, in the southern part of the field, with colors that do not correspond to a YSO. Does not seem related to the HHs of the field. 
  
    \item[\textit{Young stellar objects:}] The \object{RNO~7} cluster of YSOs, with several low and inter\-mediate-mass YSOs surrounded by nebular emission.

    \item[\textit{Proposed counterpart of IRAS 02236+7224:}] A low-mass, H$\alpha$-emission star \citep{Kun16a}
  
    \item[\textit{Optical outflows/Herbig-Haro objects:}]~
    \begin{description}
        \item[Near IRAS 02236+7224:]  Jet in  H$\alpha$ and \sii{} emerging southwards from IRAS \citep{Nan02}. Knots of the jet \citep{Mag03}:
            \begin{description}
                \item[\object{HH 671}A and B:]  most probably associated with the cluster, but not clearly related to IRAS.
                \item[Knot 3:]  located $\sim1'$ southwest of HH~671B, close to a bright H$\alpha$ emission star.
                \item[Knot 5:]  \object{HH 672} in the \citet{Rei00b} catalogue, located $\sim3'$ southwest of IRAS.
            \end{description}

        \item[HH 488:]
             Chain of emission knots, in the east--west direction, south of IRAS 02238+7222, observed in \sii{}.
        \begin{itemize}
            \item  Exciting source not established up to now.
            \item Different nomenclatures of the knots, as reported by \citet{Nan02}, \citet{Mag03}, and in the \citet{Rei00b} catalogue (see Table \ref{HH488} for the cross-identifications based on the astrometry of the present work).
            \item  IR counterparts of HH 488 and J optical knots and new knots HH 488E to G without optical counterpart from \emph{Spitzer} images \citep{Kun16b}.
        \end{itemize}
        \item 
        Near-IR counterparts (H$_2$ 2.12 $\mu$m line) catalogued by \citet{Wal16} as MHO objects (see Table \ref{HH488}).
    \end{description}
 
\end{description}

\begin{figure}[htb]
\resizebox{\hsize}{!}{\includegraphics{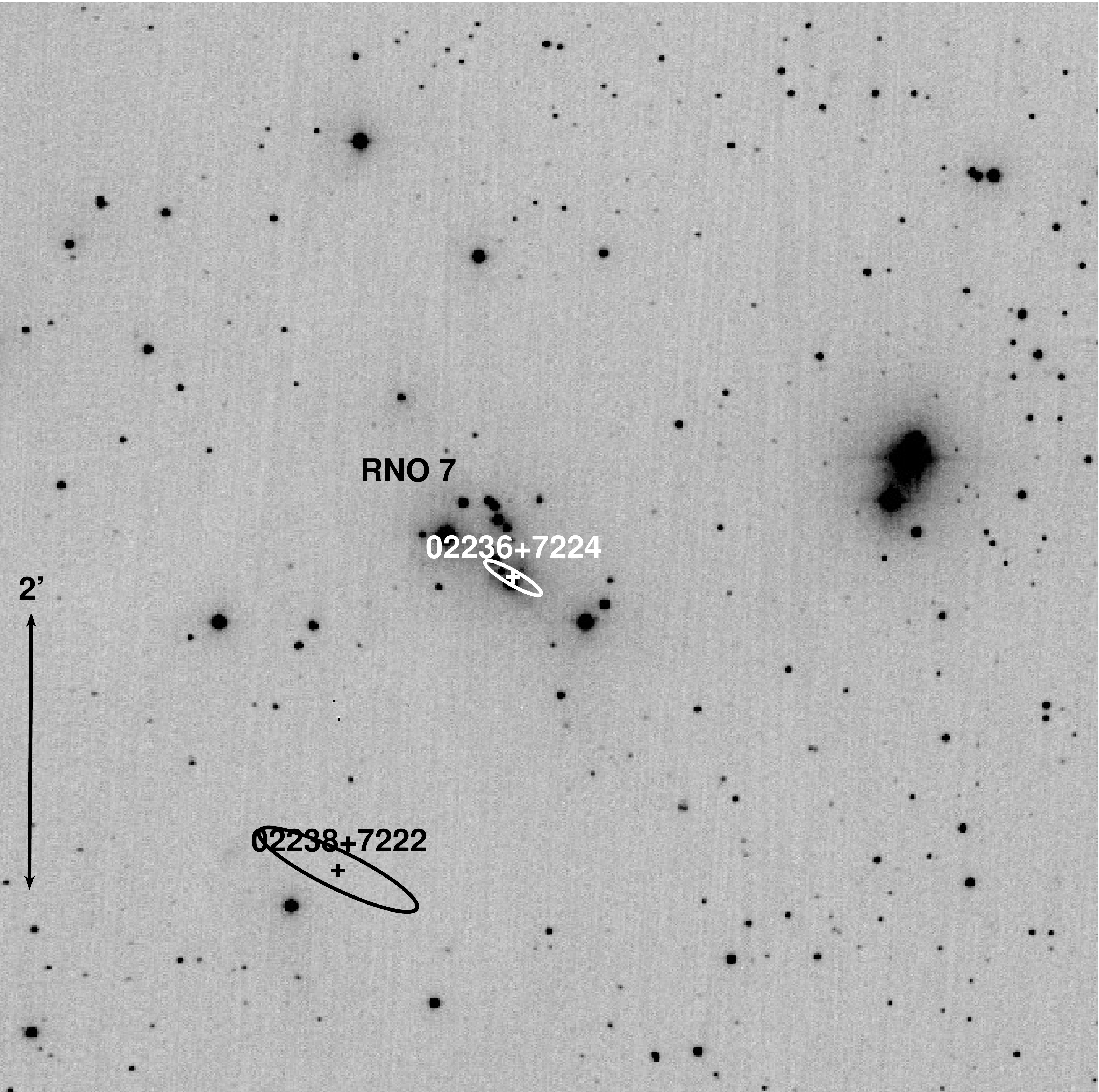}}
\caption{
IRAS 02236+7224: CAHA image of the field in the off-line filter. 
Another IRAS source (IRAS 02238+7222) lies inside the field mapped.
\label{28_cont}} 
\end{figure}

\begin{figure}[htb]
\resizebox{1.\hsize}{!}{\includegraphics{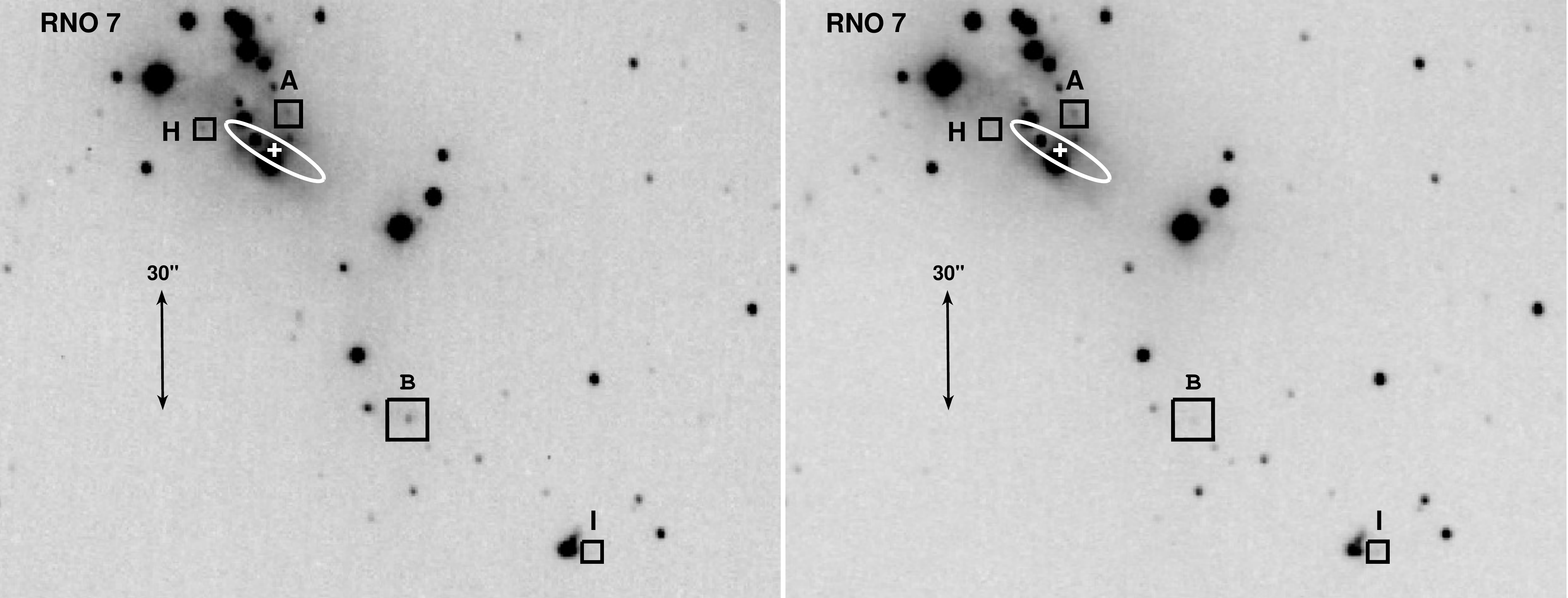}}
\caption{
IRAS 02236+7224: Close-up of the CAHA images through the H$\alpha$ (left) and \sii{} (right) filters. showing the field around the RNO 7 cluster, where the IRAS source is located. 
The position of the IRAS source and its error ellipse are marked. 
The knots are enclosed in boxes.
\label{28_line}} 
\end{figure}

\begin{figure}[htb]
\resizebox{\hsize}{!}{\includegraphics{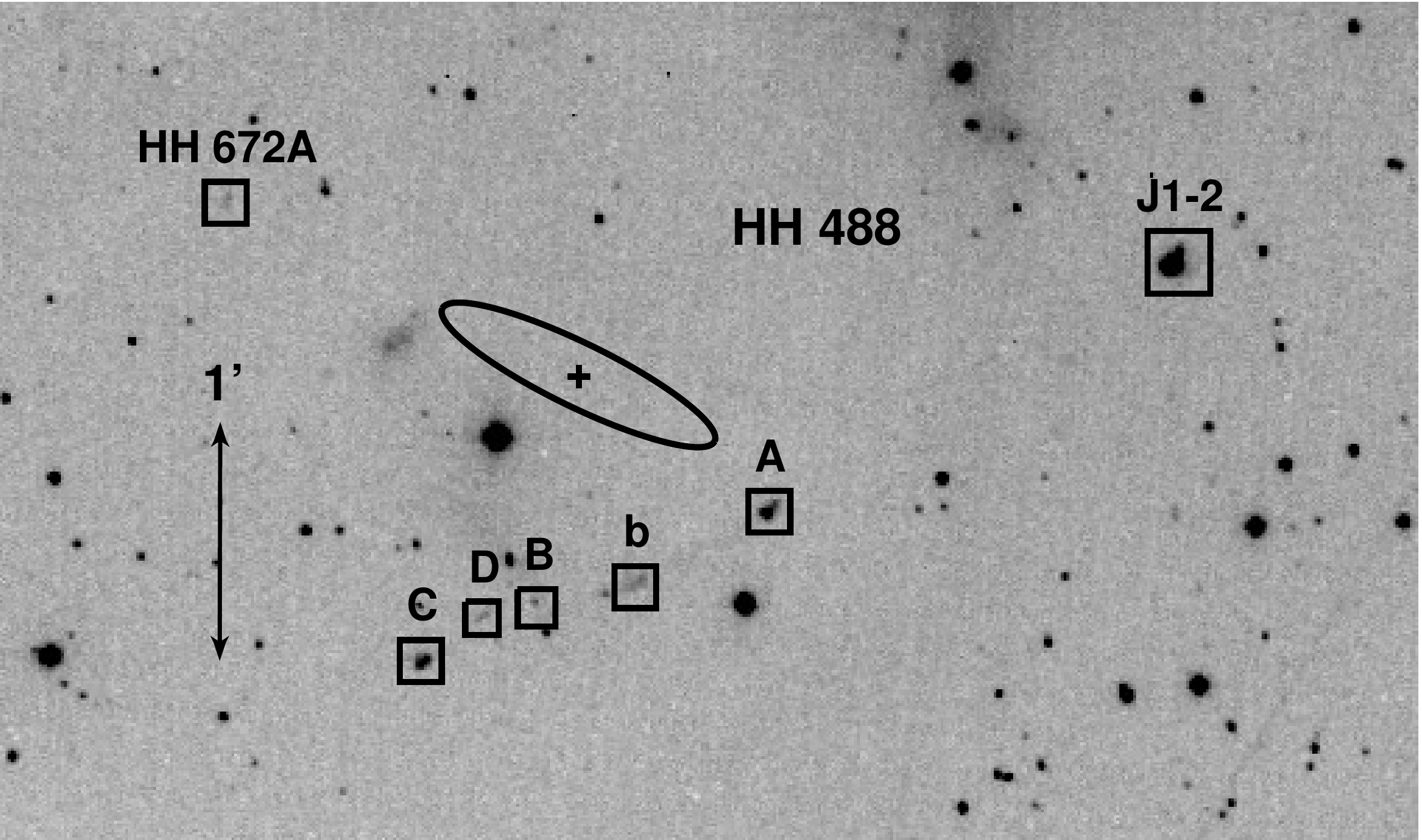}}
\resizebox{\hsize}{!}{\includegraphics{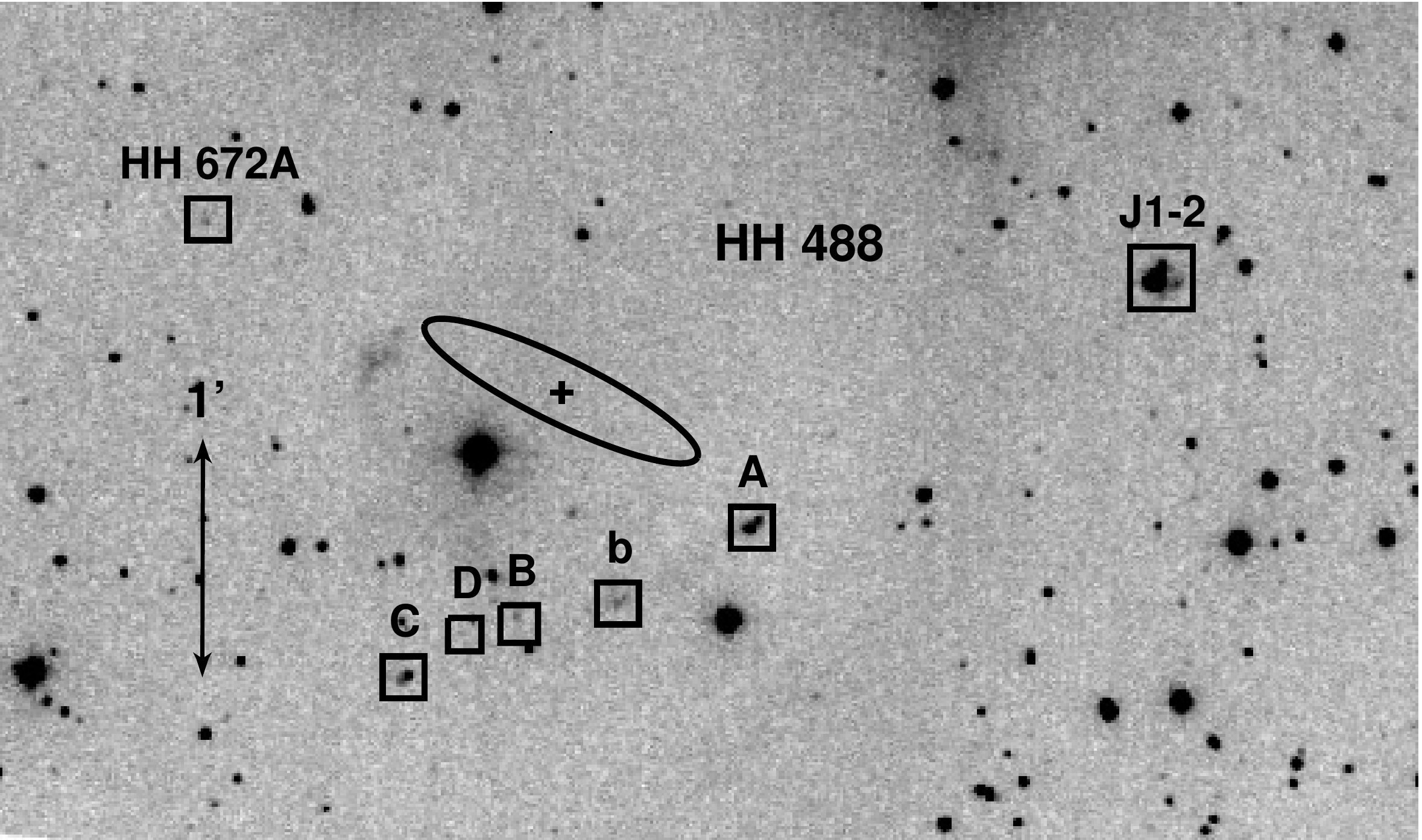}}
\caption{
HH 488: Close-up of the CAHA images in the H$\alpha$ (top) and \sii{} (bottom) filters of the chain of knots of HH 488, close to IRAS 02238+7222. 
The emission features identified are enclosed in boxes. 
\label{28_HH488}} 
\end{figure}

\begin{figure}[htb]
\resizebox{\hsize}{!}{\includegraphics{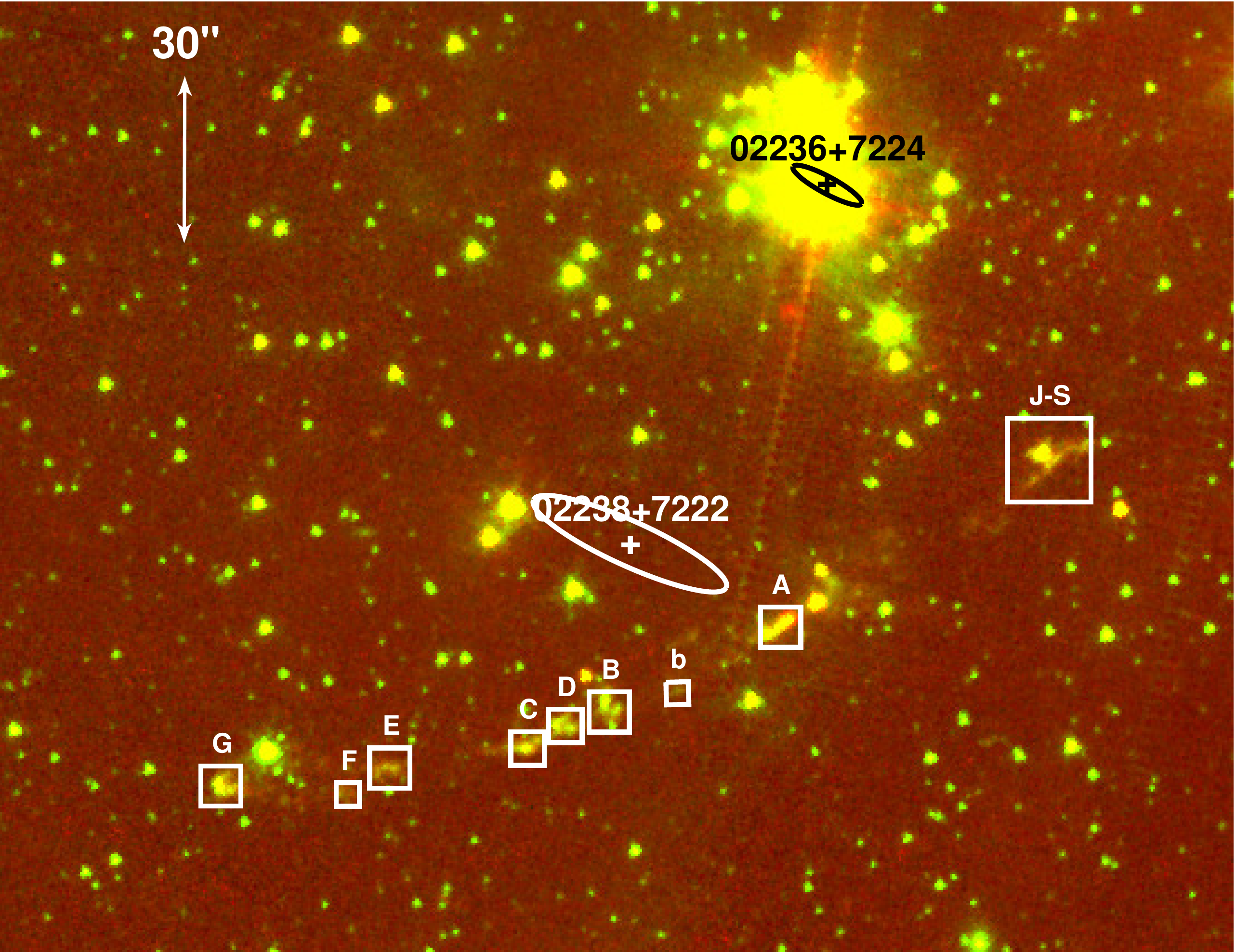}}
\caption{
IRAS 02236+7224:
Two-color image composed from \emph{Spitzer IRAC} 5.8~$\mu$m (red) and 3.6~$\mu$m (green) images with a FOV including the two IRAS sources and the HH~488 jet.
\label{28_nir}}
\end{figure}

\begin{table}[htb]
\centering
\caption{Emission line features in  the IRAS 02236+7224 field}
\label{HH488}
\begin{tabular}{llllll}  
\hline\hline
& \multicolumn{2}{c}{Position} \\
\cline{2-3}
& 
\multicolumn{1}{c}{$\alpha_{2000}$} & 
\multicolumn{1}{c}{$\delta_{2000}$} & 
\multicolumn{2}{c}{Other id.} & 
Near-IR\\
\cline{4-5}
Knot$^1$ & 
\multicolumn{1}{c}{($^\mathrm{h~m~s}$)} & 
\multicolumn{1}{c}{($^\circ$ $\arcmin$ $\arcsec$)} & 
(2) & 
(3) & 
MHO$^4$ \\
\hline
\multicolumn{5}{l}{HH 671} \\
\hline
A    & 02 28 15.65 & +72 37 45.3 & A        & 1        & 2932  \\
B    & 02 28 09.03 & +72 36 28.9 & B        & 2        & 2930  \\
\hline
\multicolumn{5}{l}{RNO7}\\
\hline
H    & 02 28 20.32 & +72 37 41.6 & $\cdots$ & $\cdots$ & 2932B \\
I    & 02 28 00.4\ & +72 35 56   & $\cdots$ & 3        & 2928D \\
\hline
\multicolumn{5}{l}{HH 672} \\
\hline
A    & 02 28 53.2  & +72 36 13   & $\cdots$ & 5        & 2936  \\
\hline
\multicolumn{5}{l}{HH 488} \\
\hline
J1-2 & 02 27 59.63 & +72 35 57.7 & J1-2     & $\cdots$ & 2928B,C\\
A    & 02 28 22.53 & +72 34 55.5 & A        & 4a       & 2928G  \\
b    & 02 28 30.01 & +72 34 36.6 & B        & 4b       & 2928J  \\
B    & 02 28 35.52 & +72 34 31.5 & $\cdots$ & $\cdots$ & 2928K  \\
D    & 02 28 38.56 & +72 34 28.9 & E        & 4d       & 2928L  \\
C    & 02 28 41.98 & +72 34 18.3 & C        & 4c       & 2928M  \\
E    & 02 28 52.98 & +72 34 12.5 & $\cdots$ & $\cdots$ & 2928N  \\
F    & 02 28 56.40 & +72 34 03.6 & $\cdots$ & $\cdots$ & 2928O  \\
G    & 02 29 06.51 & +72 34 08.1 & $\cdots$ & $\cdots$ & 2928P  \\
\hline
\end{tabular}
\tablebib{
(1) This work (see Figs.{} \ref{28_line}, \ref{28_HH488}, and \ref{28_nir});
(2) \citet{Rei00b, Nan02};
(3) \citet{Mag03};
(4) \citet{Wal16}. 
}
\end{table}

Figure \ref{28_cont} displays the image of the field around IRAS 02236+7224 mapped through the continuum filter. 
The RNO~7 cluster can be seen at the center of the image, and the location of the two IRAS sources are also marked.

Our narrow-band images allowed us to identify the optical knots HH~671A and B, the chain of knots HH~488A to D, the J complex northwest of IRAS~02238+7222, and the HH~672 knot, $\sim3'$ southeast of IRAS02236+7224. In addition, we detected a new knot $\sim20''$  west of HH~671A, labelled H in our images, which corresponds to the optical counterpart of  MHO 2932B.  

Figure \ref{28_line} shows a close-up of the field around the RNO~7 cluster of our H$\alpha$ and \sii{} images.
HH~671A and B  are detected in both lines, while knot H is only detected in H$\alpha$, and knot I only in \sii{}.
Based on geometrical arguments, its is not clear whether all these knots form part of a single stellar jet powered by  IRAS 02236+7224. 
Most probably HH~671A and knot H are tracing shocked emission produced by the stellar wind ejected from any of the YSOs of the cluster, at some places of the nebulosity surrounding the cluster.

Figure \ref{28_HH488} shows a close-up of the H$\alpha$ and \sii{} images 
of the southern part of the field around IRAS 02238+7222. They show the HH~672A knot, northwest of IRAS 02238+7222, the  
curved string of emission features lying from west to east, south of the IRAS source (HH~488), and the feature $\sim2'$ northwest of HH~488A 
labeled J1-2 in Fig.~\ref{28_HH488}.

Regarding the cross-identification of the optical and near-IR knots of HH~488, and the J complex, we show in Fig.\  \ref{28_nir} the color-composed image from the \emph{Spitzer}  archive, where these knots have been marked with the labels listed in the first column of Table \ref{HH488}.

We made accurate astrometry of all the knots detected in our images, necessary to determine the positions of the jet knots and the relationship between the optical and near-IR knot emissions in a consistent way, including the MHOs features reported by \citet{Wal16}. 
Table~\ref{HH488} lists the positions for all the HHs knots mentioned here. 
Since we found discrepancies in the identifications of several knots of the field, the table displays the knots identification of \citet{Nan02} and \citet{Mag03}, and the identification in the \citet{Rei00b} catalogue. 
The table also displays the H$_2$ MHO counterpart of each knot \citep{Wal16}. 


\begin{figure}[htb]
\resizebox{\hsize}{!}{\includegraphics{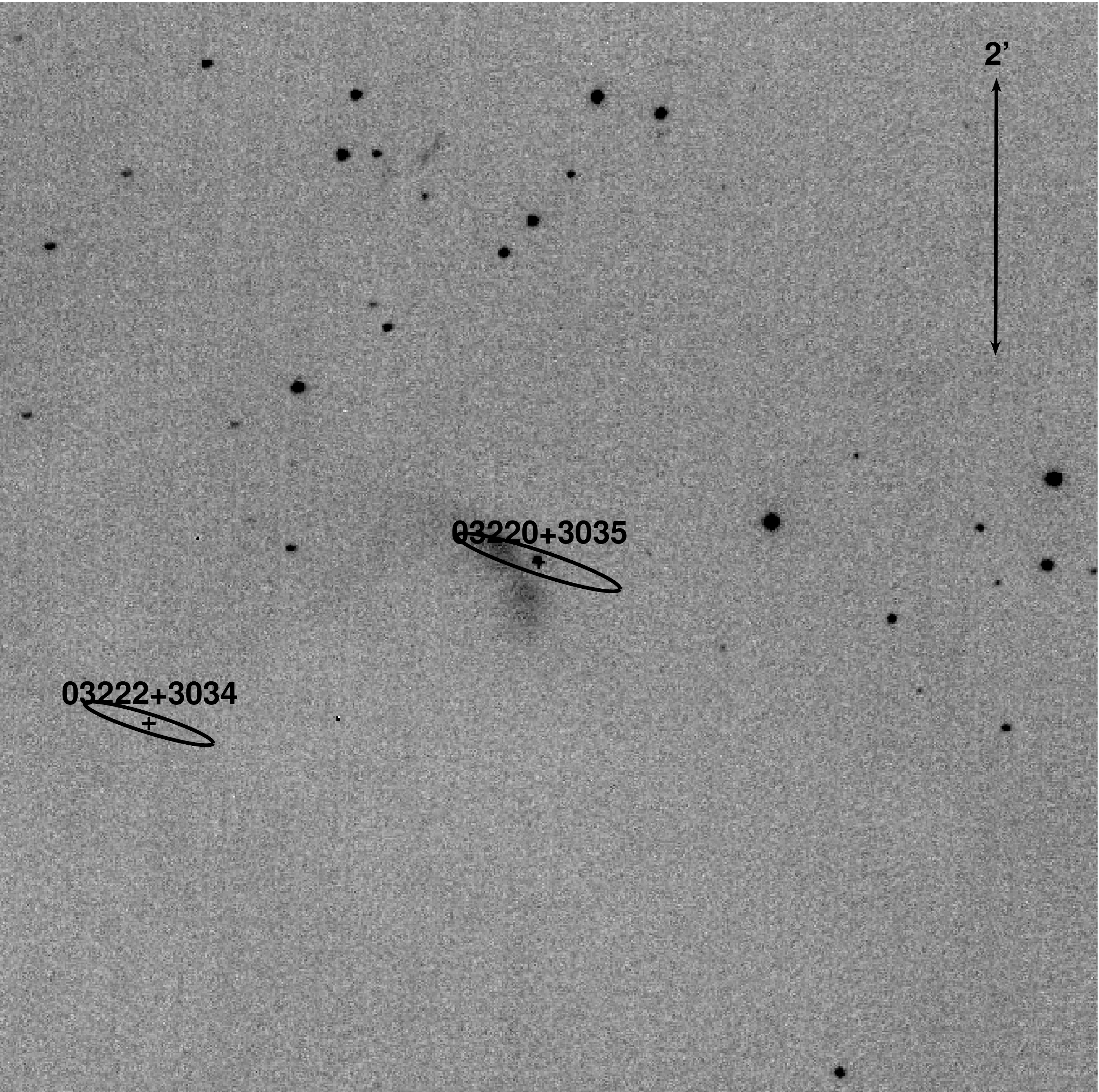}}
\caption{
IRAS 03220+3035: 
CAHA image in the continuum filter of the field. 
Another IRAS source (IRAS 03222+3034) lies inside the field mapped. 
The positions and error ellipses of the IRAS sources are marked.
\label{39_cont}} 
\end{figure}

\begin{figure}[htb]
\resizebox{\hsize}{!}{\includegraphics{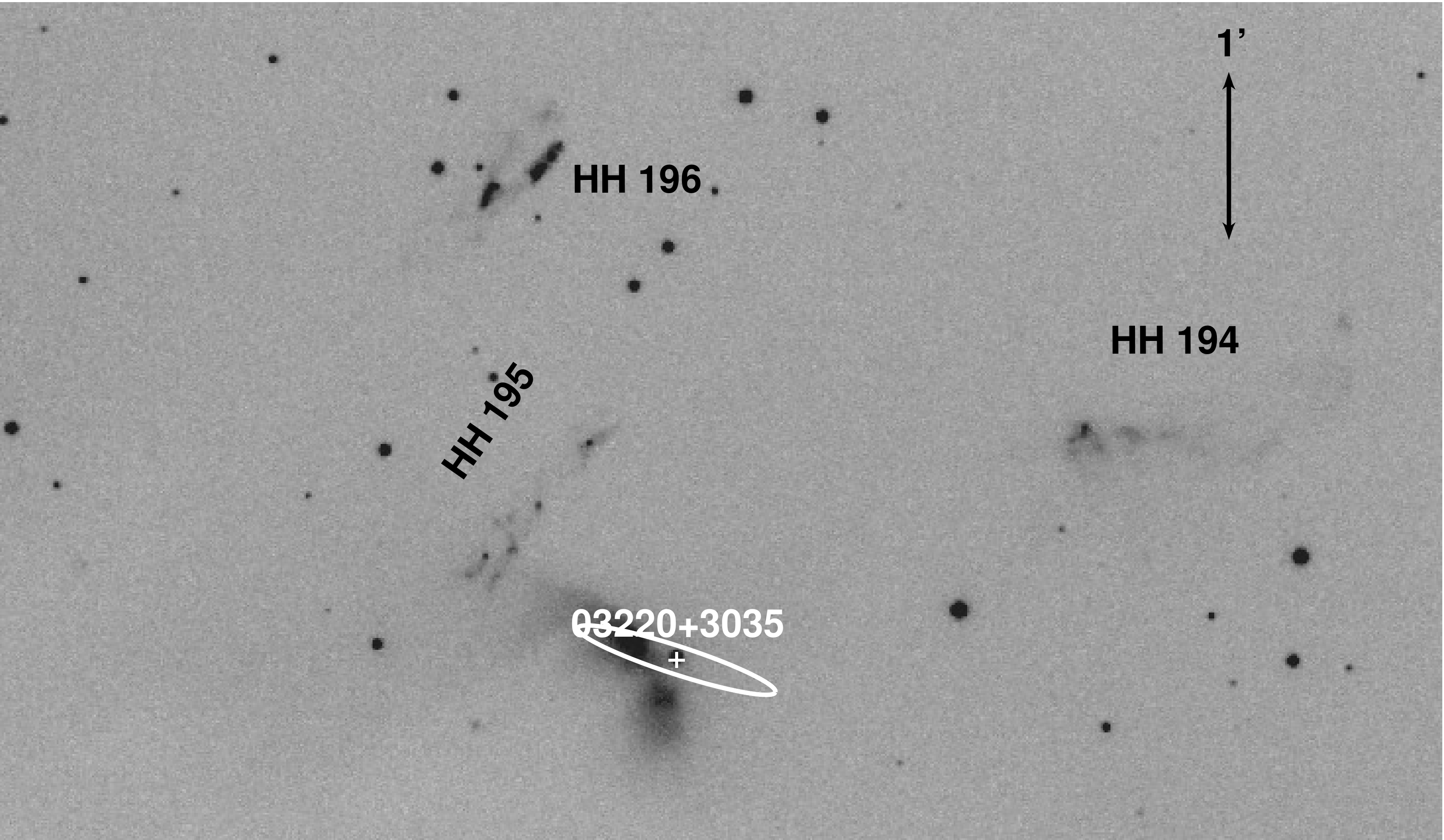}}
\resizebox{\hsize}{!}{\includegraphics{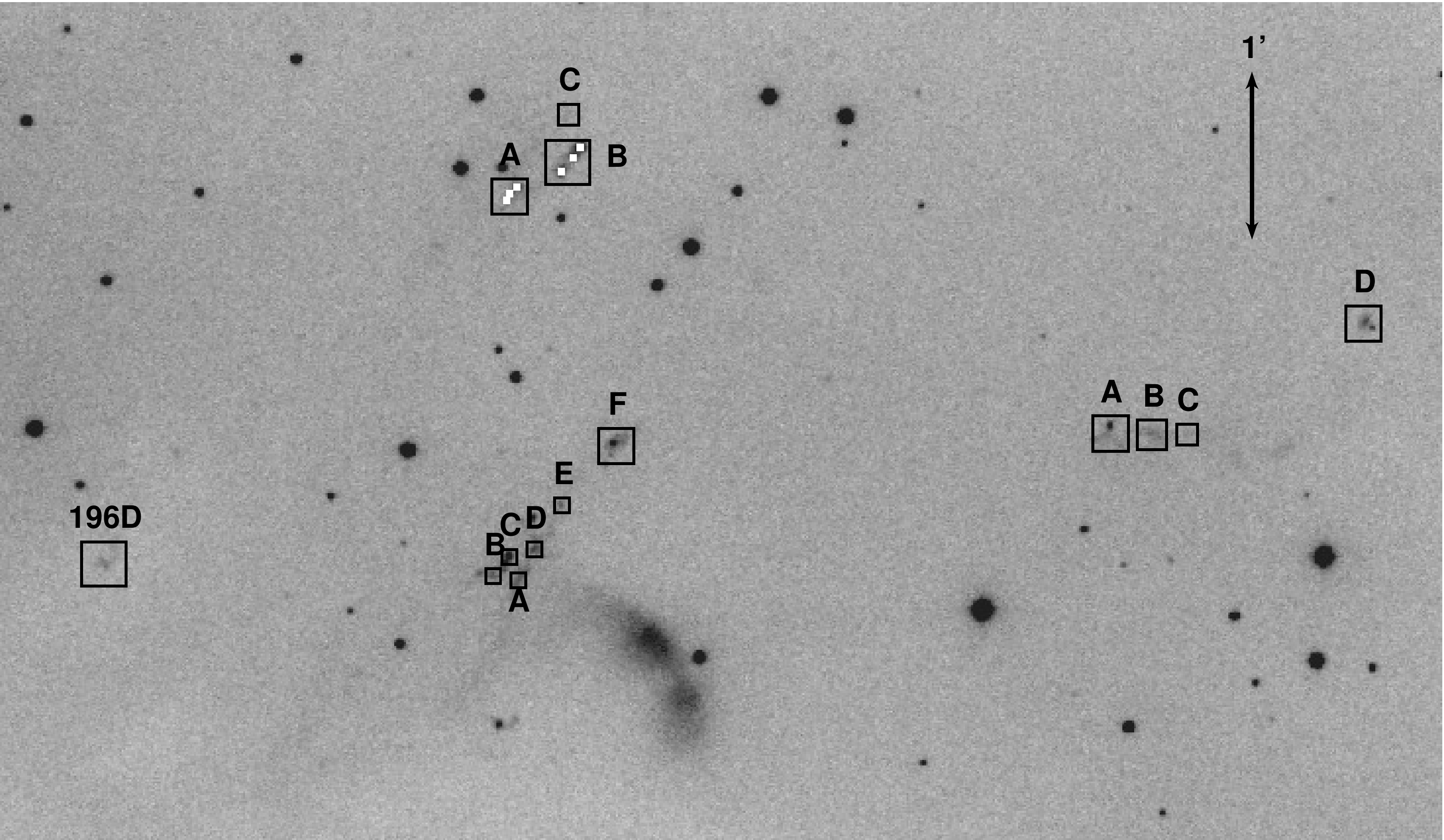}}
\caption{
IRAS 03220+3035: 
Close-up of the CAHA image through the H$\alpha$ filter \emph{(top)} and \sii{} filter \emph{(bottom)}, showing the field around the IRAS source. 
The HH objects detected (194, 195, 196) are identified \emph{(top)}, and the knots listed in Table~\ref{39_knots_pos} are labeled, enclosed in boxes \emph{(bottom)}.
\label{39_lines}}
\end{figure}

\subsubsection{\object{IRAS 03220+3035}}

\begin{table}[htb]
\centering
\caption[ ]{HH knot positions of the IRAS 03220+3035 field}
\label{39_knots_pos}
\begin{tabular}{lcc}  
\hline\hline
& 
\multicolumn{2}{c}{Position} \\
\cline{2-3}
& 
$\alpha_{2000}$ & 
$\delta_{2000}$ \\
Knot & 
($^\mathrm{h~m~s}$) & 
($^\circ$ $\arcmin$ $\arcsec$) \\
\hline
\multicolumn{3}{l}{HH 194} \\
\hline
A    & 03 24 58.0  &+30 47 41\\
B    & 03 24 56.9  &+30 47 41\\
C    & 03 24 55.9  &+30 47 41\\
D    & 03 24 51.0  &+30 48 20\\
\hline
\multicolumn{3}{l}{HH 195} \\
\hline
A    & 03 25 14.4  &+30 48 49\\
B    & 03 25 15.1  &+30 46 51\\
C    & 03 25 14.7  &+30 46 58\\
D    & 03 25 14.0  &+30 47 00\\
E    & 03 25 13.2  &+30 47 16\\
F    & 03 25 11.7  &+30 47 37\\
\hline
\multicolumn{3}{l}{HH 196} \\
\hline
A1   & 03 25 14.7  &+30 49 05\\
A2   & 03 25 14.6  &+30 49 07\\
A3   & 03 25 14.4  &+30 49 10\\
B1   & 03 25 13.2  &+30 49 15\\
B2   & 03 25 12.9  &+30 49 20\\
B3   & 03 25 12.7  &+30 49 24\\
C    & 03 25 13.0  &+30 49 36\\
D    & 03 25 25.8  &+30 46 56\\
\hline
\end{tabular}
\end{table}

IRAS 03220+3035 is located in \object{L1448}, in the Perseus cloud complex, at a distance of $240\pm12$~pc \citep{Zuc20}. In the following we present a short description of the field around the source.
\begin{description}

    \item[\textit{Proposed association of IRAS with other objects:}]
    \begin{description}\item[]
        \item[\object{RNO~13}:] Red reflection nebula \citep{Coh80}, and
        \item[\object{L1448~IRS1}:] IR source \citep{Hod94}.
    \end{description}
  
    \item[\textit{Other IRAS in the region:}] 
    \begin{description}\item[]
        \item[\object{IRAS 03222+3034},] southeast of IRAS 03220+3035.
        \begin{itemize}
            \item Associated with \object{L1448 IRS2}.
            \item Very embedded, with no optical nor near-IR counterpart.
            \item Classified as a Class 0 protostar, from far-IR and submillimeter data \citep{Oli99}.
        \end{itemize}
        \item[\object{IRAS 03225+3034 IRS3},] associated with YSOs \citep{Ang89}, but outside the field mapped.
    \end{description}
    
    \item[\textit{Binarity:}]
        L1448~IRS1 is a close ($1\farcs37$) binary system \citep[$L$ band images of ][]{Con08}.
        Characteristics of each component from near-IR spectroscopy \citep{Con10}:
    \begin{itemize}
        \item Northern component: line emission characteristic of a cavity or a young stellar jet very close to the exciting source.
        \item Southern component: line emission compatible with a weak TTauri star, thus in an older evolutionary stage than the northern component.
    \end{itemize}
    
    \item[\textit{Molecular outflows:}]
    Low-velocity CO outflow, with blue\-shifted emission east of the source \citep{Lev88}.

    \item[\textit{Optical outflows/Herbig-Haro objects:}]~
    \begin{description}
        \item[\object{HH 194}:] H$\alpha$+\sii{} emission \citep{Bal97}.
        \item[\object{HH 195},] \object{HH 196}: 
        H$\alpha$+\sii{} \citep{Bal97} and \hmol{} emission \citep{Eis00}.
    \end{description}
\end{description}

Figure~\ref{39_cont} shows the wide field mapped in the continuum filter. The location of the IRAS sources included in the field are marked.
Figure~\ref{39_lines} is a close-up of the field mapped through the H$\alpha$ and \sii{} narrow-band filters.
The images show the location of HH 194, HH 195, and HH 196 (H$\alpha$ image), and the knots identified in our \sii{} image, which have been labeled beginning with A, the knot closest to the proposed driving source.

None of these HH objects is aligned with IRAS 03222+3034 (our target), so there is no clear geometric argument to associate the driving source of any of the HH objects with the IRAS source.

Let us discuss each HH object in more detail.
\begin{description}
\item[HH 194:]
Its knots are aligned along the northern edge of the east-west CO outflow \citep{Lev88}, most probably powered by the IRAS source. 
Thus, HH 194 can be tracing the cavity wall of the CO outflow and is probably driven by IRAS 03222+3034.
We identified four knots (labeled A to D in the \sii{} image). 
Three knots, A, B, and C, are nearly aligned in the east-west direction, have at bow-shaped morphology with the apex pointing northwards, and are brighter in H$\alpha$ than in \sii{}. 
In addition, there is an arc-like diffuse emission beyond knot C, ending $\sim1'$ northwest of HH 194C in a compact knot, D, brighter in \sii{} than in H$\alpha$, which  probably forms part of the same outflow. 
\item[HH 195:]
Consists of a ``V''-shaped string of knots located $\sim1'$ northeast of IRAS 03220+3035, pointing toward IRAS 03222+3034 (see Fig.\ \ref{39_lines}). It coincides with the northern knots of the \hmol{} outflow centered on IRAS 03222+3034 \citep{Eis00}.
Thus, IRAS 03222+3034 is most probably the driving source of HH~195.
The HH 195 knots are brighter in \sii{} than in H$\alpha$. 
\item[HH 196:]
Consists of an arc-shaped string of knots $\sim3'$ northeast of IRAS 03220+3035, with the apex oriented toward IRAS 03225+3034, outside the FOV of our images,
which has been  proposed as the driving source of the outflow \citep{Bal97}. 
Our images also show HH 196D, $\sim~3\farcm2$ southeast of HH 196A, in the direction of its apex,
identified as a part of the same optical outflow \citep{Bal97}, and reported in the near-IR too \citep{Eis00}. 
Our images show that both HH 196A and B could be resolved into several compact substructures engulfed in nebular emission. 
We labeled these knotty substructures by adding a number to the letter identifying the knot.
\end{description}
Astrometry of the HHs 194, 195 and 196 knots, including the newly identified substructures, is given in Table~\ref{39_knots_pos}.


\subsubsection{\object{IRAS 04073+3800}}

IRAS 04073+3800 is located in the dark cloud \object{L1473} in the Perseus complex, at a distance of $\sim350$ pc \citep{Coh83}. In the following we present a short description of the field around the source.
\begin{description}

    \item[\textit{Proposed association of IRAS with other objects:}]
    \object{PP~13}, an optical red nebula with a position closely matching that of IRAS \citep{Par79}.

    \item[\textit{Young stellar objects:}]
    The PP~13 nebula consists of two pre-main-sequence stars, 
    classified from optical, near-IR and submillimeter data \citep{Asp01}: 
    \begin{description}
        \item[PP~13N,] a TTauri star, 
        \item[PP~13S,] a variable FU Ori star, also detected at 2.2 $\mu$m  \citep{Con07}.
    \end{description}
    
    \item[\textit{Optical outflows/Herbig-Haro objects:}]
    Three chains of HH objects around IRAS 04073+3800 \citep{Asp00},
    also detected at 2.2 $\mu$m  \citep{Con07}:
    \begin{description}
        \item[\object{HH 463}:] a cometary-like emission $\sim10\arcsec$ long, arising from PP~13S southwestward, with five \sii{} knots (A to E), plus a compact isolated \sii{} emission at $\sim1'$ in the direction of the jet axis
        \item[\object{HH 464}:] a chain of five knots (A to E) along a curved path in the north--south direction, beginning east of PP~13S and pointing toward the PP~13N tail.
        \item[\object{HH 465}:] an isolated knot north of  PP~13N. 
    \end{description}
    HH 464 and HH 465 form a bipolar jet driven by PP~13N \citep{Asp00}. 

\end{description}

\begin{figure}[htb]
\resizebox{\hsize}{!}{\includegraphics{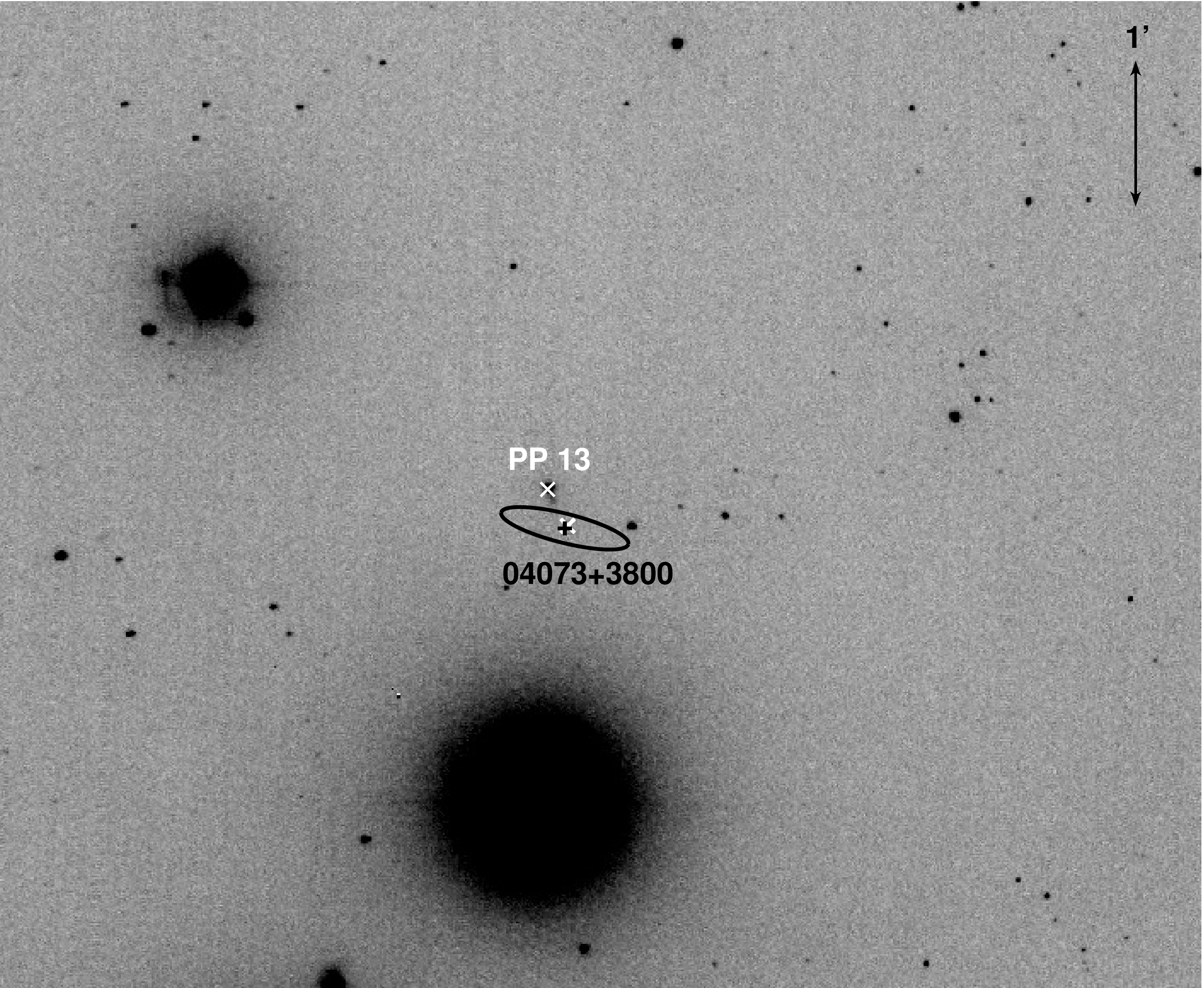}}
\caption{
IRAS 04073+3800: 
CAHA image through the continuum filter. 
The IRAS source position and its error ellipse are shown. 
The positions of the reflection nebulae PP~13N and S have been marked with white ``$\times$''. \label{51_cont}} 
\end{figure}

\begin{figure}[htb]
\resizebox{\hsize}{!}{\includegraphics{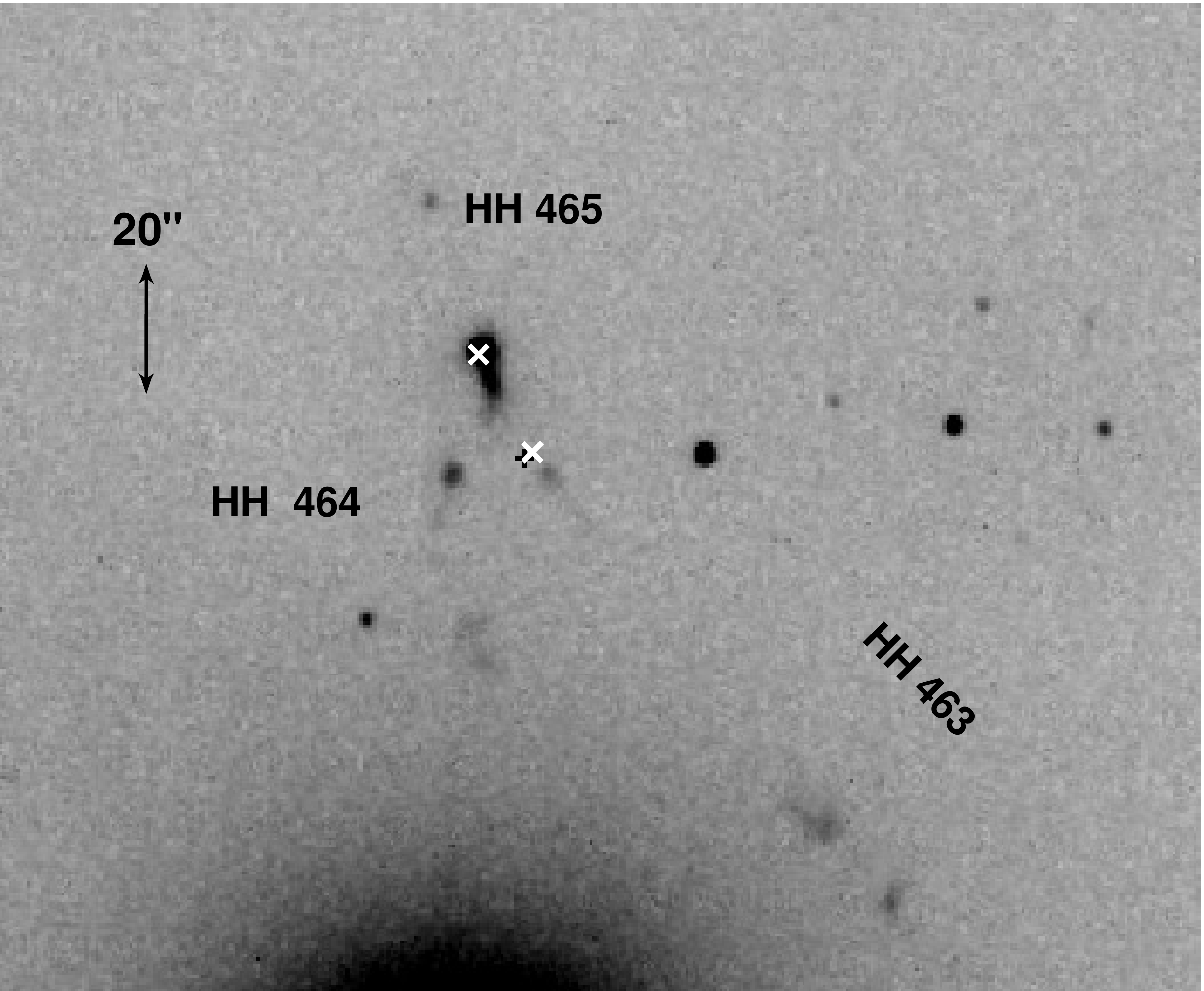}}
\resizebox{\hsize}{!}{\includegraphics{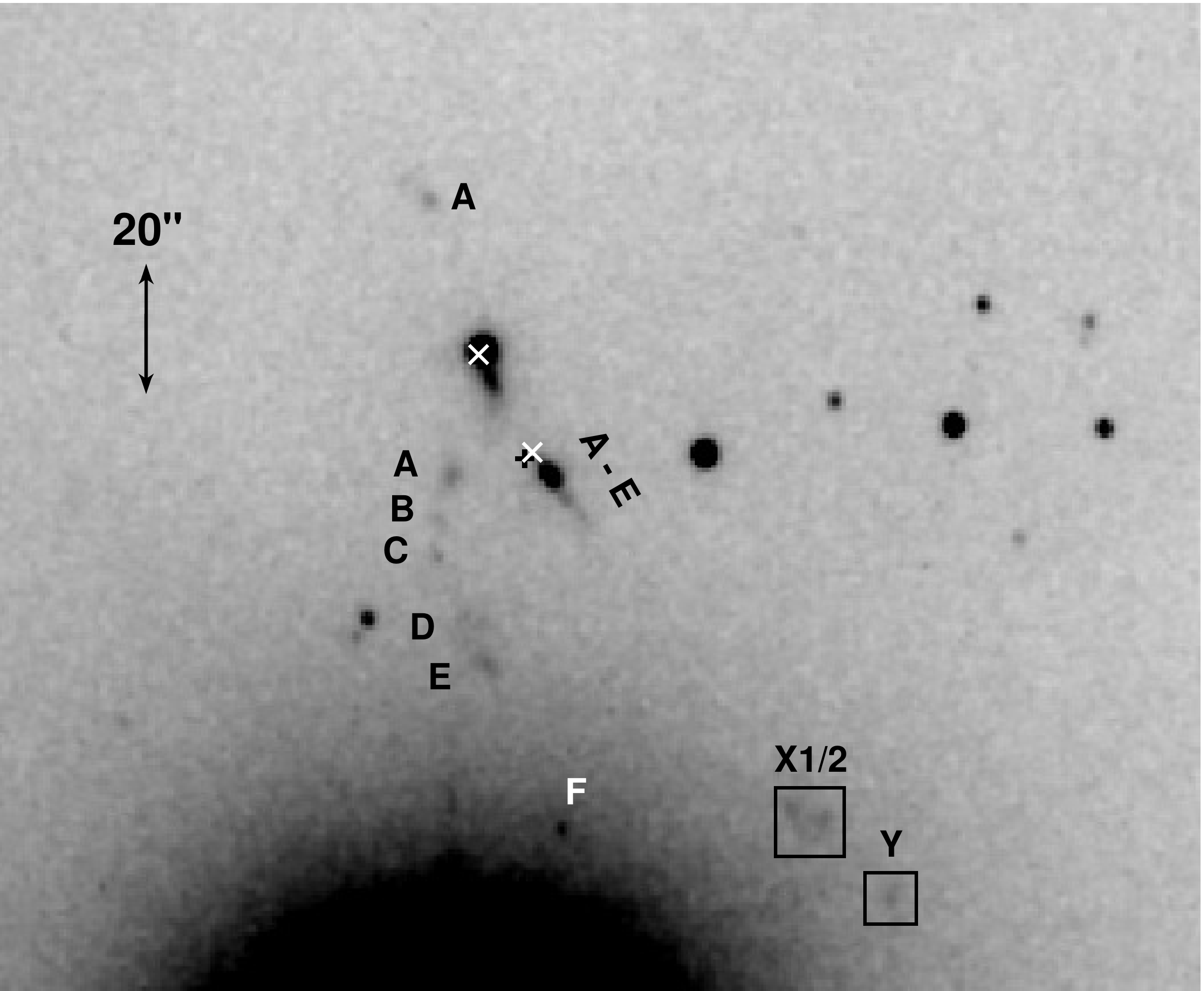}}
\caption{
IRAS 04073+3800:
Close-up of the CAHA image through the H$\alpha$ filter \emph{(top)} and \sii{} filter \emph{(bottom)}, showing the field around the IRAS source. 
The HH objects detected (463, 464, 465) are identified \emph{(top)}, and the knots listed in Table~\ref{51_knots_op} are labeled \emph{(bottom)}.
The positions of the two red nebulous objects (PP~13N and S) are marked as in Fig.~\ref{51_cont}.
\label{51_lines}}
\end{figure}

\begin{figure}[htb]
\resizebox{\hsize}{!}{\includegraphics{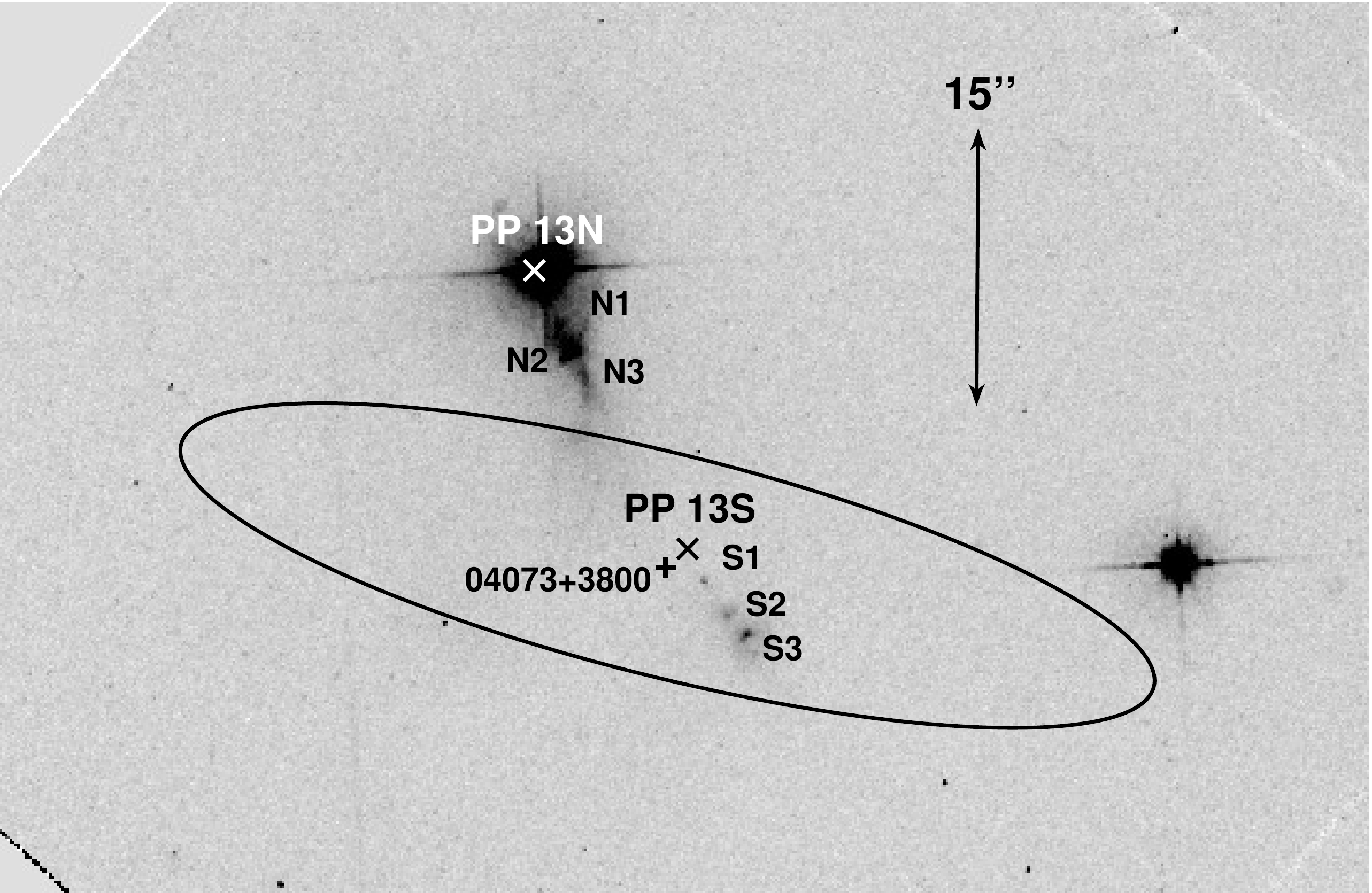}}
\caption{
IRAS 04073+3800: 
\emph{HST} image of the field. 
The position of the IRAS source and its error ellipse are marked. 
The positions of the reflection nebulae PP~13N and S are marked with a ``$\times$''. 
\label{51_HST}} 
\end{figure}

\begin{table}[htb]
\centering
\caption{HH objects in the IRAS 04073+3800 field}
\label{51_knots_op}
\begin{tabular}{lcc}  
\hline\hline
& 
\multicolumn{2}{c}{Position} \\
\cline{2-3}
& 
$\alpha_{2000}$ & 
$\delta_{2000}$  \\
Knot & 
($^\mathrm{h~m~s}$) & 
{($^\circ$ $\arcmin$ $\arcsec$)}\\
\hline
\multicolumn{3}{l}{HH 463} \\
\hline
A    & 04 10 41.1 &+38 07 53\\
B    & 04 10 40.8 &+38 07 51\\
C/D  & 04 10 40.7 &+38 07 47\\
E    & 04 10 40.5 &+38 07 45\\
X1   & 04 10 37.7 &+38 07 00\\
X2   & 04 10 37.3 &+38 06 59\\
Y    & 04 10 36.5 &+38 06 46\\
\hline
\multicolumn{3}{l}{HH 464} \\
\hline
A    & 04 10 42.2 &+38 07 51\\
B    & 04 10 42.3 &+38 07 44\\
C    & 04 10 42.4 &+38 07 39\\
D    & 04 10 42.0 &+38 07 29\\
E    & 04 10 41.7 &+38 07 22\\
F    & 04 10 40.7 &+38 06 58\\
\hline
\multicolumn{3}{l}{HH 465} \\
\hline
A    & 04 10 42.5 &+38 08 34\\
\hline
\end{tabular}
\end{table}

\begin{table}[htb]
\centering
\caption{\emph{HST} knots in the IRAS 04073+3800 field}
\label{51_HST_red_pos}
\begin{tabular}{lccc}  
\hline\hline
&
\multicolumn{2}{c}{Position} \\
\cline{2-3}
& 
$\alpha_{2000}$ & 
$\delta_{2000}$  \\
Knot & ($^\mathrm{h~m~s}$) & 
{($^\circ$ $\arcmin$ $\arcsec$)} &
 Identif. \\
\hline
N1    & 04 10 41.69  & +38 08 06.7 \\
N2    & 04 10 41.63  & +38 08 05.4 \\
N3    & 04 10 41.57  & +38 08 04.1 \\
S1    & 04 10 41.02  & +38 07 53.2 & HH 463A \\
S2    & 04 10 40.92  & +38 07 51.4 & HH 463B \\
S3    & 04 10 40.83  & +38 07 50.3 & HH 463B \\
\hline
\end{tabular}
\end{table}

Fig.\ \ref{51_cont} displays the image of the IRAS~04073+3800 mapped through the continuum filter, where the location of the red nebulae and the IRAS source are marked. 
As can be seen in the figure, PP 13S is most probably the counterpart of IRAS 04073+3800 because their positions are nearly coincident, while PP 13N does not seem to be related to the IRAS source because it lies outside its error ellipse.

A close-up of the field through the H$\alpha$ and \sii{} line filters is shown in Fig.~\ref{51_lines}. 
The three HH jets (HH~463, 464, 465) are identified in the H$\alpha$ image.
Several knots have been identified in the jets and are labeled in the \sii{} image according to previous \sii{} observations of \citet{Asp00}. 
The astrometry of the knots of the \sii{} image is given in Table \ref{51_knots_op}.
As can be seen in Fig.\ \ref{51_lines}, most of the knots are brighter in \sii{} than in H$\alpha$. 
The exceptions are knot A of HH~464, the knot closest to the exciting source, and 
the line emission features X1/2 and Y, associated with HH~463, which are brighter in H$\alpha$ than in \sii{}. 
In the case of HH 463X1/2 and Y, their position far away from the exciting source and their disordered morphology suggest that they may be tracing bow shocks from older mass-ejection episodes.

The field close to IRAS~04073+3800 was imaged by the \emph{HST} with the WPC2 camera through the F814W filter (PI D. Padgett. Program ID 9160. Cycle 10) and with the NIC2 camera through the filter F110W, F160W and F205W (PI D. Padgett. Program ID 10603. Cycle 14). 
Figure~\ref{51_HST} shows a close-up of the F814W image.

We identified three knotty structures (N1, N2, N3) south of PP~13N at a $\mathrm{PA}\simeq25\degr$, and another three knots (S1, S2, S3) south of PP~13S at a $\mathrm{PA}\simeq35\degr$. 
The astrometry of these knots, derived from the \emph{HST} image, is given in Table~\ref{51_HST_red_pos}. 
The S knots are part of the HH~463 jet. 
Given their positions, we could identify S1 with HH~463A, while S2 and S3 correspond to HH~463B, resolved in different substructures because of the better resolution of the \emph{HST} image. 

Regarding the N knots, they are well aligned with HH~465A ($\mathrm{PA}\simeq20\degr$), so they could be part of the jet/counterjet system powered by PP~13N. 
However, they are also aligned with the S knots of the HH~463 jet, which shows a curved morphology at large scales. 
Thus, they could also trace the counterjet of HH~463, projected onto the PP~13N nebula.


\begin{figure}[htb]
\resizebox{\hsize}{!}{\includegraphics{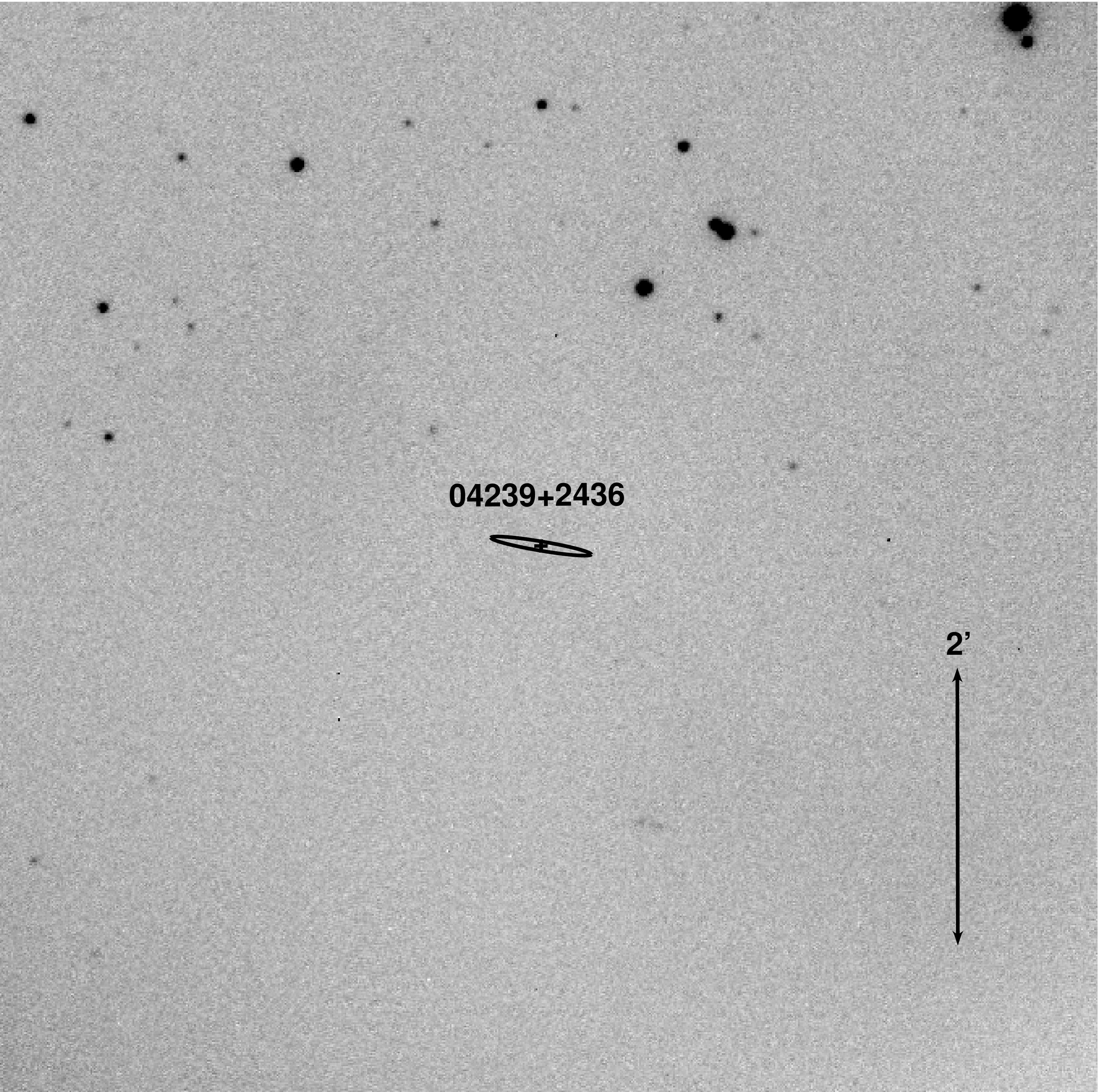}}
\caption{
IRAS 04239+2436:
CAHA image through the continuum filter. 
The IRAS source position and error ellipse are shown.  
\label{66_cont}} 
\end{figure}

\begin{figure}[htb]
\resizebox{\hsize}{!}{\includegraphics{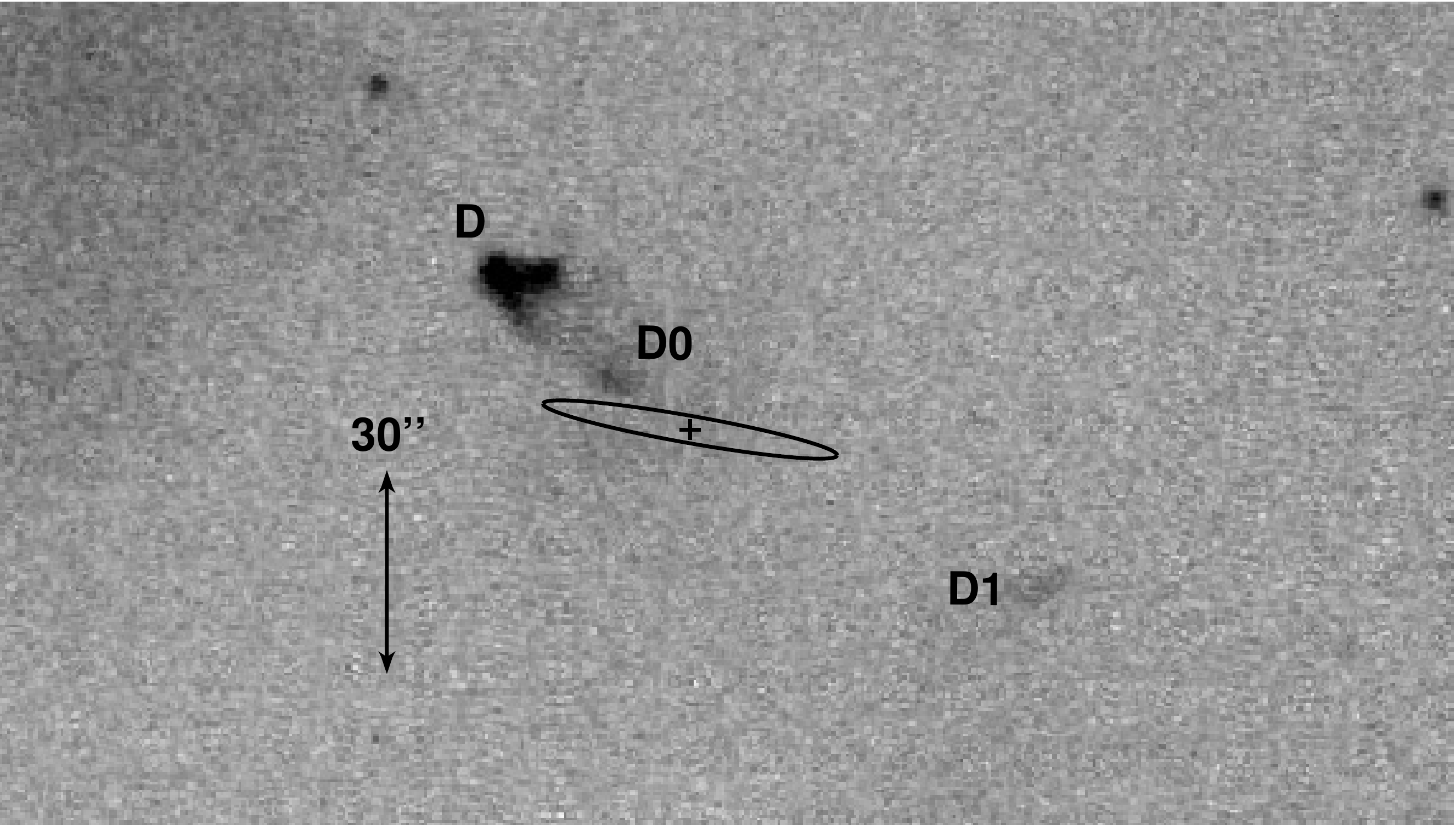}}
\resizebox{\hsize}{!}{\includegraphics{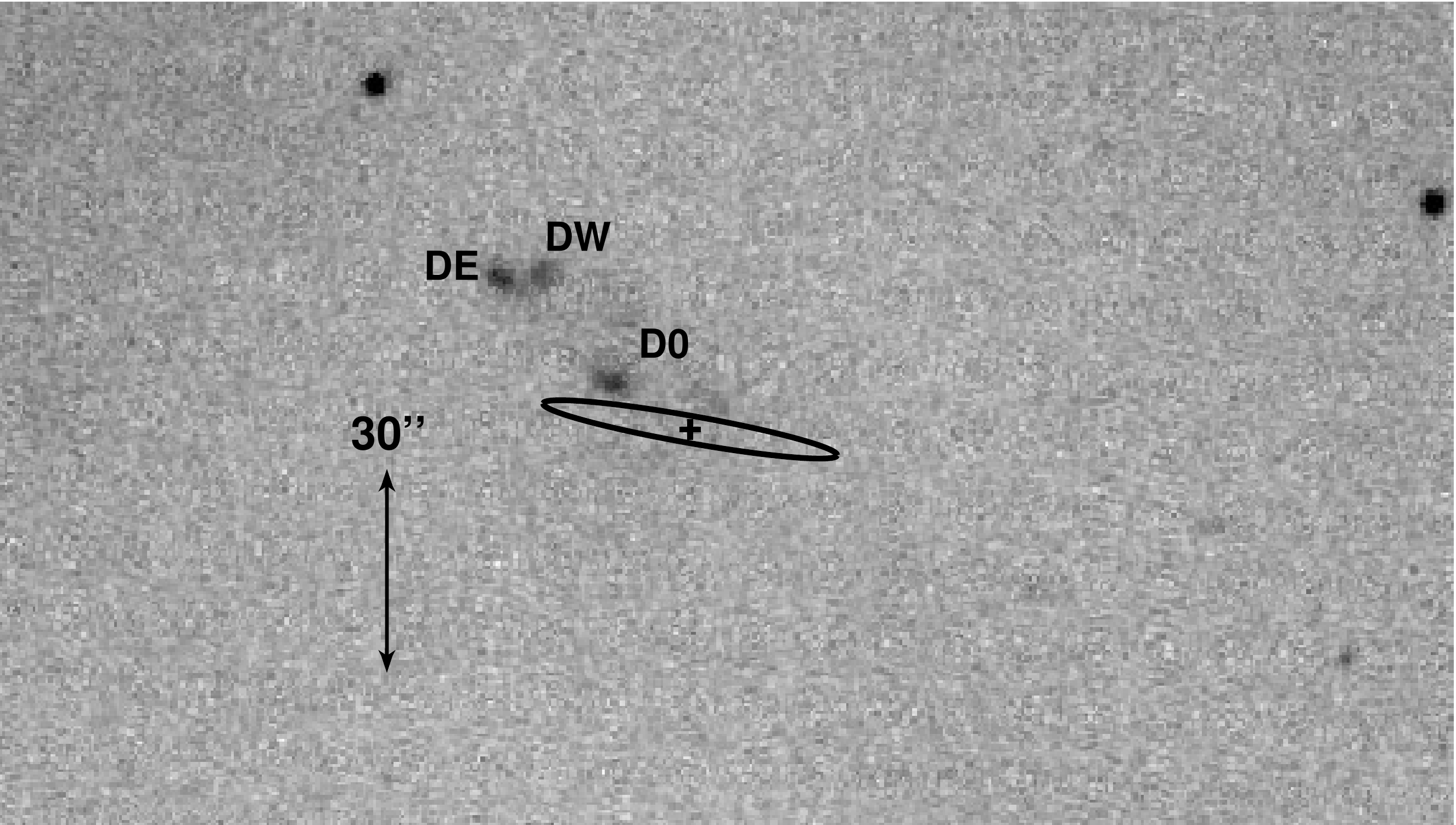}}
\caption{
IRAS 04239+2436:
Close-up of the CAHA image through the H$\alpha$ filter \emph{(Top)} and through the \sii{} filter \emph{(Bottom)}. 
\label{66_lines}}
\end{figure}

\subsubsection{\object{IRAS 04239+2436}}

IRAS 04239+2436 is located in the \object{B18} cloud of Taurus,  at a distance of $129.0\pm0.8$~pc \citep{Gal19}.
In the following we present a short description of the field around the source.
\begin{description}

    \item[\textit{Classification:}]
      Low-luminosity Class I protostar, from its near-IR spectrum \citep{Gre96}.
     
    \item[\textit{Binarity:}]
      Close binary (separation $0\farcs3$, 42 au in projection) \citep{Rei00a}.
     
    \item[\textit{Molecular outflows:}]
    CO outflow driven by IRAS \citep{Mor92}.
     
    \item[\textit{Optical outflows/Herbig-Haro objects:}]
    Giant HH bipolar outflow \object{HH 300} \citep{Rei97}, driven by IRAS.
    \begin{description}
        \item[HH 300A, B and C:] redshifted, bright H$\alpha$+\sii{} knots with a bow-shock morphology, $\sim30'$ southwest of IRAS.
        \item[HH 300D:] blueshifted, compact and fainter knot, $\sim30''$ northeast of IRAS.
    \end{description}
    
    \item[\textit{Near-IR emission:}]~
    \begin{itemize}
        \item Cometary nebula surrounding the system (\emph{HST}  \citep[NICMOS F160W and F205W;][]{Rei00a}
      \item Jet emission in \fe{} collimated and bipolar \citep{Dav11}, with axis coincident with the optical HH 300A--C knots axis \citep{Rei00a}.
      \item Jet emission in H$_2$ collimated and bipolar \citep{Dav11}.
      \item Jet emission in Br$\gamma$ isotropical \citep{Dav11}.
    \end{itemize}

\end{description}

\begin{table}[htb]
\centering
\caption{HH 300 knots in the IRAS 04239+2626 field}
\label{knots}
\begin{tabular}{lcc}  
\hline\hline
&\multicolumn{2}{c}{Position}\\
\cline{2-3}
HH 300 & 
$\alpha_{2000}$ & 
$\delta_{2000}$  \\
Knot & 
($^\mathrm{h~m~s}$)  &
{($^\circ$ $\arcmin$ $\arcsec$)} \\
\hline
D1  & 04 26 53.4 &+24 43 12\\
D   & 04 26 59.2 &+24 43 59\\
DE  & 04 26 59.1 &+24 43 58\\
DW  & 04 26 58.6 &+24 43 59\\
D0  & 04 26 57.9 &+24 43 42\\
\hline
\end{tabular}
\end{table}

The field of our images centered on IRAS 04239+2436 did not include the bright redshifted HH 300A, B and C knots. 
Figure~\ref{66_cont} shows the field imaged in the continuum filter with the location of the IRAS source. 
Figure~\ref{66_lines} shows a close-up of the field surrounding IRAS~04239+2436 through the H$\alpha$ and \sii{} line filters. 
As can be seen in the figure we detected emission in the two lines from two knots of the blueshifted HH 300 jet. 
In the H$\alpha$+\sii{} image of \citet{Rei97}, HH~300D shows a conical shape, reminiscent of a bow-shock. 
Our images revealed  that the morphology of knot D changes from H$\alpha$ to \sii{}. 
In the H$\alpha$ line, knot D shows the bow-shock shape reported before, while in the \sii{} line knot D appears split in two (labeled DE and DW in our \sii{} image): 
knot DE coincides with the apex of the H$\alpha$ emission of knot D; and 
knot DW coincides with the northern tip of knot D. 
A new knot, D0, $\sim15''$ northeast of the IRAS position was found in the two lines. 
Finally, we found knotty, faint H$\alpha$ emission without \sii{} counterpart $\sim55''$  southwest of the IRAS position (labeled D1) most probably corresponding to the redshifted jet component. 
We made astrometry of the knots from our images and their positions are given in Table~\ref{knots}.


\begin{figure}[htb]
\resizebox{\hsize}{!}{\includegraphics{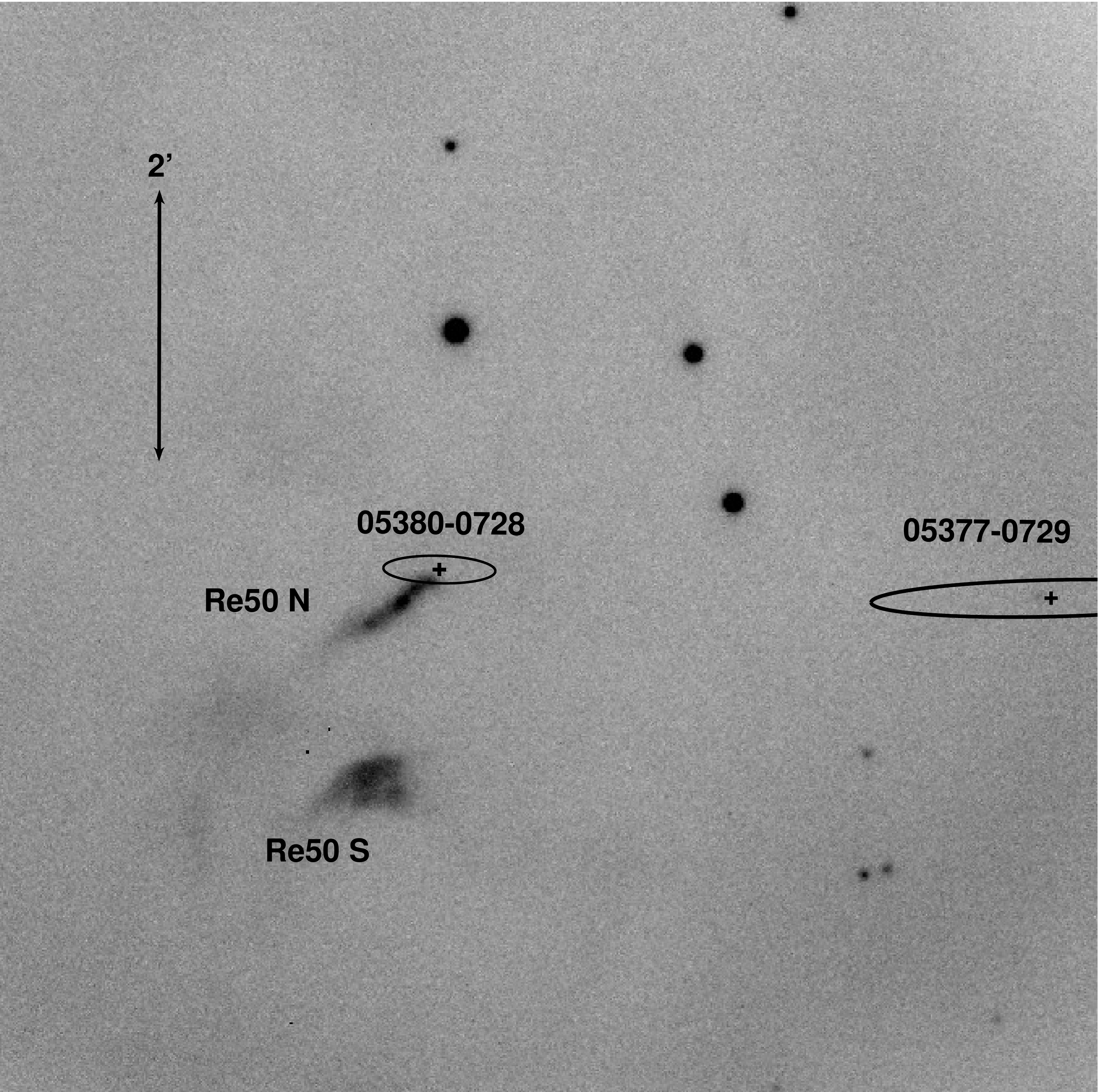}}
\caption{
IRAS 05380$-$0728:
CAHA image through the continuum filter. 
The IRAS source position and error ellipse are shown.  
The positions of the other IRAS source and the Red Nebulous Objects in the field are also shown. 
\label{116_cont}} 
\end{figure}

\begin{figure}[htb]
\resizebox{\hsize}{!}{\includegraphics{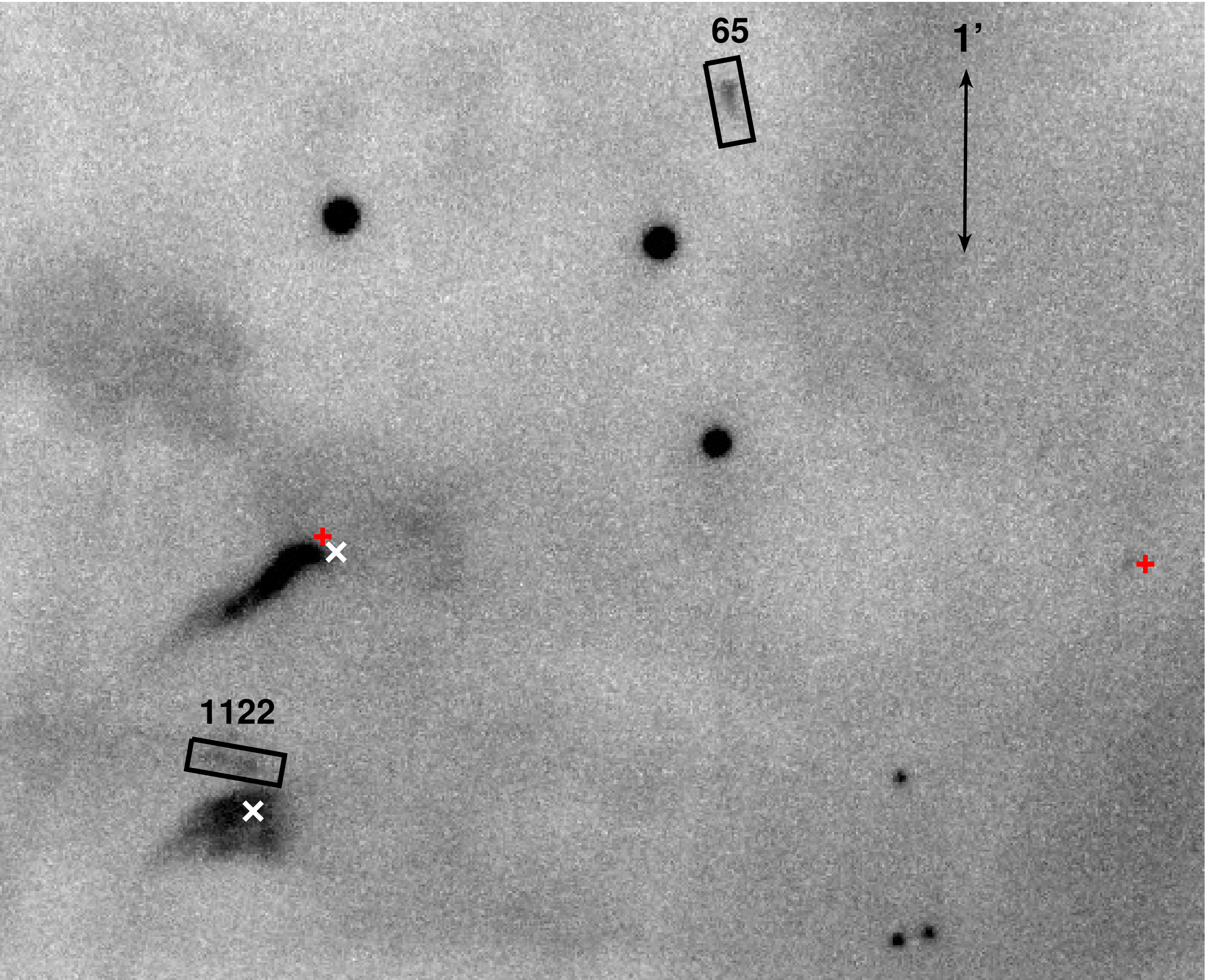}}
\resizebox{\hsize}{!}{\includegraphics{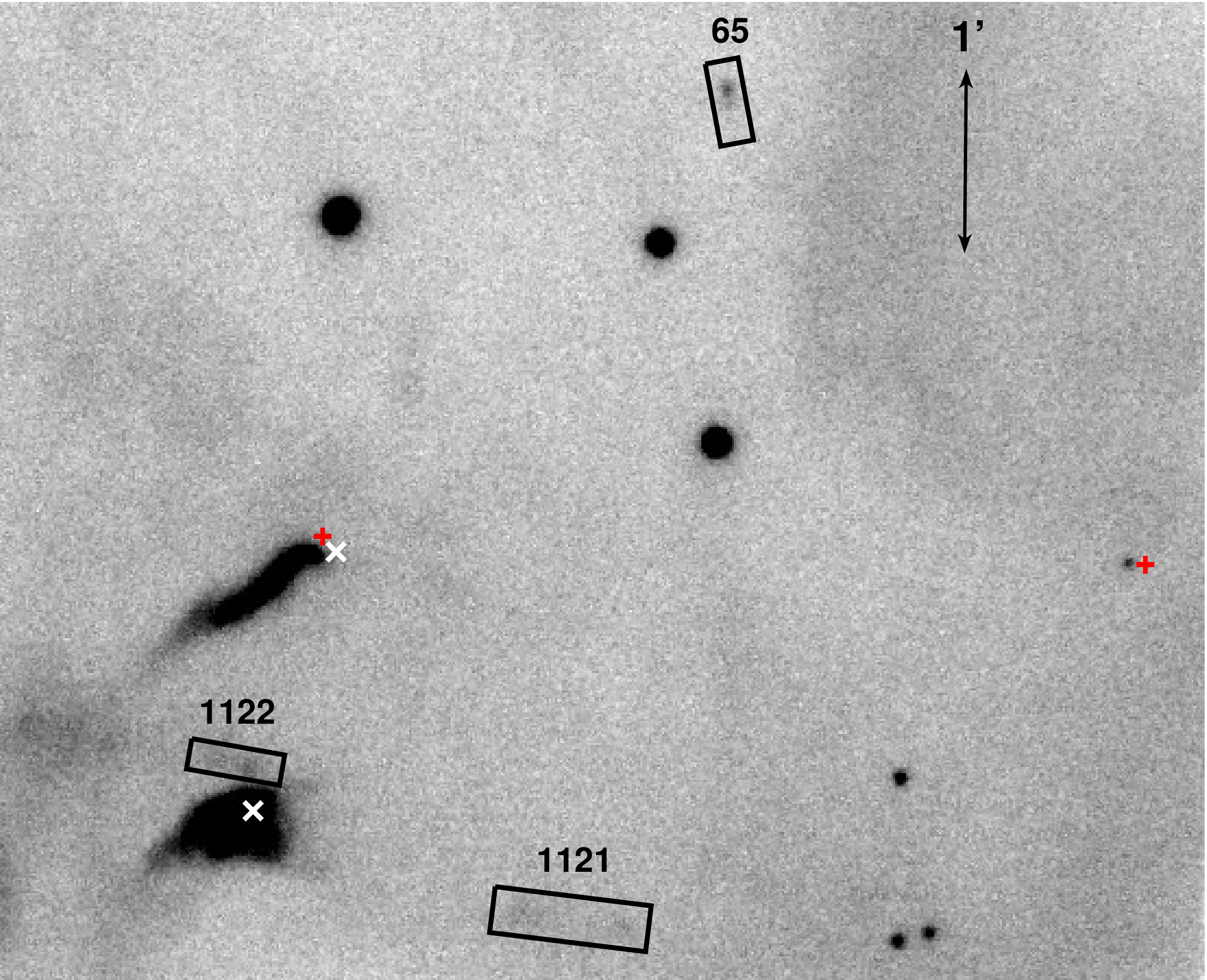}}
\caption{
IRAS 05380$-$0728:
Close-up of the CAHA image through the H$\alpha$ \emph{(top)} and \sii{} \emph{(bottom)} filters.
The positions of the Red Nebulous Objects are marked with white ``$\times$'',
and that of the IRAS sources with red ``+''. 
Black rectangles mark the locations of the emission line features reported in the field.
\label{116_lines}}
\end{figure}

\subsubsection{\object{IRAS 05380$-$0728}}

IRAS 05380$-$0728 is located in the southern region of the \object{L1641} molecular cloud, at a distance of 460 pc \citep{Coh90}.
In the following we present a short description of the field around the source.
\begin{description}

    \item[\textit{Luminosity:}] 
    One of the most luminous objects of the cloud. Estimated total luminosity of 250 $L_\odot$ \citep{Rei86}.

    \item[\textit{Association of IRAS with other objects:}]
          IRS1, in the western edge of the reflection nebula Re50N, a very red point-like source \citep{Cas91}.
    \begin{itemize}
        \item Detected in the near-IR ($J$, $H$, $K$, $L$ bands).
        \item Coinciding with the IRAS position.
    \end{itemize}

    \item[\textit{Reflection nebulae:}]
    \object{Re50}S and \object{Re50N} \citep{Rei85}.
    \begin{itemize}
        \item Associated with YSOs.
        \item Extended emission.
        \item Variability: from 2006 to 2014, Re50N increased its brightness while Re50S faded significantly, probably caused by dusty material orbiting the sources \citep{Chi15}.
    \end{itemize}
    
    \item[\textit{Optical outflows/Herbig-Haro objects:}]
    \begin{description}\item[]
        \item[\object{SMZ9}-4, 5 and 6:] a large-scale strand of knots and filaments in the H$_2$ line at 2.12~$\mu$m \citep{Sta00}. Driven by a different source, IRAS 05380-0731.
        \item[\object{HH 65}:] faint, single knot in \sii{}  \citep{Rei88}, located at the red lobe of a bipolar CO outflow. It is the optical counterpart of SMZ9-6. 
        \item[\object{HH 1121}:] a knotty chain north of the Re50S nebulosity \citep{Chi15}.
        \item[\object{HH 1122}:] two faint knots $\sim1'$ southwest of Re50S \citep{Chi15}.
    \end{description}
    
\end{description}

Fig.~\ref{116_cont} displays the field centered on IRAS 05380$-$0728 imaged trough the continuum filter. 
Another IRAS source, \object{IRAS 05380-0729}, appears also in the image. 
IRAS 05380$-$0728 is located at the tip of Re50N. 
In our image, both Re50N and Re50S appear with a similar brightness. 
Re50N has an `S' shape, while  Re50S has an arrowhead shape. 
The same morphology is found in the H$\alpha$ and \sii{} filter images (Fig.~\ref{116_lines}). 
 
Regarding the variability of Re50N and Re50S, both nebulosities keep the same shape as observed in the \sii{} image of 2014 \citep{Chi15}.
However, in our \sii{} image (Fig.~\ref{116_lines}) Re50N and Re50S have a similar brightness, and Re50S is not dimmer than Re50N as reported in 2014 \citep{Chi15}. 
Concerning the HH objects of the region, we detected HH~65 in the H$\alpha$ and \sii{} images. 
We found a slightly different morphology in H$\alpha$ and \sii{}: the knot is more compact and brighter in \sii{} than in H$\alpha$. 
HH~1121 was only detected in the \sii{} image. 
In contrast, HH~1122 was detected in both emission lines, being brighter in H$\alpha$ than in \sii. 
We did not detect any additional emission line features in our deep narrow-band images of this region.


\begin{figure}[htb]
\resizebox{\hsize}{!}{\includegraphics{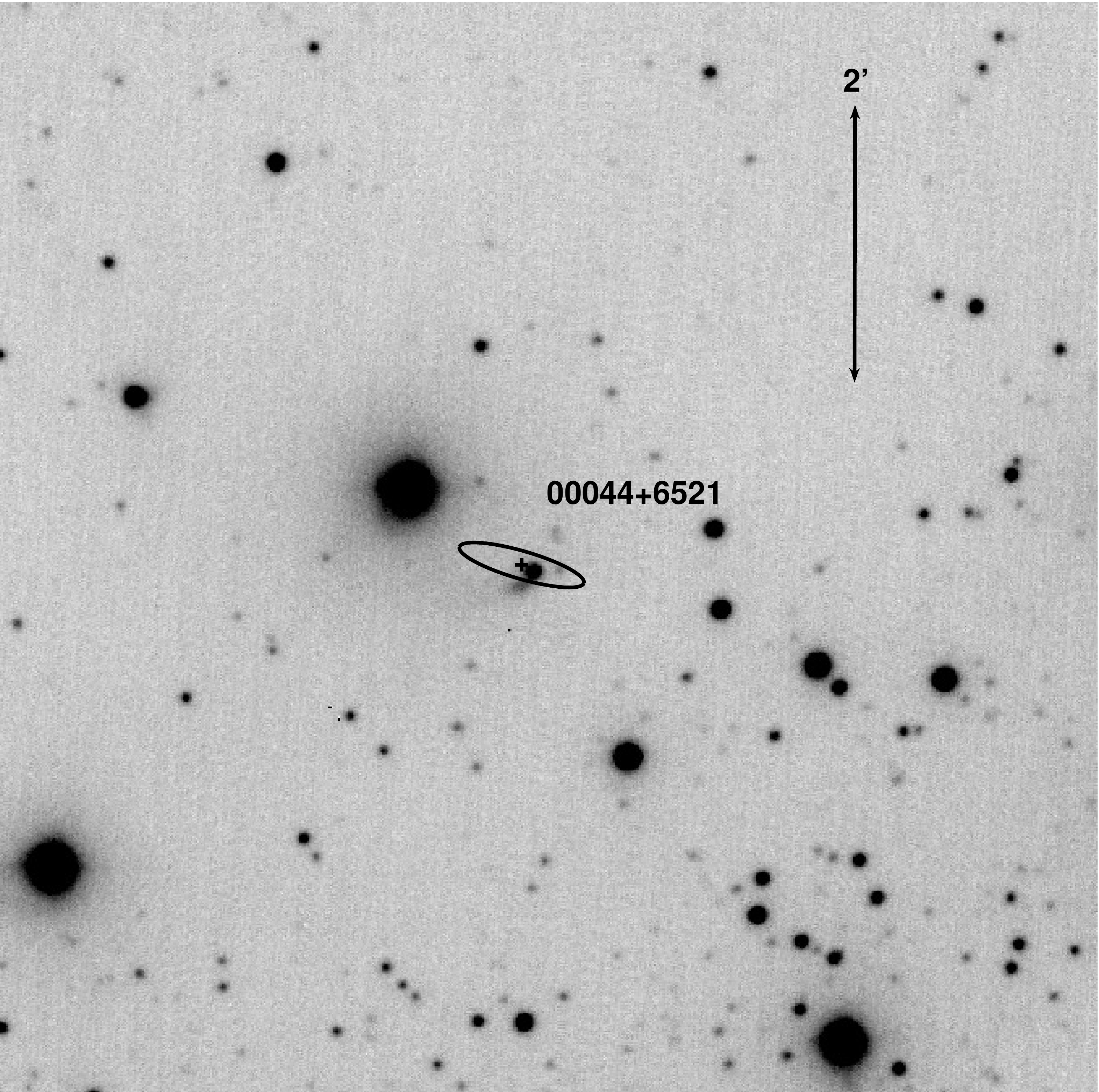}}
\caption{
IRAS 00044+6521:
CAHA image through the continuum filter. 
The IRAS source position and error ellipse are shown.  
\label{1_cont}} 
\end{figure}

\begin{figure}[htb]
\resizebox{\hsize}{!}{\includegraphics{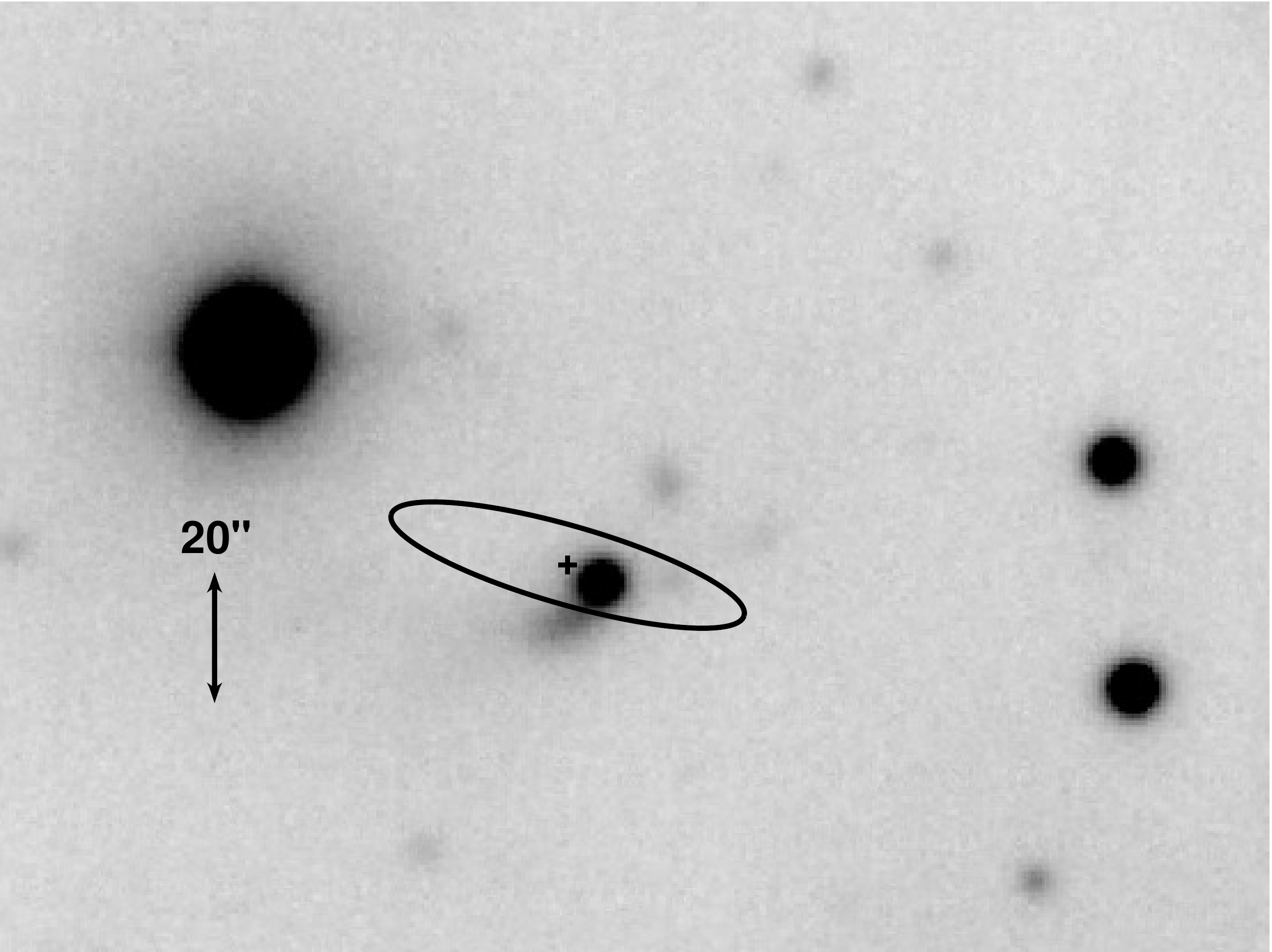}}
\resizebox{\hsize}{!}{\includegraphics{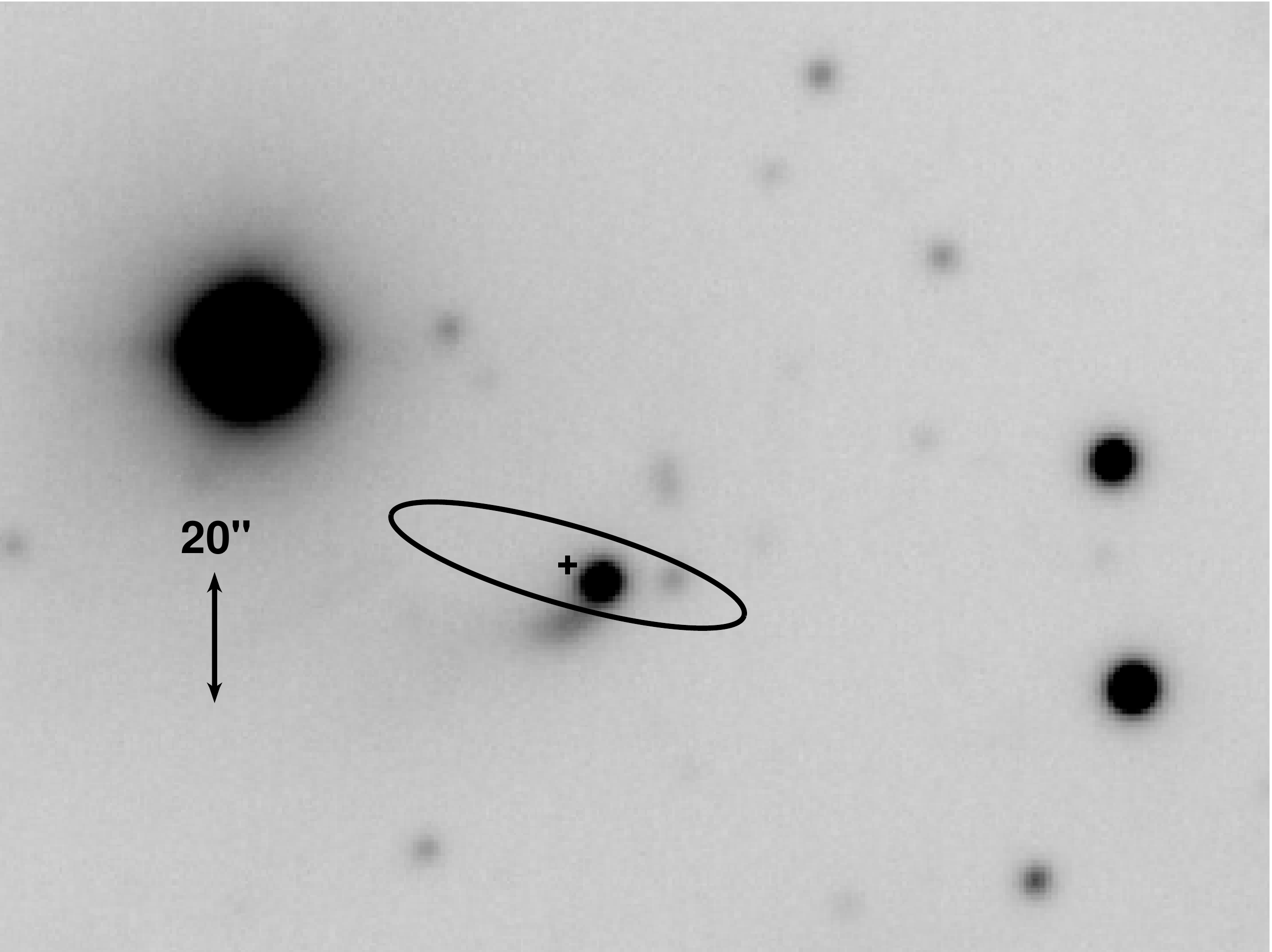}}
\caption{
IRAS 00044+6521:
Close-up of the CAHA images through the H$\alpha$ \emph{(top)} and \sii{} \emph{(bottom)} filters.
\label{1_lines}}
\end{figure}

\begin{figure}[htb]
\resizebox{\hsize}{!}{\includegraphics{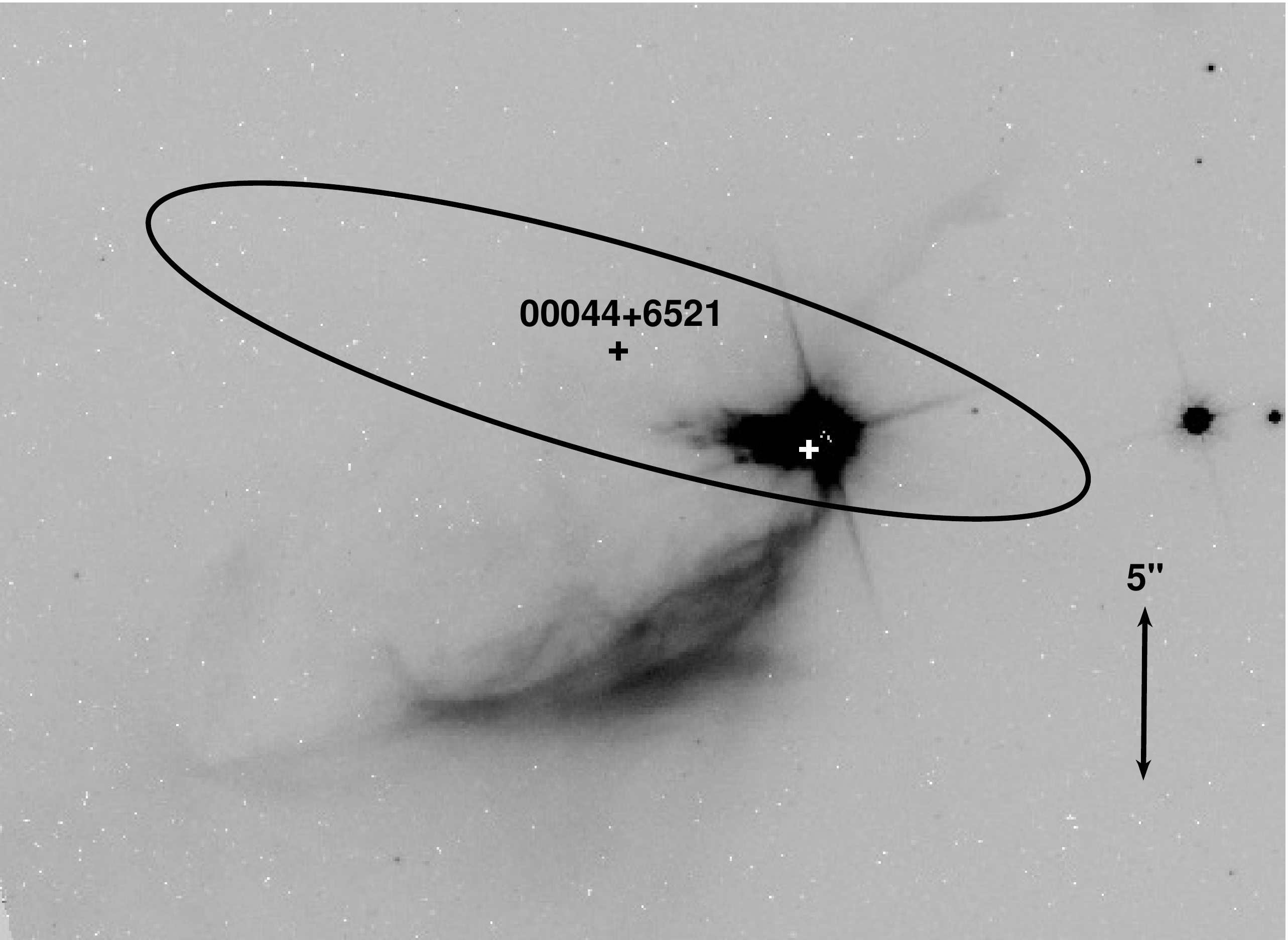}}
\caption{
IRAS 00044+6521:
\emph{HST} ACS/WFC image through the F814W filter of the field around the IRAS source. 
The white `'+'' marks the position of the 2MASS source J00070260+6538381. 
The IRAS position and its error ellipse are also shown. 
\label{1_HST}} 
\end{figure}

\subsection{Group II: Sources associated with extended nebular emission}

\subsubsection{\object{IRAS 00044+6521}}

IRAS 00044+6521 belongs to the \object{Cepheus IV} association, at a distance of $845\pm110$ pc \citep{Mac68}.
In the following we present a short description of  the source.
\begin{description}

    \item[\textit{Proposed IRAS counterpart:}]~
    \begin{itemize}

    \item The optical object appears as a Herbig Ae/Be of the \citet{Her88} catalogue as \object{HBC~1}.

    \item Also known as the emission star \object{MacC H12} \citep{Mac68}, with H$\alpha$ line emission \citep{Coh76}.

    \item Included in the Herbig Ae/Be survey of  \citet{The94}, and  classified with a spectral type around F4 \citep{Her04}.

    \item Broad-band ($R$, $I$, $J$, $H$, $K$, $L$) images  \citep{Ori90} show a red stellar component and a nebula, consistent with a TTauri star with an accretion disk and a cold dust envelope.
        
    \item Its position is offset $\sim5''$ from the IRAS position.

    \end{itemize}
\end{description}

Figure \ref{1_cont} shows the image of the field around IRAS 00044+6521 through the continuum filter, and
Fig.{} \ref{1_lines} shows close-ups of the H$\alpha$ and \sii{} lines  filter images.

As can be seen in the figures, the target shows a compact emission, inside the position error ellipse of IRAS 00044+6521, plus a cometary tail $\sim10''$ long, extending southeastward.
The extended emission has the same shape in continuum and line emission, thus confirming that the origin of the emission is most probably scattered light, and not shock excitation from a stellar microjet. 
We measured an offset of $6\farcs5$ between the nominal position of the IRAS source and the 2MASS source \object{J00070260+6538381}, the 2MASS source coinciding with the photocenter of the compact optical counterpart.
 
Figure~\ref{1_HST} shows a close-up of the field around IRAS 00044+6521 from the \emph{HST} ACS/WFC image through the F814W filter, extracted from the \emph{HST} Legacy Archive (PI Sahai. Program ID: 10536). 
The extended emission can be seen in more detail, showing an arc-shaped morphology, with its center toward the IRAS position.
The proposed optical/near-IR counterpart of IRAS lies at the northwest tip of the nebulosity.


\begin{figure}[htb]
\resizebox{\hsize}{!}{\includegraphics{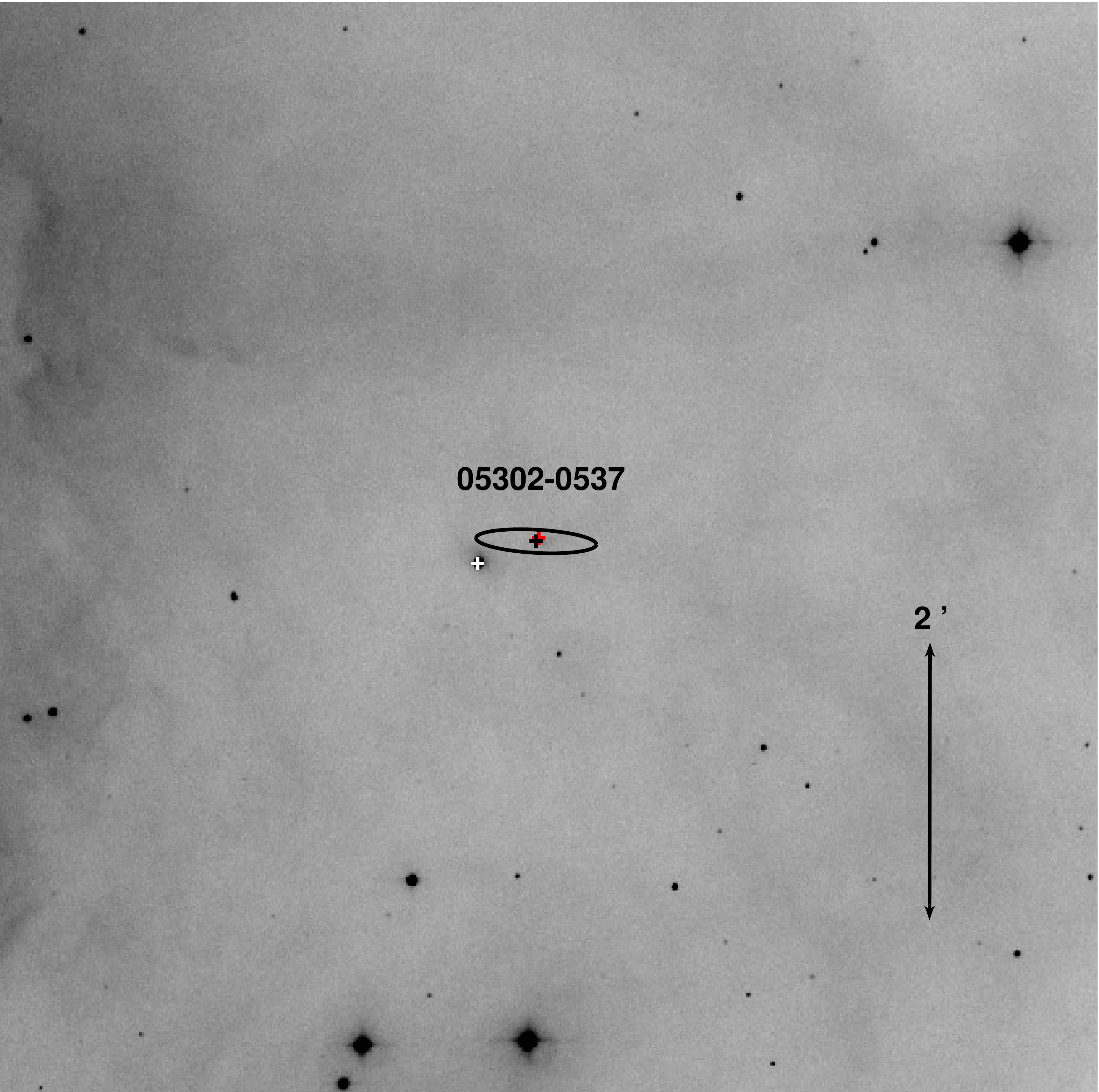}}
\caption{
IRAS 05302$-$0537:
CAHA image in the narrow-band \sii{} filter. 
The IRAS source position and error ellipse are shown in black.  
Haro~4-145 is marked with a white ``+''. 
The IR source Vision J5324165$-$0535461 \citep{Mei16} is marked with a red ``+'', nearly overlapping the black ``+'' of the IRAS source. 
\label{103}} 
\end{figure}

\subsubsection{\object{IRAS 05302$-$0537}}

IRAS 05302$-$0537 is located at the southern part of \object{Orion A}, at a distance of $319\pm17$~pc \citep{Gai18}. 
In the following we present a short description of the field around the source.
\begin{description}

    \item[\textit{First proposed counterpart of IRAS:}]
        \object{Haro~4-145}, an H$\alpha$ emission star \citep{Par82}.
    \item[\textit{Present proposed counterpart of IRAS:}]~
    \begin{description}
        \item[$K$-band reflection nebula:] point-like counterpart with a diffuse nebula extending northwards from the near-IR source \citep{Con07}.
        \item[\object{J05324165-0535461}:] Near-IR source  \citep[Orion A VISTA catalogue;][]{Mei16}, with position coincident with IRAS. 
        Source slightly elongated in the northeast--southwest direction (consistent with its binary nature, see below), surrounded by diffuse, arc-shaped emission extending from northwest to southeast of the compact source.
    \end{description}
    \item[\textit{Binarity:}]
    Binary system with an angular separation of $0\farcs65$ \citep[$L$-band images;][]{Con08}.
    \item[\textit{Tracers of YSOs:}]~
    \begin{itemize}
        \item Bipolar CO outflow \citep[Orion A-west;][]{Fuk86}.
        \item High-density ammonia clump centered on IRAS \citep{Har93}.
    \end{itemize}
\end{description}

Due to weather conditions, only the narrow-band \sii{}  image of the IRAS 05302$-$0537 field was obtained in our CAHA survey (Fig.~\ref{103}). 
We barely detected a faint, compact emission coinciding with the IRAS source, but we did not detect any extended emission associated with it.
Our image also shows another YSO, Haro~4-145. This object is in  a more evolved stage than J05324165-0535461, as indicated by its near-IR colors, and the lower extinction allowing to be detected at optical wavelengths. 
Haro~4-145 lies $\sim25''$ southeast of the IRAS source position,  outside the IRAS error ellipse. Thus, it is unlikely to be the optical counterpart of the IRAS source. Instead, J05324165-0535461 is  more likely to be the IRAS counterpart.


\begin{figure}[htb]
\resizebox{\hsize}{!}{\includegraphics{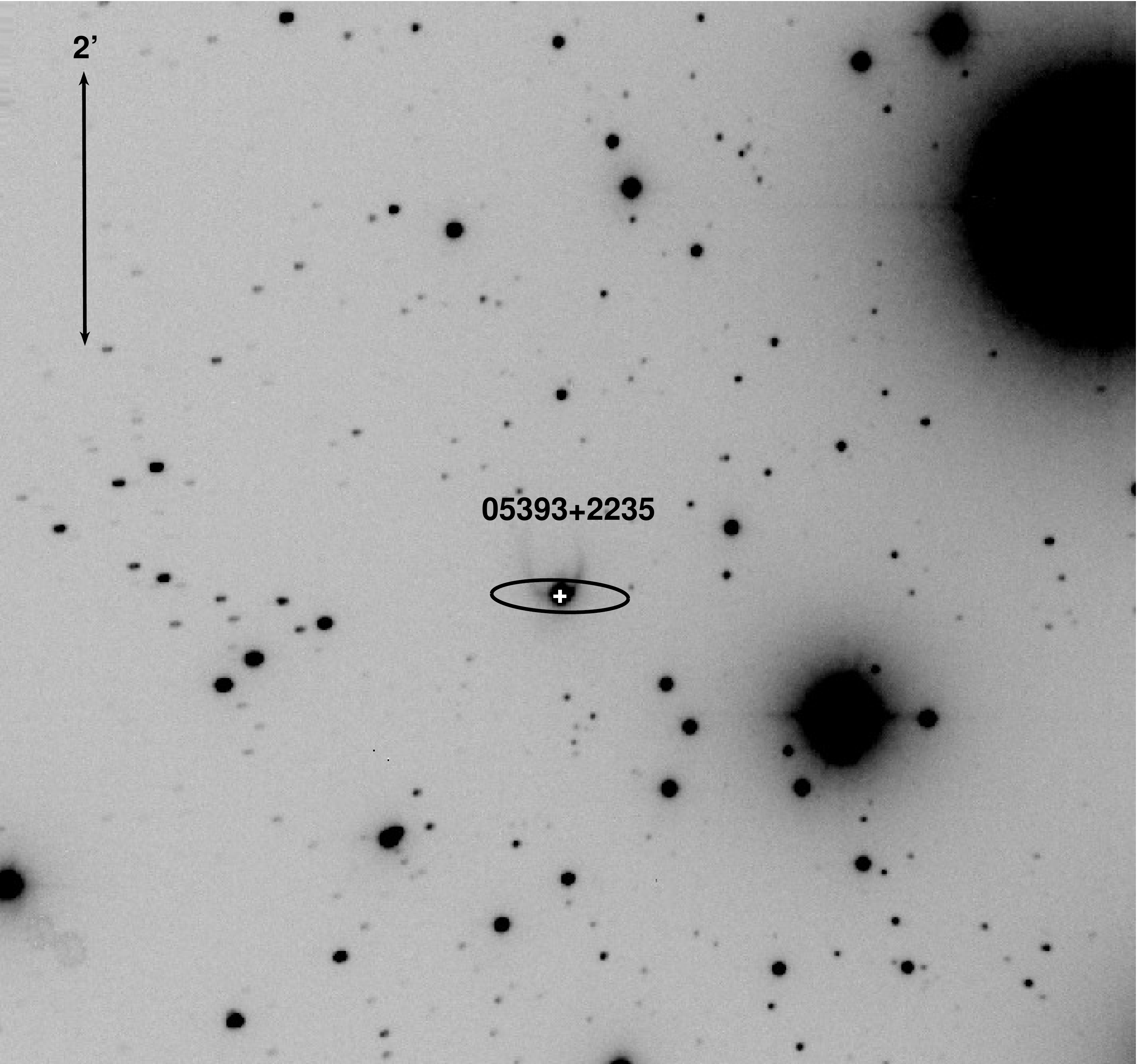}}
\caption{
IRAS 05393+2235:
CAHA image through the continuum filter. 
The IRAS source position and error ellipse are shown.  
\label{119_cont}} 
\end{figure}

\begin{figure}[htb]
\resizebox{\hsize}{!}{\includegraphics{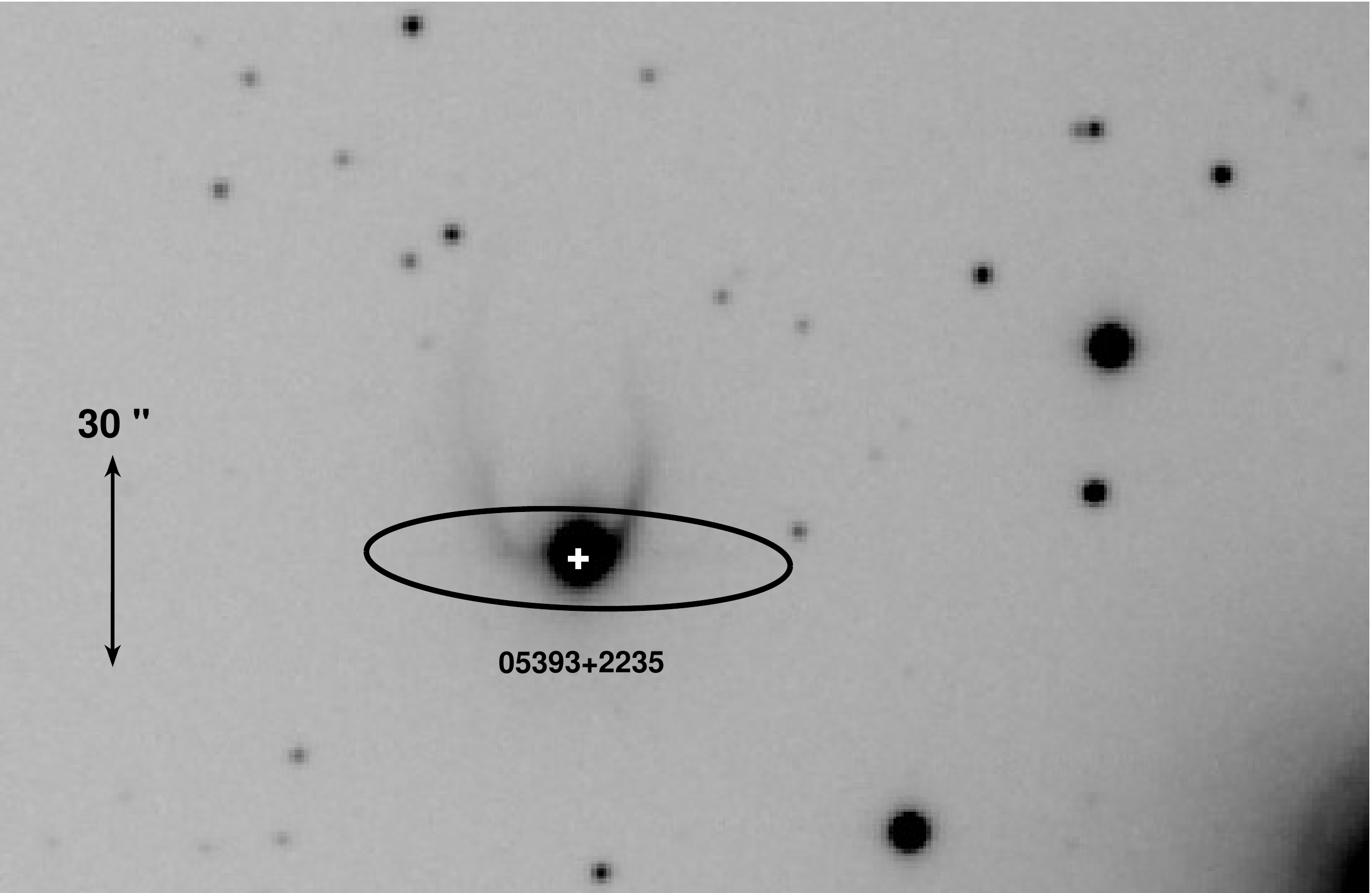}}
\resizebox{\hsize}{!}{\includegraphics{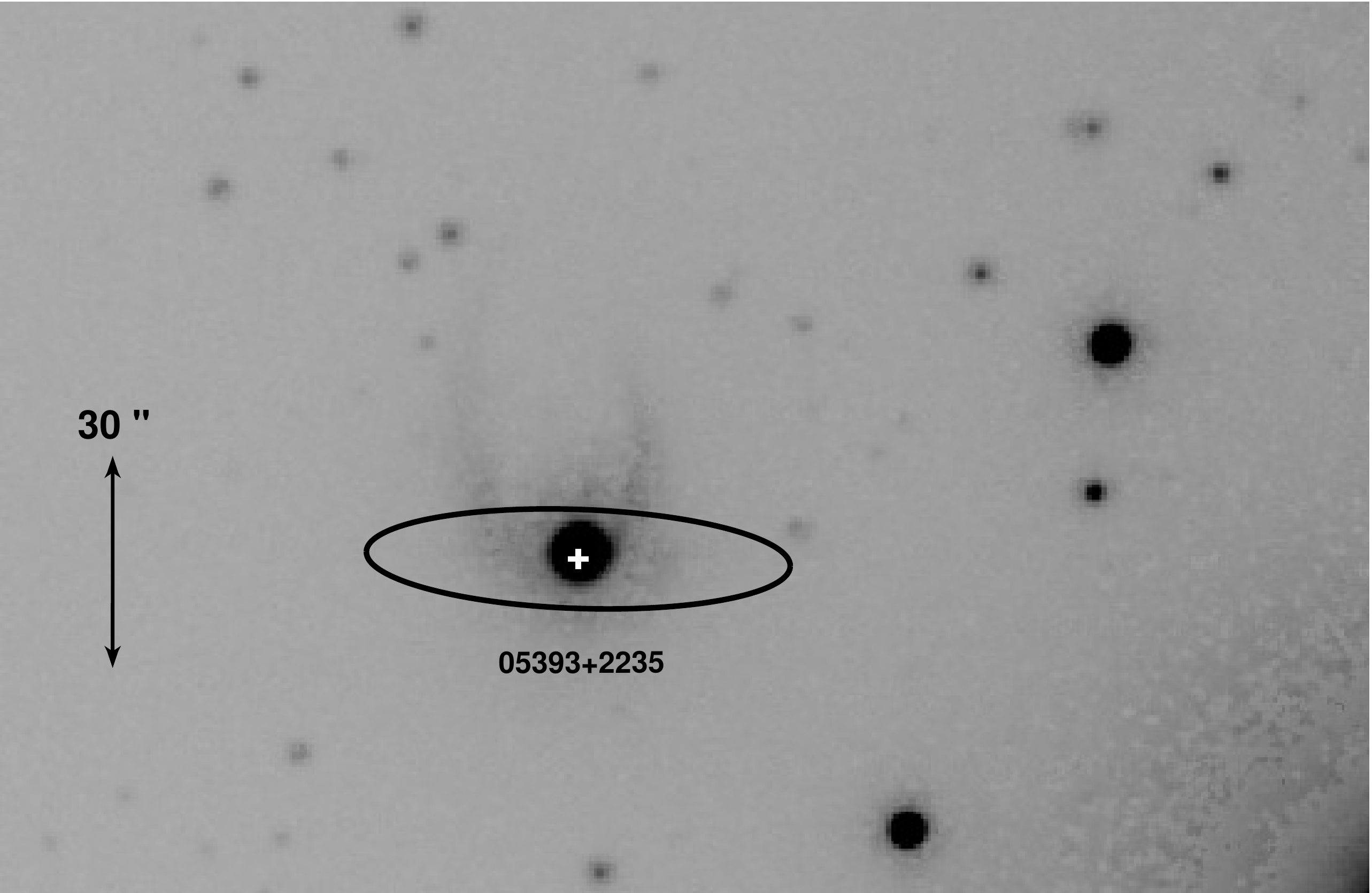}}
\caption{
IRAS 05393+2235:
Close-up of the CAHA images through the H$\alpha$ \emph{(top)} and \sii{} \emph{(bottom)} filters.
\label{119_lines}}
\end{figure}

\subsubsection{\object{IRAS 05393+2235}}

IRAS 05393+223 is located at a distance of $1540\pm106$~pc \citep{Gai18}.
In the following we present a short description of the source.
\begin{description}
    \item[\textit{IRAS counterpart:}] \object{RNO~54}, a red nebulous object \citep{Coh80}.
    \begin{itemize}
        \item Associated with extended emission \citep{Coh80} with a cometary-shape morphology \citep[broad-band, $R$ filter image;][]{Goo87},
        \item FU Ori star, with a spectral type F5~II \citep{Goo87}.
        \item Probable post-FU Ori star \citep[double Li absorption profile;][]{Tor95}.
    \end{itemize}
\end{description}

Figure \ref{119_cont} shows the IRAS 05393+2235 field in the continuum filter and Figure~\ref{119_lines} shows close-ups in the H$\alpha$ and \sii{} line filters. 
As can be seen in the figures, the extended, arc-shaped  nebulosity surrounding the compact counterpart  of the IRAS source presents the same morphology in the continuum and in the line images, indicating that  the extended emission is most probably a reflection nebula. 
No evidence of shocked gas was found in our narrow-band images.


\begin{figure}[htb]
\resizebox{\hsize}{!}{\includegraphics{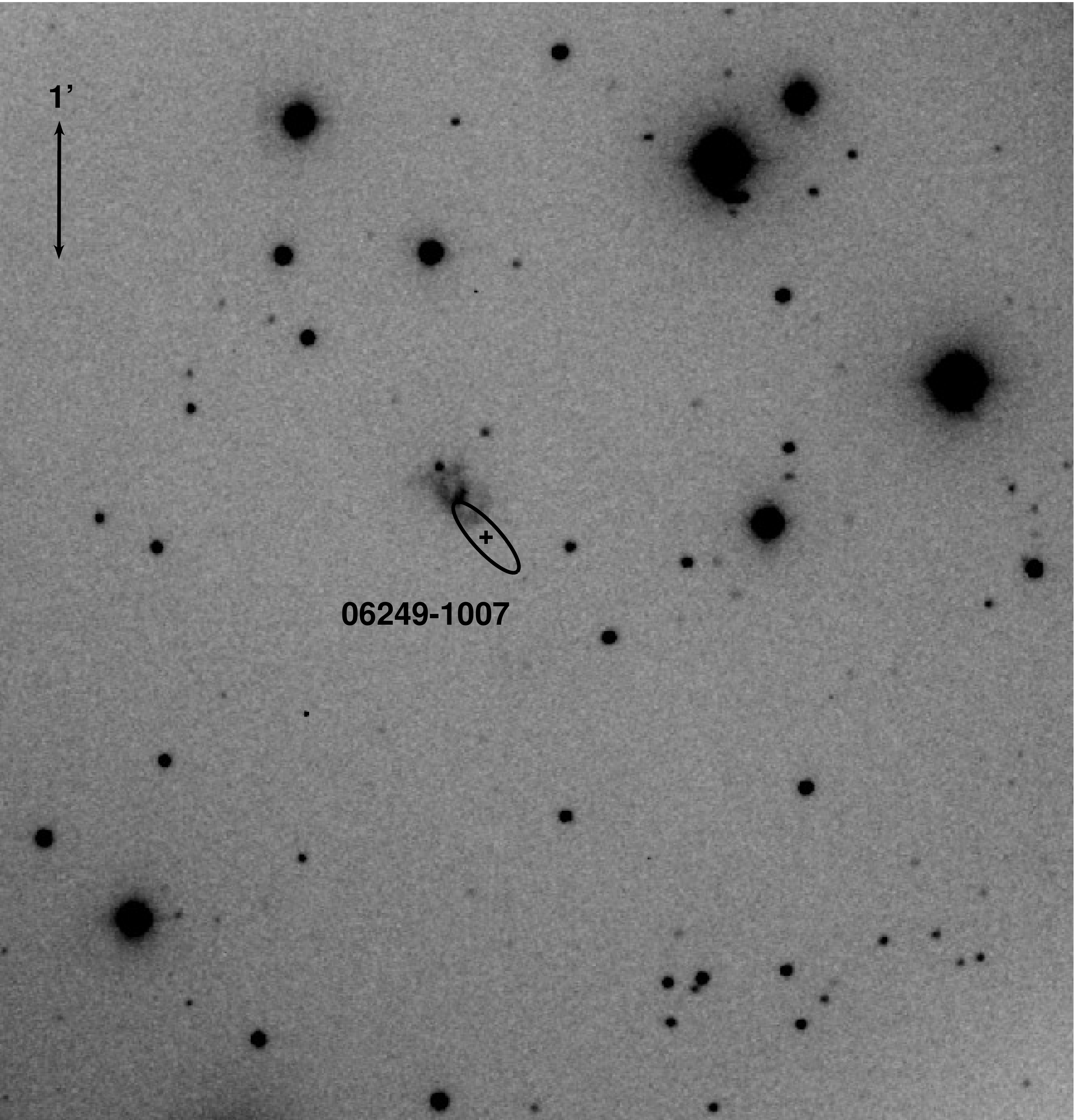}}
\caption{
IRAS 06249$-$1007:
CAHA image through the continuum filter. 
The IRAS source position and error ellipse are shown.  
\label{148_cont}} 
\end{figure}

\begin{figure}[htb]
\resizebox{\hsize}{!}{\includegraphics{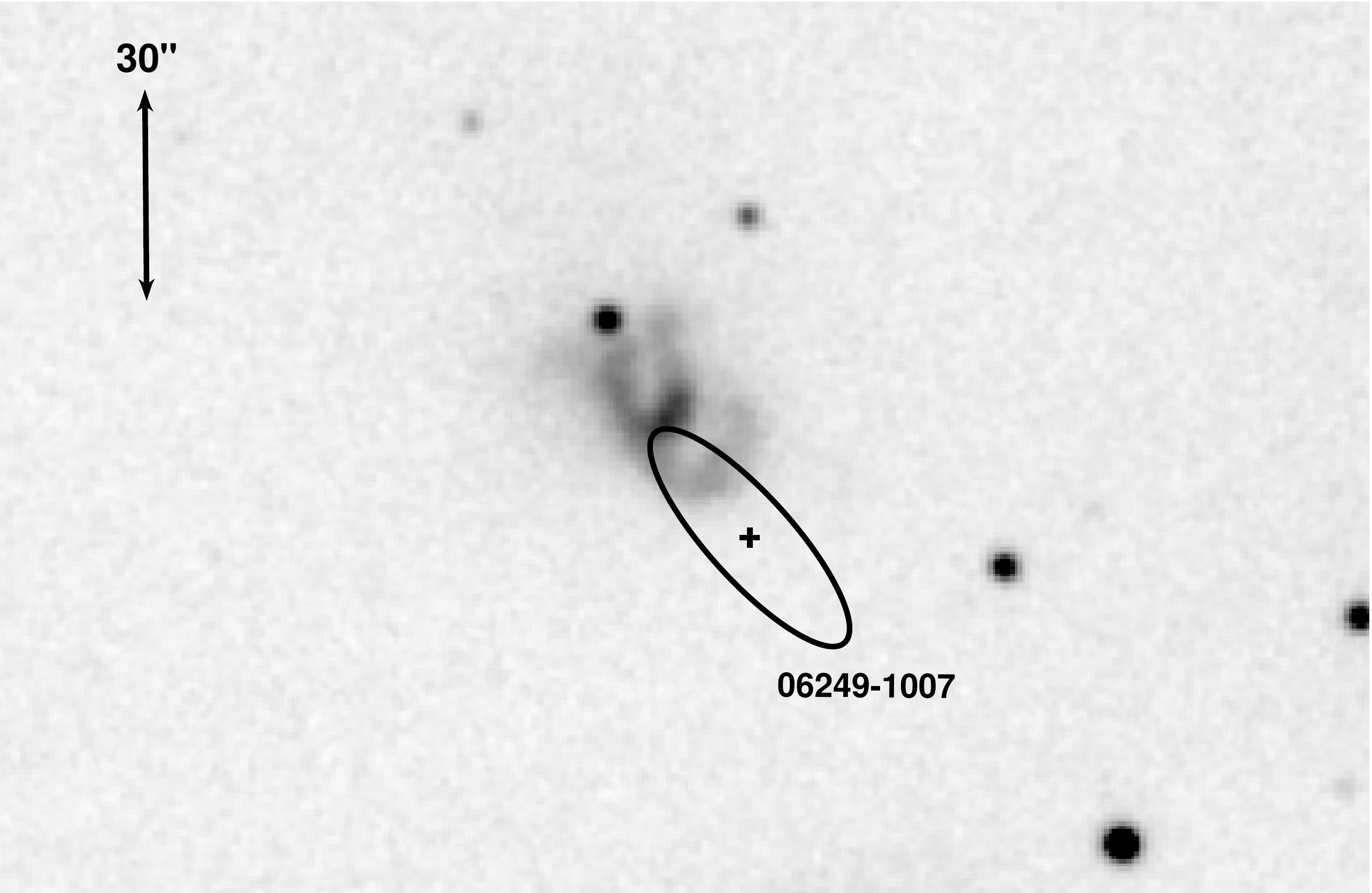}}
\resizebox{\hsize}{!}{\includegraphics{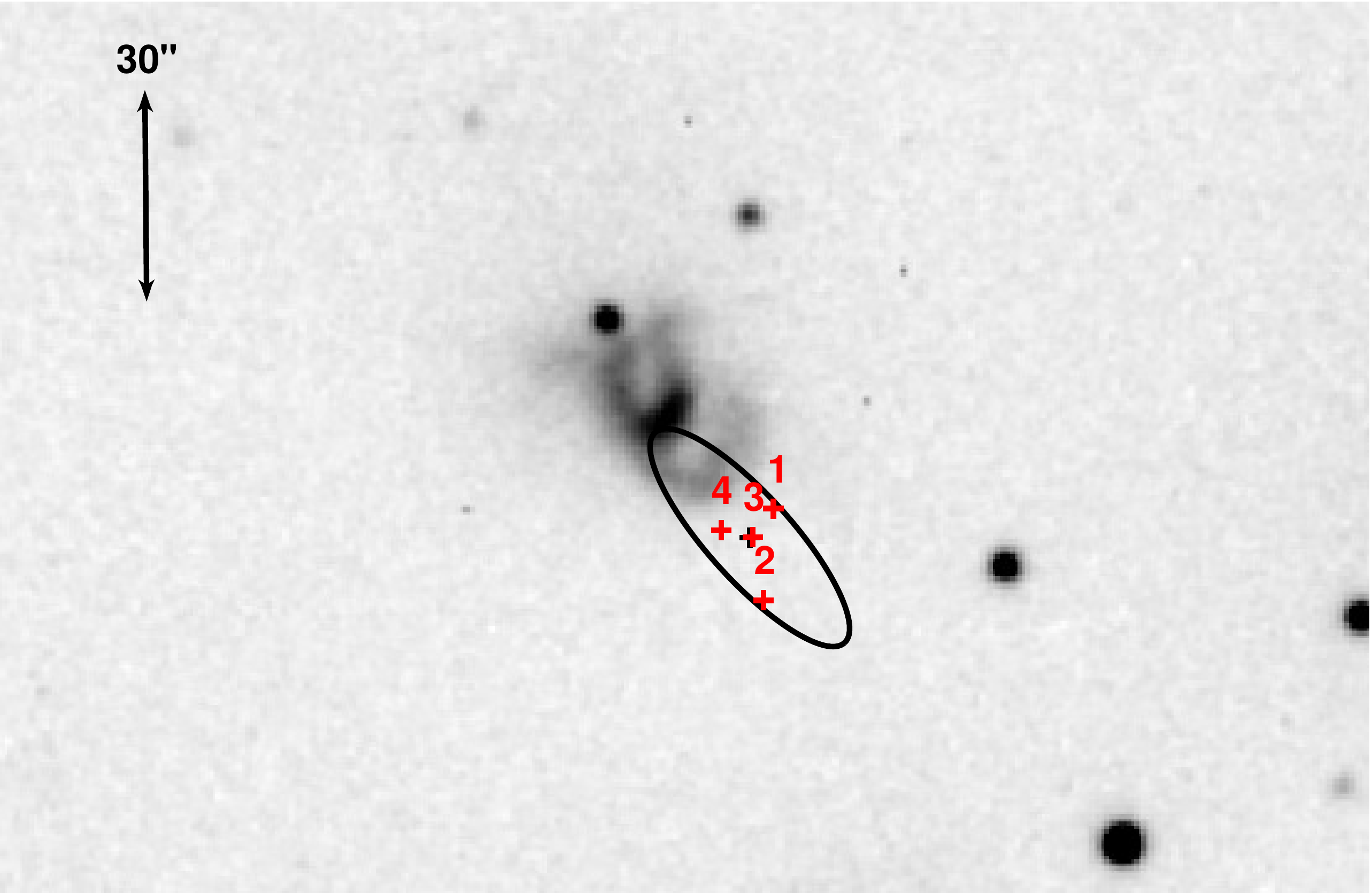}}
\caption{
IRAS 06249$-$1007:
Close-up of the CAHA image through the H$\alpha$ \emph{(top)} and \sii{} \emph{(bottom)} filters showing the field centered on the IRAS source
The stars of the IR cluster reported by \citet{Tap97} are marked with ``+''.
\label{148_lines}}
\end{figure}

\subsubsection{\object{IRAS 06249$-$1007}}

IRAS 06249$-$1007 is located at a kinematic distance of $0.86\pm0.07$ kpc (see Footnote \ref{footnote:distance}), determined from its radial velocity, $V_\mathrm{LSR}=12.2$ \kms{} \citep{Wil89}. In the following we present a short description of the field around the source.

\begin{description}
    \item[\textit{Optical and IR emission:}]~
    \begin{itemize}
        \item IRAS located $\sim10''$ from the southwest edge of the loop-shaped nebula \object{HHL~43} \citep{Gyu84}.
        \item HHL 43 has the same loop-shaped morphology in broad-band optical ($I$) and near-IR ($J$, $H$, $K$) images \citep{Tap97}.
        \item Emission at 100 $\mu$m peaking at the nominal position of IRAS \citep{Fra98}.
    \end{itemize}
    \item[\textit{Young stellar objects:}]~
    \begin{itemize}
        \item A cluster of four stars, all inside the IRAS error ellipse, with near-IR colors characteristic of embedded TTauri stars \citep{Tap97}.
        \item The position of one of the stars (star 3) coincides with the IRAS position and with the 2MASS source \object{J06271812$-$1009387}.
    \end{itemize}
    \item[\textit{Molecular outflows:}]
    A CO outflow toward the IRAS position, most probably driven by the source \citep{Wil89}.
\end{description}

Figure \ref{148_cont} shows the image of IRAS 06249$-$1007 field through the continuum filter. 
Close-ups of the narrow-band images in the H$\alpha$ and \sii{} line filters are shown in Fig.{} \ref{148_lines}. 
As can be seen in the figures, the nebula shows the same morphology in continuum and in the line images, indicating that the origin of the emission is mainly reflection (dust illuminated by the cluster of stars), without signs of shocked emission.
In our images, the cluster of YSO including the IRAS counterpart lies outside the nebular emission. 
We did not detect any optical counterpart of any of the YSO of the cluster. 
This is consistent with an extinction increasing from north to south along the nebula \citep{Tap97}, with the YSO cluster being embedded in extended $K$-band emission.

\begin{figure}[htb]
\resizebox{\hsize}{!}{\includegraphics{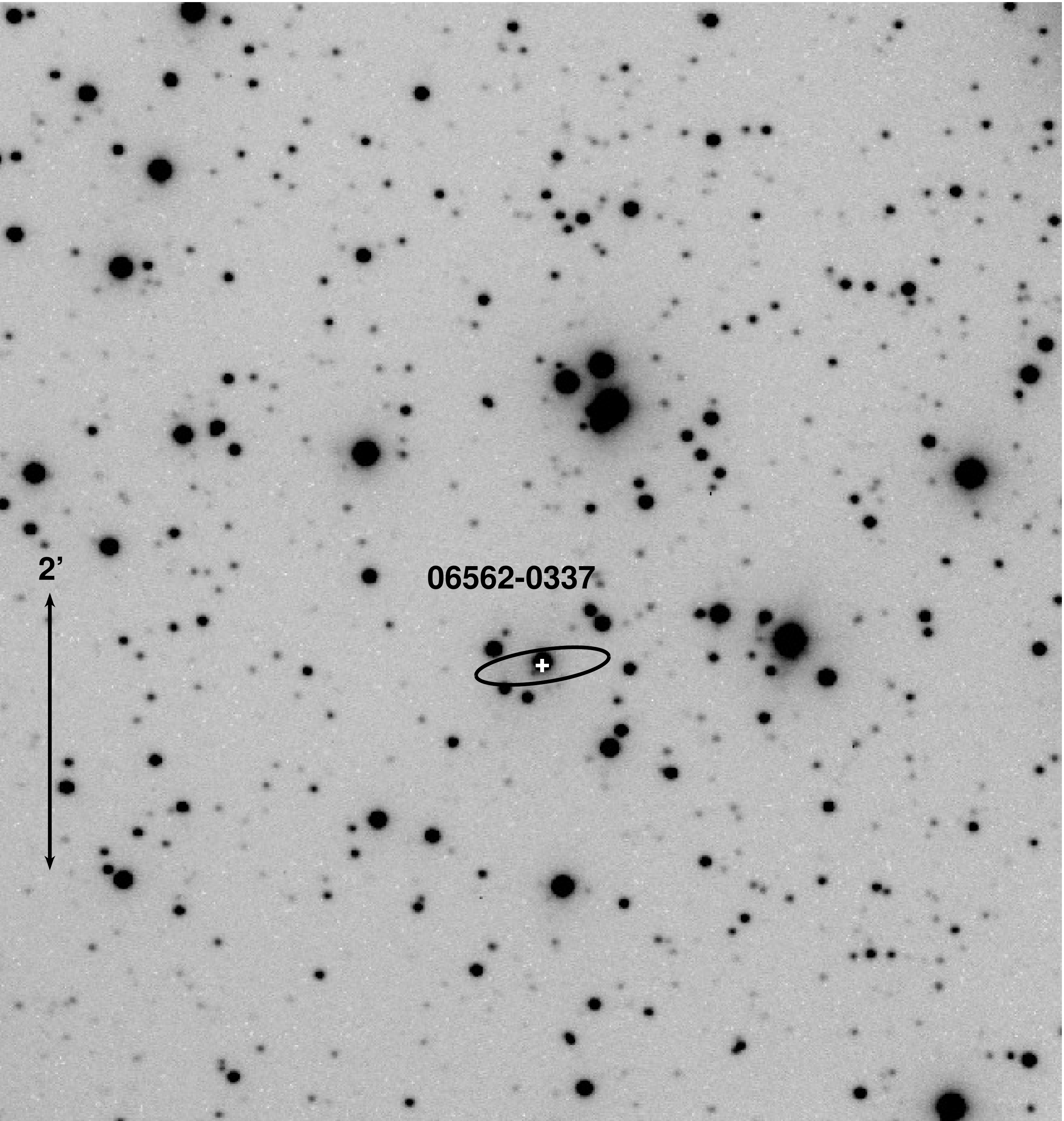}}
\caption{
IRAS 06562$-$0337:
CAHA image through the continuum filter. 
The IRAS source position and error ellipse are shown.  
C\label{165_cont}} 
\end{figure}

\begin{figure}[htb]
\resizebox{\hsize}{!}{\includegraphics{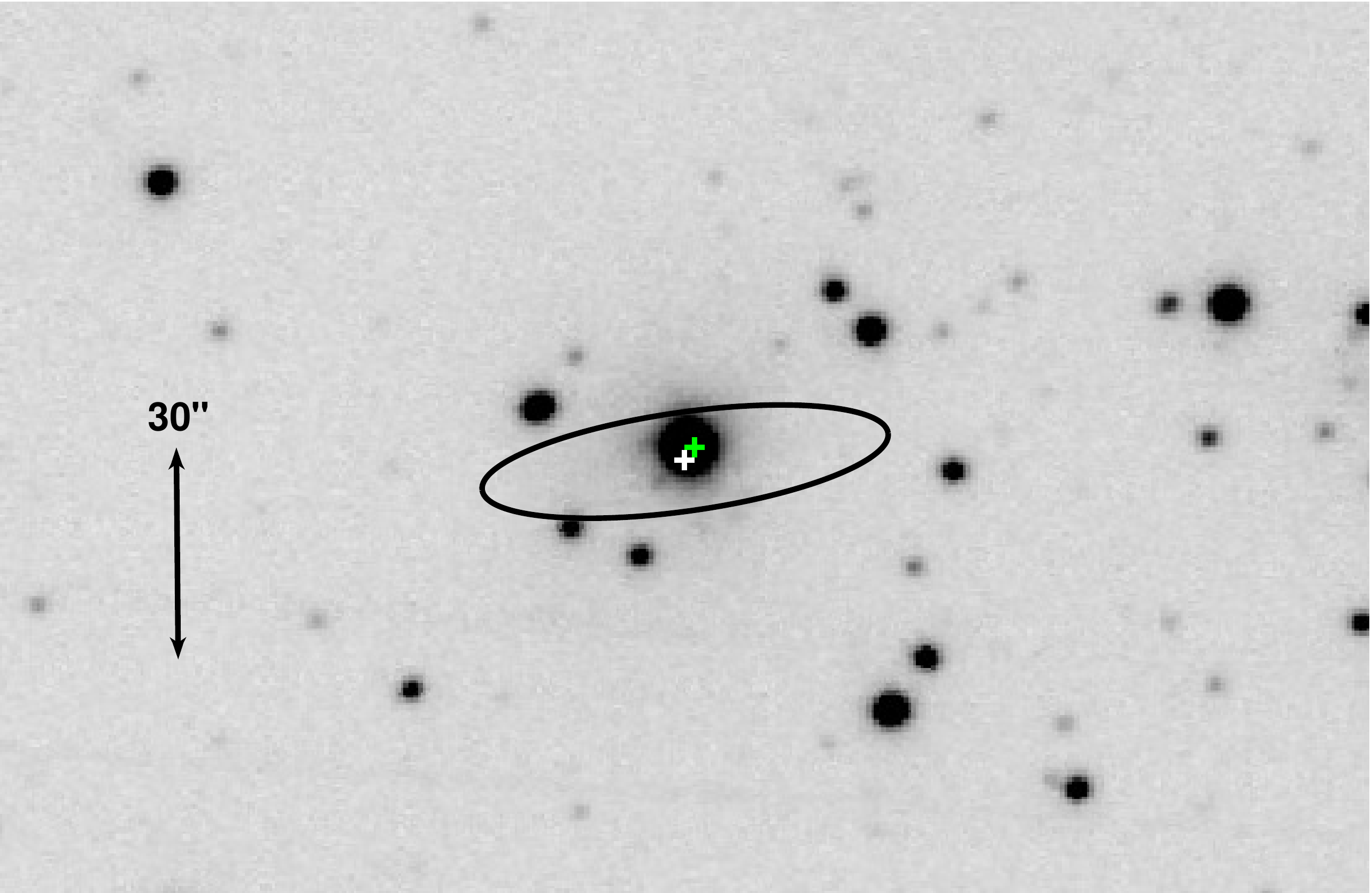}}
\resizebox{\hsize}{!}{\includegraphics{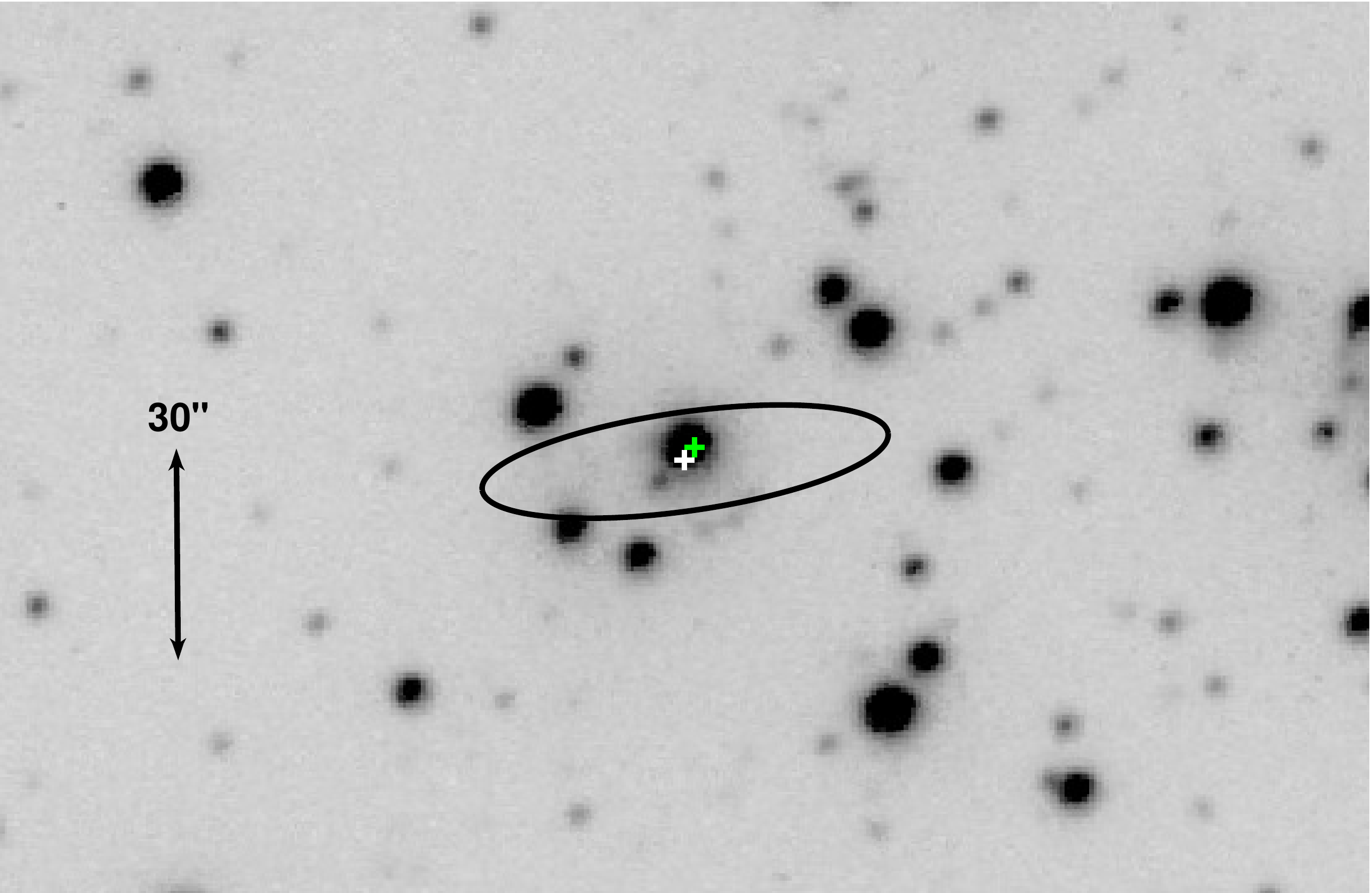}}
\caption{
IRAS 06562$-$0337:
Close-up of the CAHA image through the H$\alpha$ \emph{(top)} and \sii{} \emph{(bottom)} filters, showing the field centered on the IRAS source. 
The positions of the IRAS source (white) and the $K$-band star reported by \citet{Alv98} (green) are marked with ``+''.
\label{165_lines}}
\end{figure}

\subsubsection{\object{IRAS 06562$-$0337}}

IRAS 06562$-$0337 is located at a kinematic distance  of $5.65\pm0.43$ kpc
(see Footnote \ref{footnote:distance}), 
determined from its radial velocity, $V_\mathrm{LSR}=54.0\pm0.2$ \kms{} \citep{Bac98}.
 In the following we present a short description of the field around the source.
\begin{description}
    
    \item[\textit{Controversial nature of IRAS:}]~
    \begin{itemize}
        \item Originally classified as a possible planetary nebula (PN) \citep{Mac78}, based on a spectrum showing the H$\alpha$ emission line and  absence of continuum emission.
        \item Renamed as \object{Iron-clad Nebula}, and classified to be in a transition phase from AGB to PN, based on its high variability and spectrum dominated by allowed and forbidden \feP{} lines \citep{Ker96}.
    \end{itemize}

    \item[\textit{Association of IRAS with molecular gas:}]
    CO, $^{13}$CO, and CS emission at millimeter and sub-millimeter wavelengths \citep{Bac98}.
    \begin{itemize}
        \item Powering source of a molecular outflow traced by the high-velocity CO emission.
        \item Non-evolved object because the CS molecule is destroyed in the proto-PN stage.
    \end{itemize}

    \item[\textit{Association of IRAS with IR emission: }]
    $K'$-band image of the field \citep{Alv98}.
    \begin{itemize}
        \item IRAS is a young, rich stellar cluster embedded in  diffuse emission.
        \item The cluster contains $\sim70$ stars within a $30''$ radius around the bright central object.
        \item The central object is likely a Herbig Be star.
        \item Spectral variability of the central object attributed to a stellar wind in its extended atmosphere.
        \item The coordinates of the central source in the $K'$-band image are in agreement with the CO emission peak.
    \end{itemize}

\end{description}

Figure \ref{165_cont} shows the IRAS 06562$-$0337 field mapped through a continuum filter,  and Fig.~\ref{165_lines} shows close-ups of the  H$\alpha$ and \sii{} line filter images, where the stars of the IR cluster have been marked.
We did not detect any nebular, shocked emission from jet structures.


\begin{figure}[htb]
\resizebox{\hsize}{!}{\includegraphics{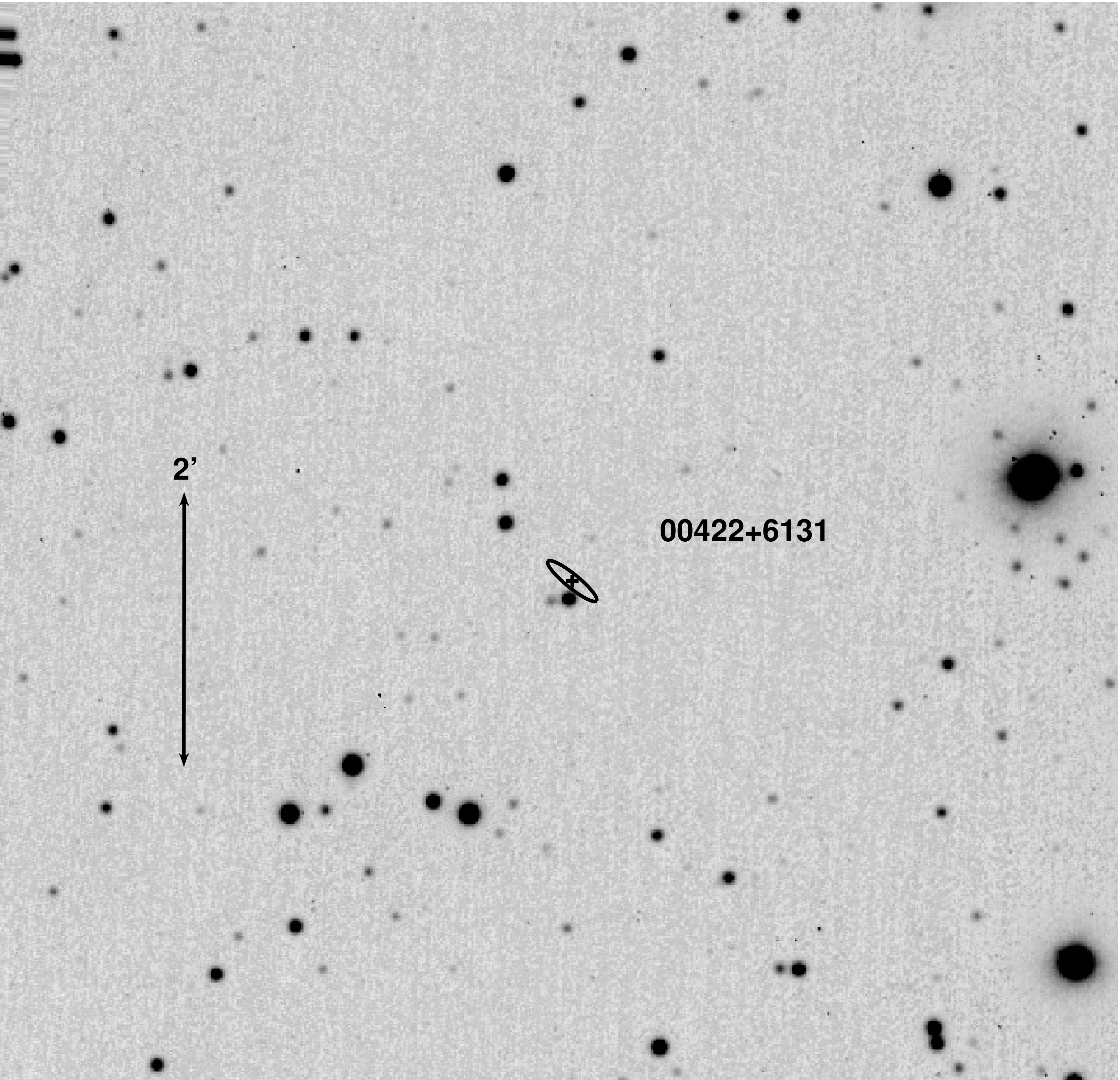}}
\caption{
IRAS 00422+6131:
CAHA image through the continuum filter. 
The IRAS source position and error ellipse are shown.  
\label{9_cont}} 
\end{figure}

\begin{figure}[htb]
\resizebox{\hsize}{!}{\includegraphics{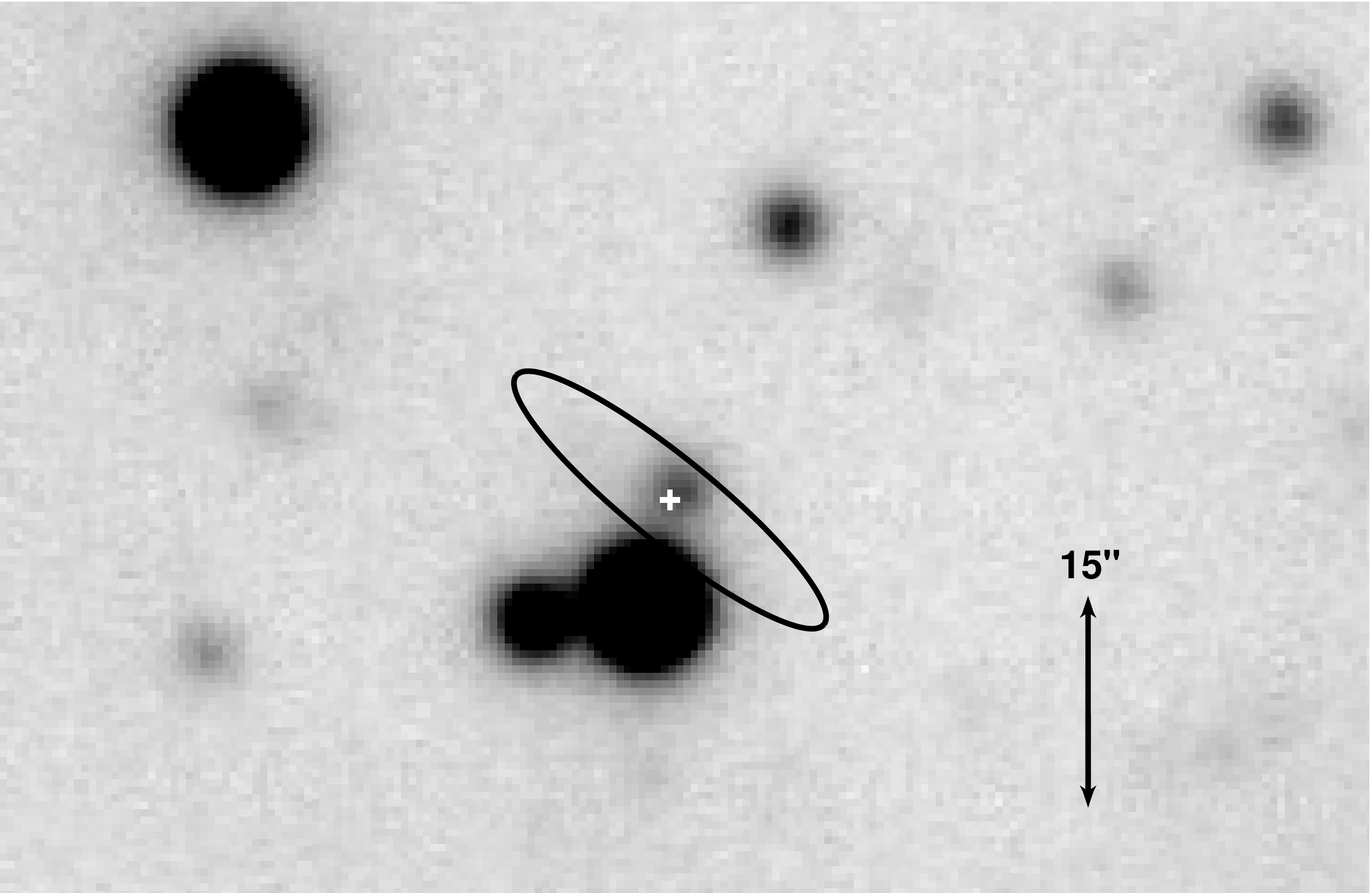}}
\resizebox{\hsize}{!}{\includegraphics{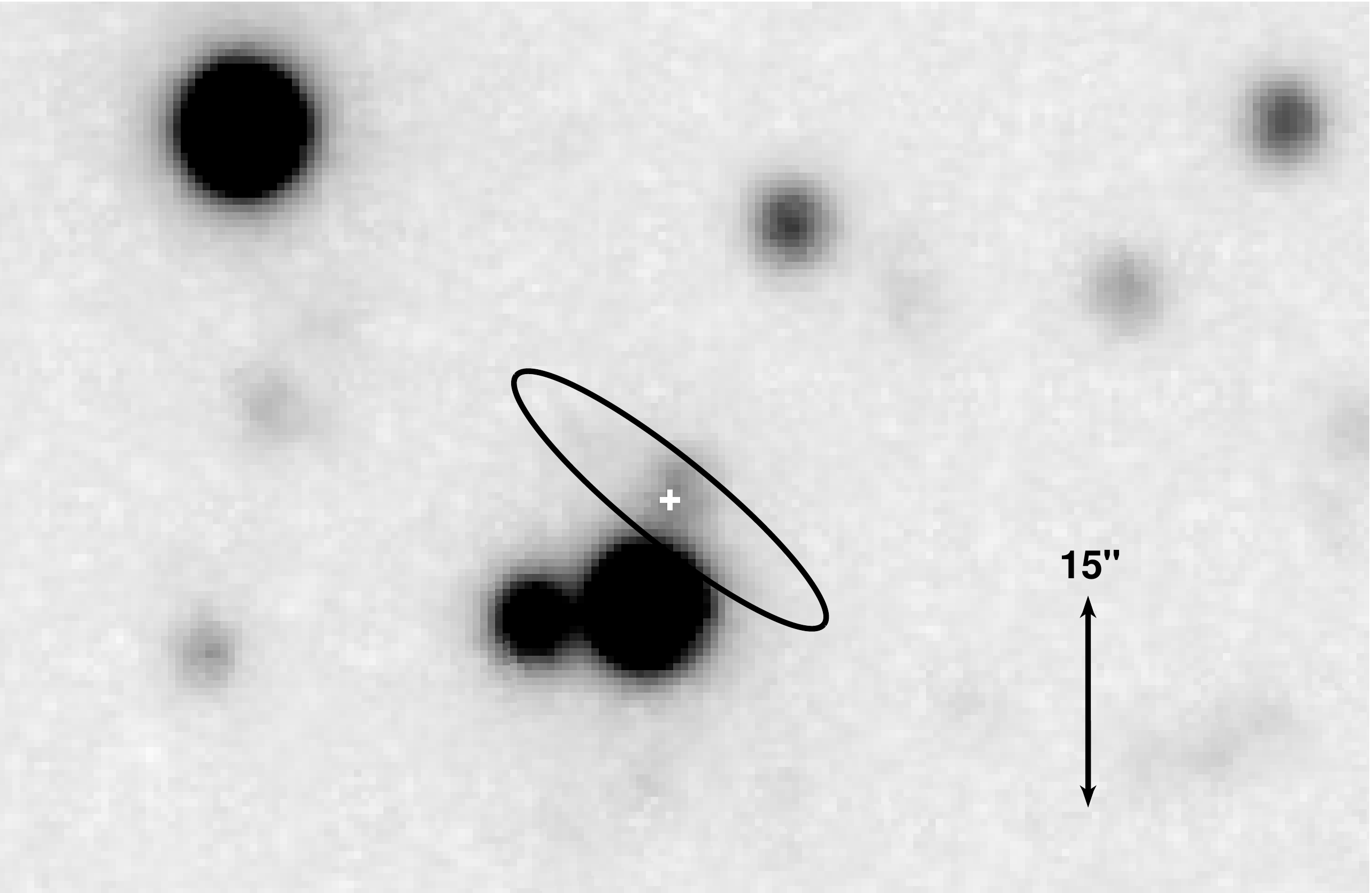}}
\caption{
IRAS 00422+6131:
Close-up of the CAHA image through the H$\alpha$ \emph{(top)} and \sii{} \emph{(bottom)} filters showing the field centered in IRAS 00422+6131.
\label{9_lines}}
\end{figure}

\subsection{Group III: Point-like sources}

\subsubsection{\object{IRAS 00422+6131}}

IRAS 00422+6131 is located at a distance of $2400^{+920}_{-520}$ pc, derived from its parallax \citep{Gai18}.
In the following we present a short description of the field around the source.
\begin{description}

    \item[\textit{Controversial nature of IRAS:}]~
    \begin{itemize}
        \item The source lies in projection toward the young open cluster \object{NGC~225}, although it is not a member of it \citep{Lat91}.
        \item First identified as a TTauri star on he basis of its IRAS colors \citep{Gar97}.
        \item Later proposed to be a G giant star, based on its optical spectrum \citep{Per07}.
  \end{itemize}

    \item[\textit{Association of IRAS with IR emission:}]
    2MASS $J$, $H$, $K$ images show extended emission surrounding the point-like source J00450982+6147574.
  
\end{description}

Figure \ref{9_cont} shows the image of the IRAS 00422+6131 field through a continuum filter. Close-ups of the narrow-band images in the H$\alpha$ and \sii{} line filters are shown in Fig.{} \ref{9_lines}.

All the images show a similar morphology of the IRAS optical counterpart. 
We did not detect in our narrow-band line images any shocked emission from jet structures.


\begin{figure}[htb]
\resizebox{\hsize}{!}{\includegraphics{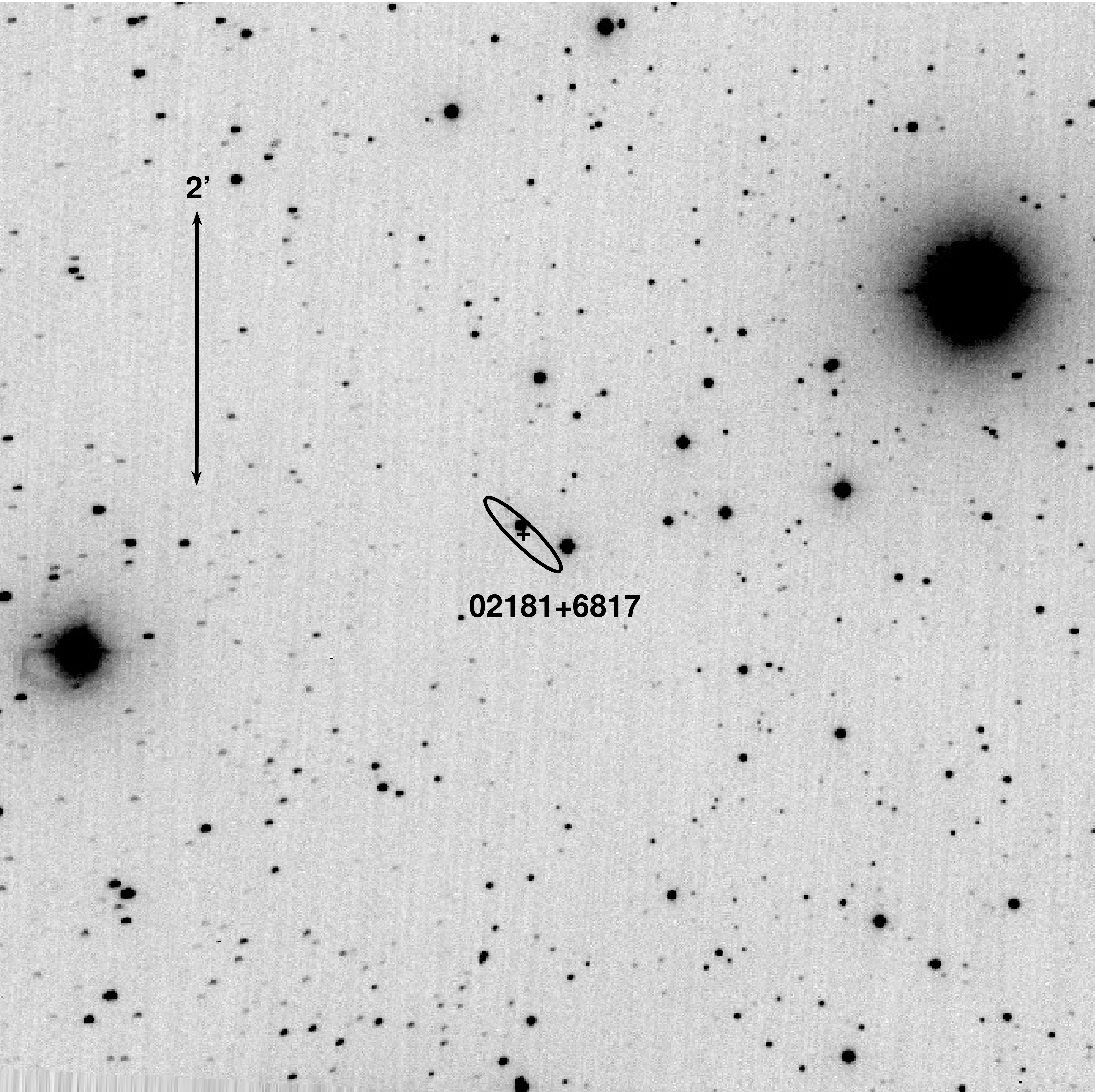}}
\caption{
IRAS 02181+6817:
CAHA image through the continuum filter. 
The IRAS source position and error ellipse are shown.  
\label{27_cont}} 
\end{figure}

\begin{figure}[htb]
\resizebox{\hsize}{!}{\includegraphics{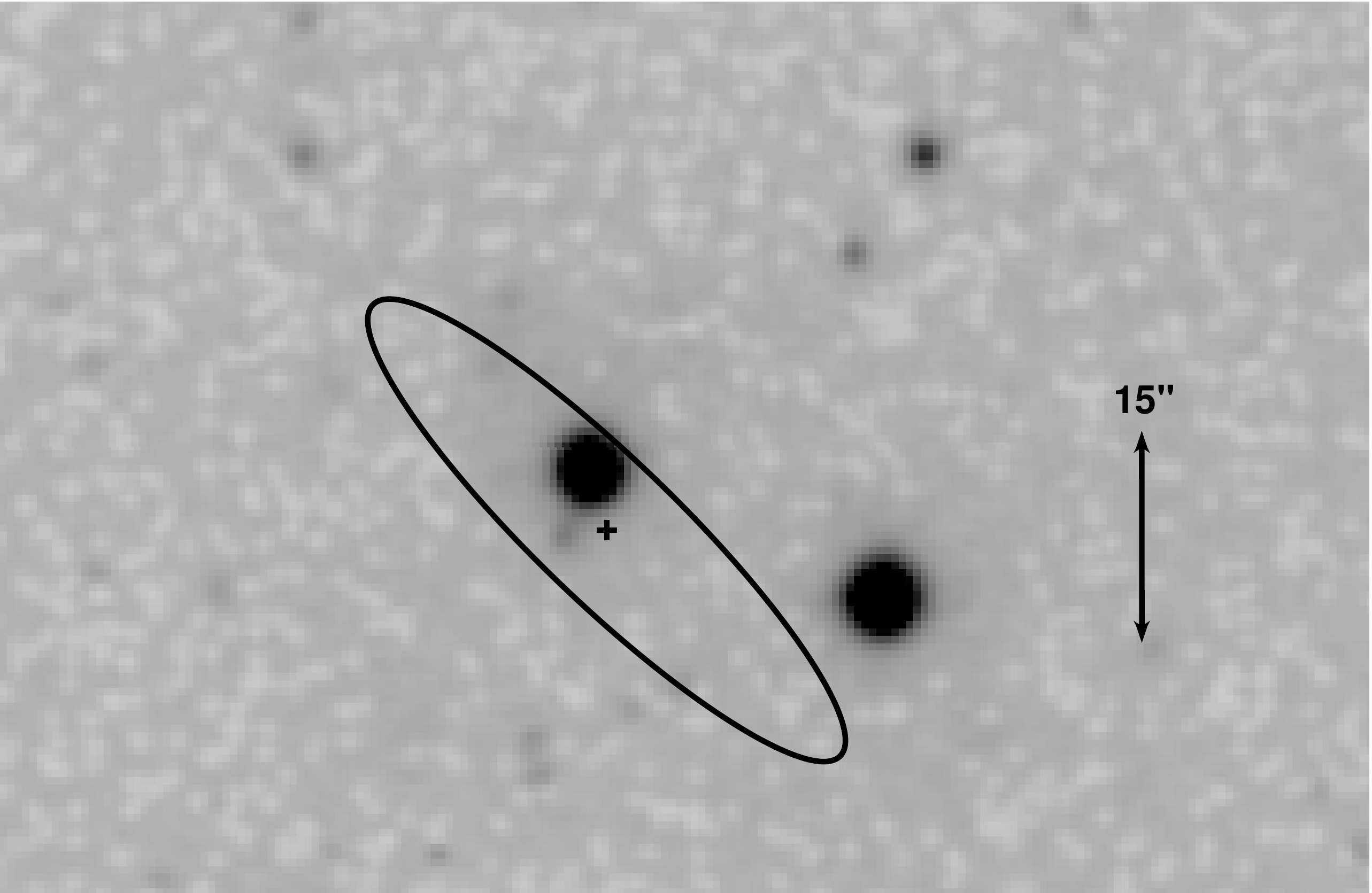}}
\resizebox{\hsize}{!}{\includegraphics{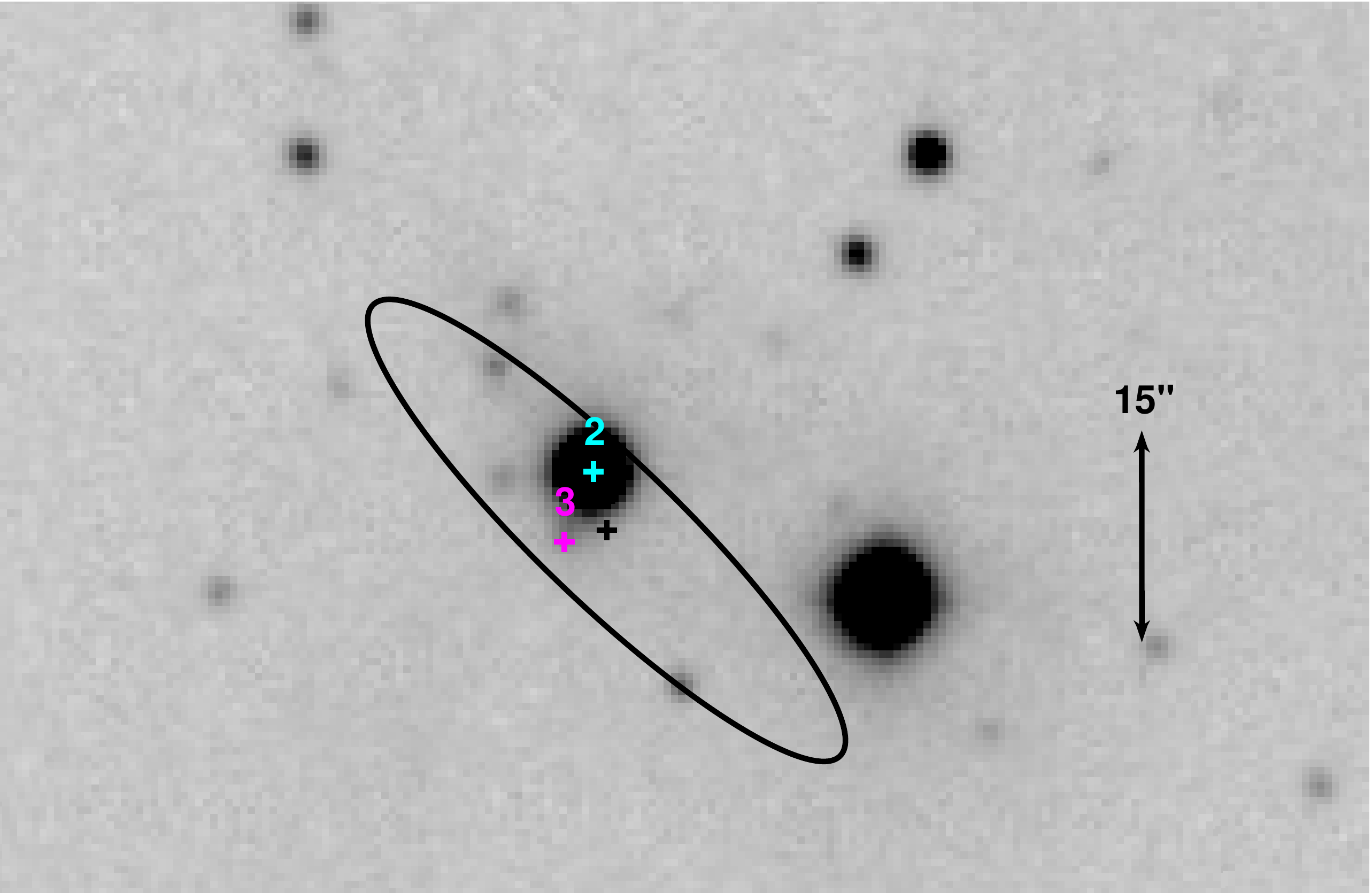}}
\caption{
IRAS 02181+6817:
Close-up of the CAHA image through the H$\alpha$ \emph{(top)} and \sii{} \emph{(bottom)} filters. 
The catalogue position of IRAS 02181+6817 
is marked in black, 
the 2MASS source J02222276+6830471 
in cyan, 
and an additional point-like object 
in magenta. 
\label{27_lines}}
\end{figure}

\subsubsection{\object{IRAS 02181+6817}}

IRAS 02181+6817 is located at a distance of $664\pm25$ pc \citep{Gai18}.
In the following we present a short description of the source.
\begin{description}

\item[\textit{Proposed IRAS counterpart:}] An optical/near-IR bright star.
    \begin{itemize}
        \item Classified as a TTauri star, based on the IRAS colors \citep{Gar91}.
        \item Reported as a non-periodic variable star, \object{CO Cas} \citep{Hof36, Sam17}, with an average $V_\mathrm{mag}=15.52$, and a variation amplitude of 1.15 mag \citep{Koc17}\footnote{
        htpps://asa-sn.osu.edu, ASAS-SN Sky Patrol}
        \item Coincides with the 2MASS star \object{J02222276+6830471}, with near-IR colors compatible with being a Herbig Ae/Be source.
        \item The 3D dust maps of \citet{Gre19}\footnote{
        http://argonaut.skymaps.info, 3D Dust Mapping with Pan-STARRS 1, 2MASS and Gaia}
        give an extinction in the $R$ band $A_R= 0.93$ at the distance of the source.
    \end{itemize}

\end{description}

Fig.{} \ref{27_cont} shows the image of the IRAS 02181+6817 field through a continuum filter.  Close-ups of the  narrow-band images in the H$\alpha$ and \sii{} line filters are shown in Fig.{} \ref{27_lines}.

We found two sources located less than $5''$ from the IRAS position, inside the IRAS error ellipse. 
The  brighter source is CO Cas, the proposed IRAS counterpart. 
The other source is located a few arcsec southeast of CO Cas (labeled as 3 in Fig.~\ref{27_lines}). The source is faint and slightly elongated in the northwest--southeast direction, and could be tracing a micro-jet powered by CO Cas.


\begin{figure}[htb]
\resizebox{\hsize}{!}{\includegraphics{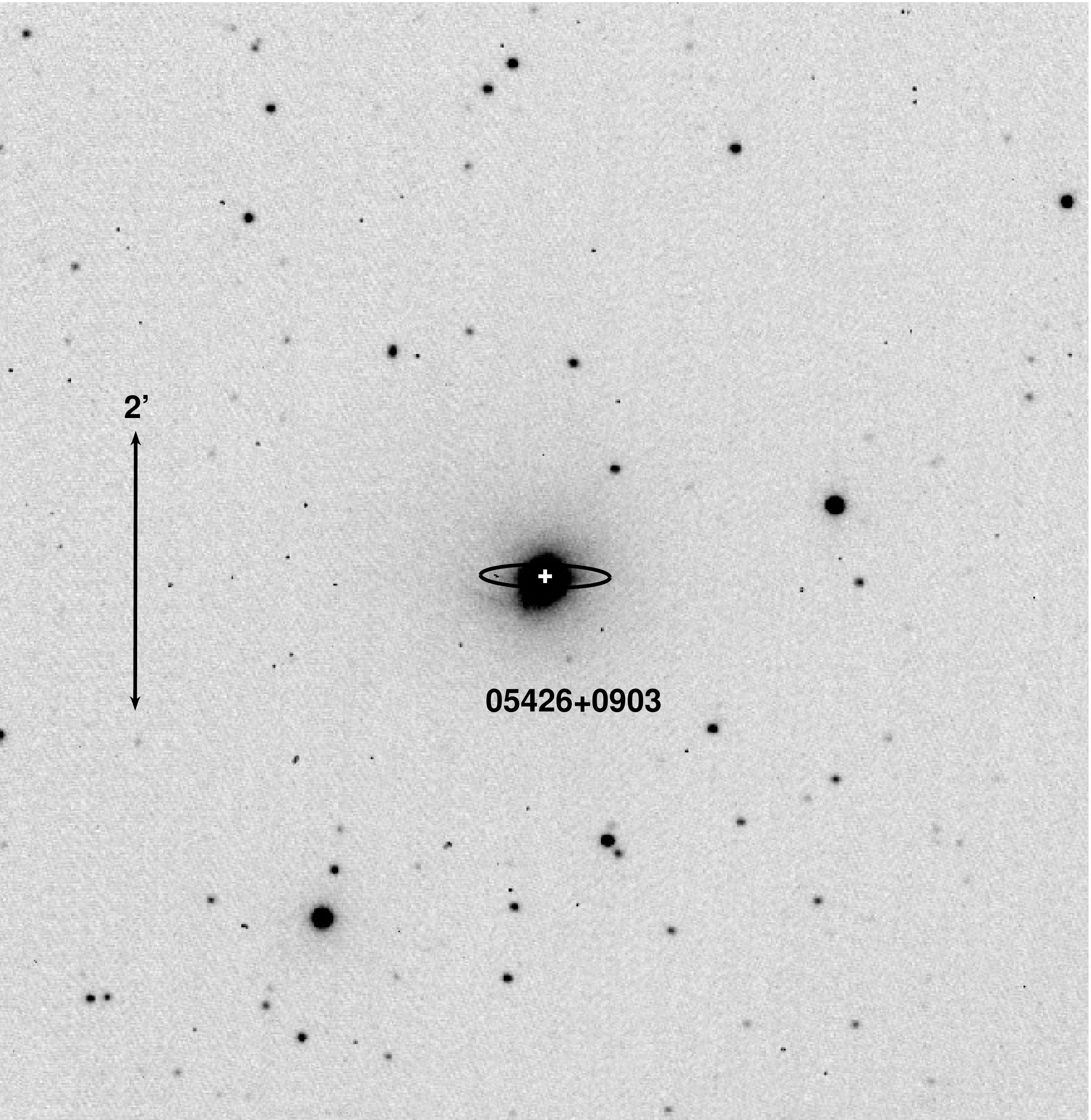}}
\caption{
IRAS 05426+0903:
CAHA image through the continuum filter. 
The IRAS source position and error ellipse are shown.  
\label{126_cont}} 
\end{figure}

\begin{figure}[htb]
\resizebox{\hsize}{!}{\includegraphics{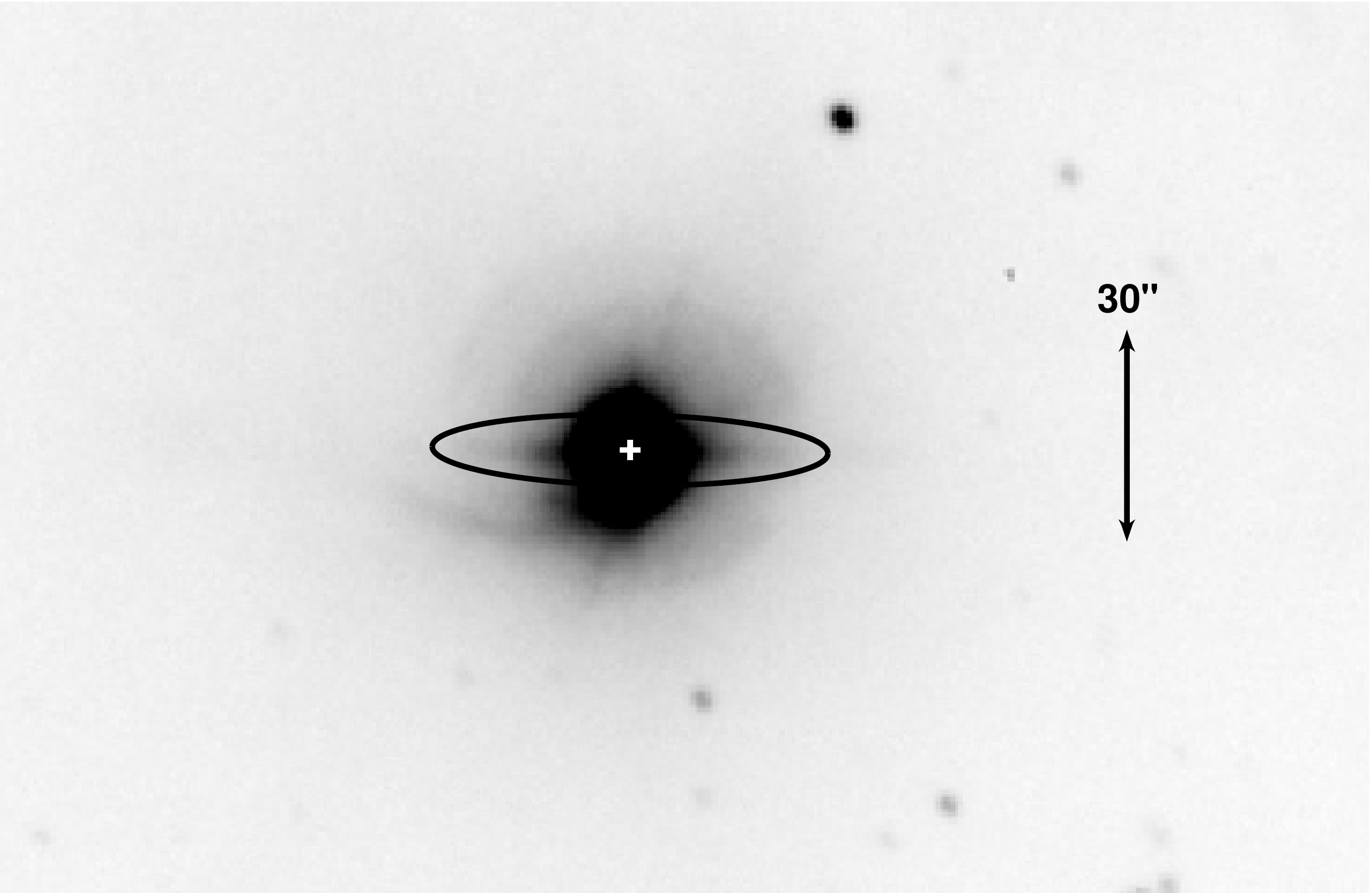}}
\resizebox{\hsize}{!}{\includegraphics{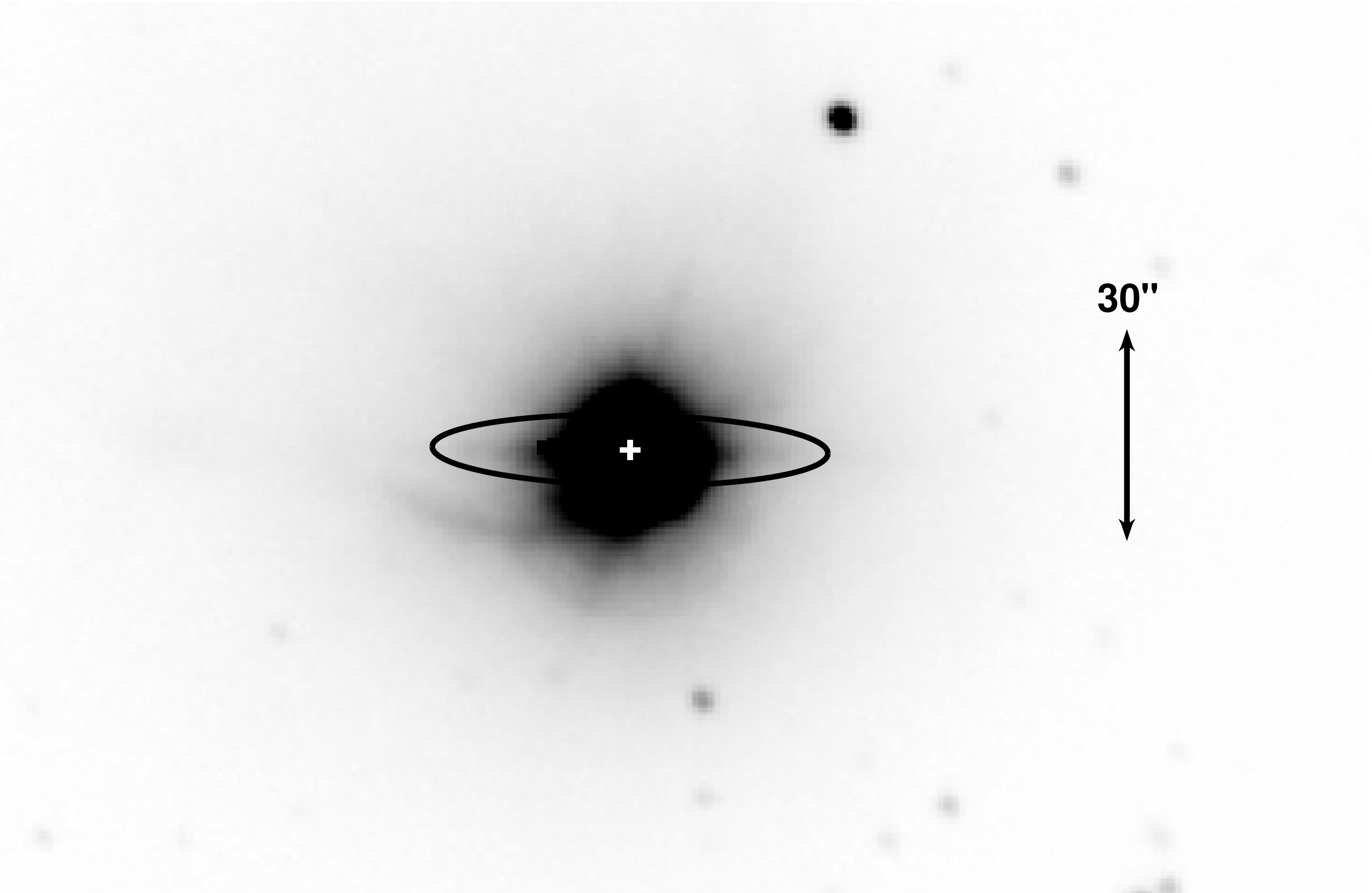}}
\caption{
IRAS 05426+0903:
Close-up of the CAHA image through the H$\alpha$ \emph{(top)} and \sii{} \emph{(bottom)} filters.
\label{126_lines}}
\end{figure}

\subsubsection{\object{IRAS 05426+0903}}

The optical counterpart of IRAS 05426+0903 is the prototypical, young pre-main sequence star \object{FU Ori}. 
It is located at a distance of $416\pm9$~pc \citep{Gai18}.
In the following we present a short description of the source.
\begin{description}

    \item[\textit{Binarity:}]~
        \begin{itemize}
            \item The source is a binary system \citep{Wan04, Rei04}.
            \item The companion is a red star, \object{FU Ori S}, $\sim0\farcs5$ south of FU Ori.
        \end{itemize}
    
     \item[\textit{FU Ori characteristics:}]~
        \begin{itemize}
            \item Optically thick accretion disk \citep[mid-IR, 8--13 $\mu$m interferometric observations;][]{Bec12}.
            \item Continuum emission at 1 mm detected from circumstellar disk \citep[ALMA Band 7 with $0\farcs6$ angular resolution;][]{Hal15}.
        \end{itemize}

    \item[\textit{FU Ori S characteristics:}]~
    \begin{itemize}
        \item Hints of a disk \citep[mid-IR;][]{Bec12}.
        \item FU Ori S is an actively accreting young star, with $\dot{M}_\mathrm{acc} \simeq \mbox{(2--3)} \times 10^{-8} M_{\odot}$~yr$^{-1}$, and is the more massive component of the binary system \citep{Bec12}.
        \item Continuum emission at 1 mm detected from circumstellar disk \citep{Hal15}.
        \item High-density tracer HCO$^{+}$ emission-peak close to FU Ori S, indicating that it is embedded in dense molecular gas \citep{Hal15}.
    \end{itemize}
   
    \item[\textit{Circumbinary reflection nebula:}]~
    \begin{itemize}
        \item Detected through near-IR, high-resolution imaging polarimetry \citep{Liu16, Tak18}.
        \item Arc-like morphology, reminiscent of a spiral arm stretching from east to northeast of the FU Ori system. 
        \item Morphology probably caused by gravitational instabilities in the accretion disks, in an unstable phase of protoplanetary disk evolution.
    \end{itemize}

\end{description}
 Figure~\ref{126_cont} shows the  IRAS 05426+0903 field mapped through a continuum filter.  
Close-ups of the narrow-band images in the H$\alpha$ and \sii{} line filters are shown in Fig.~\ref{126_lines}.
All the images show a similar morphology of the IRAS optical counterpart. 
We did not detect shocked emission from jet structures in our narrow-band line images.


\begin{figure}[htb]
\resizebox{\hsize}{!}{\includegraphics{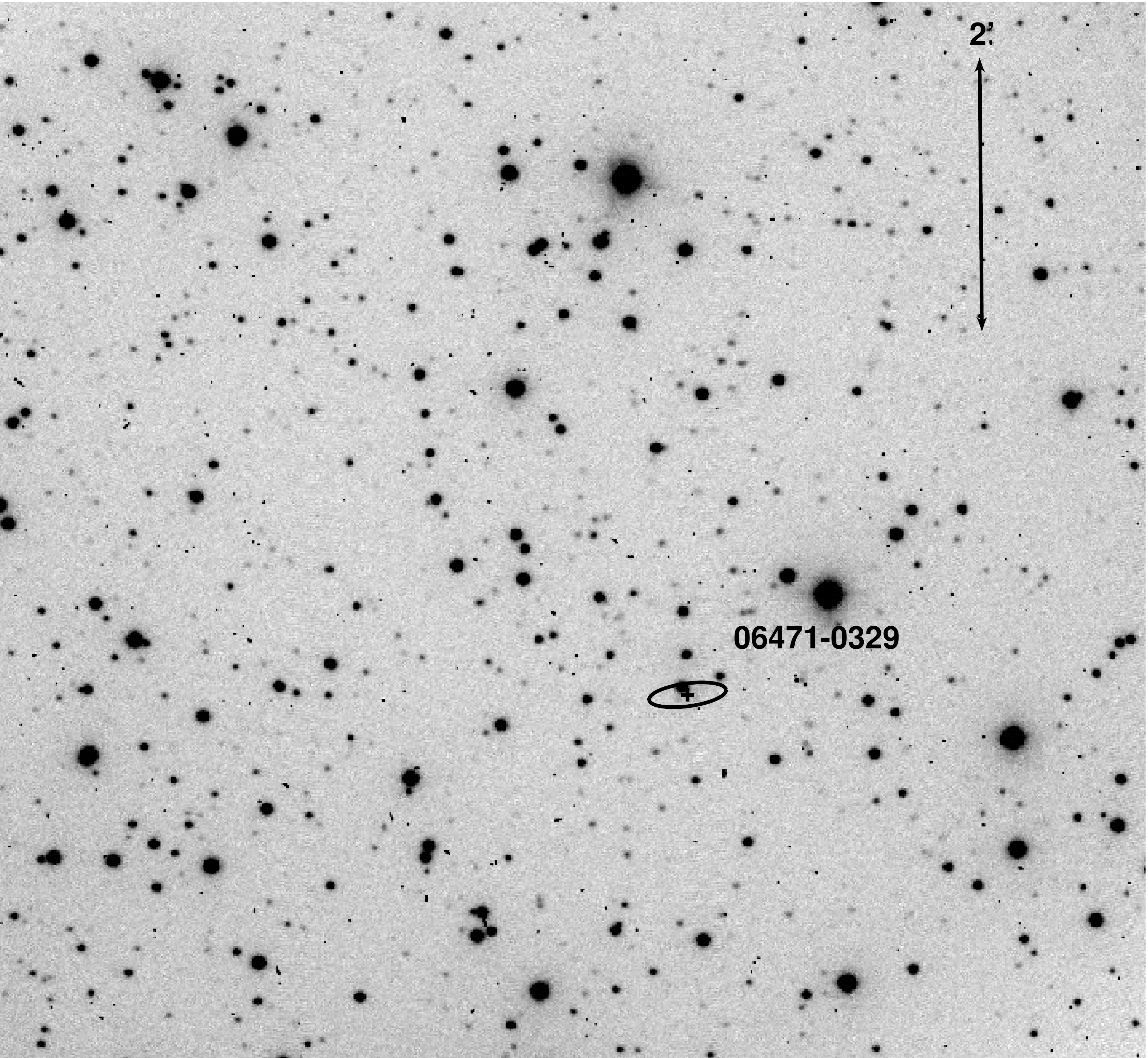}}
\caption{
IRAS 06471$-$0329:
CAHA image through the continuum filter. 
The IRAS source position and error ellipse are shown.  
\label{156_cont}} 
\end{figure}

\begin{figure}[htb]
\resizebox{\hsize}{!}{\includegraphics{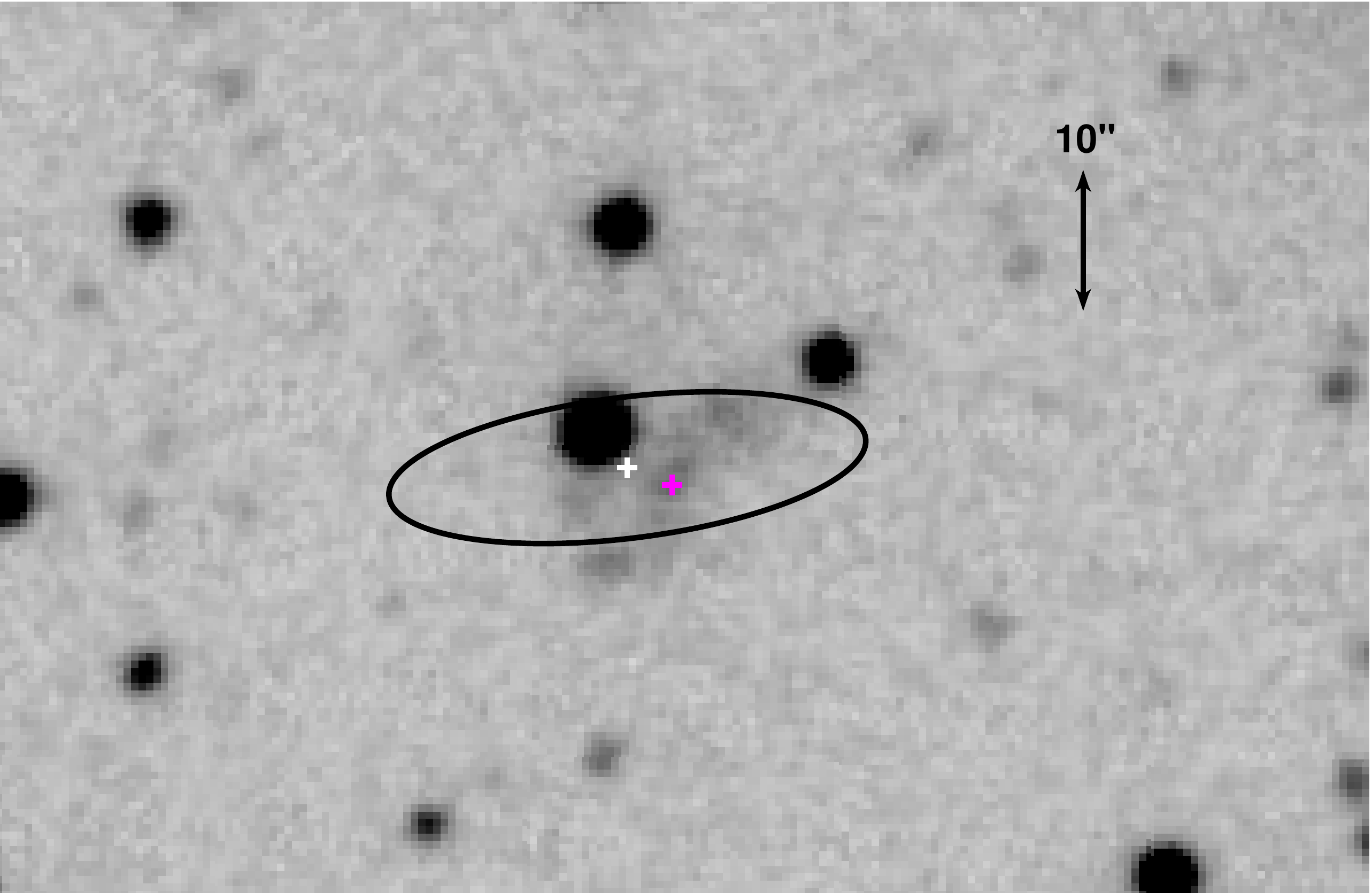}}
\resizebox{\hsize}{!}{\includegraphics{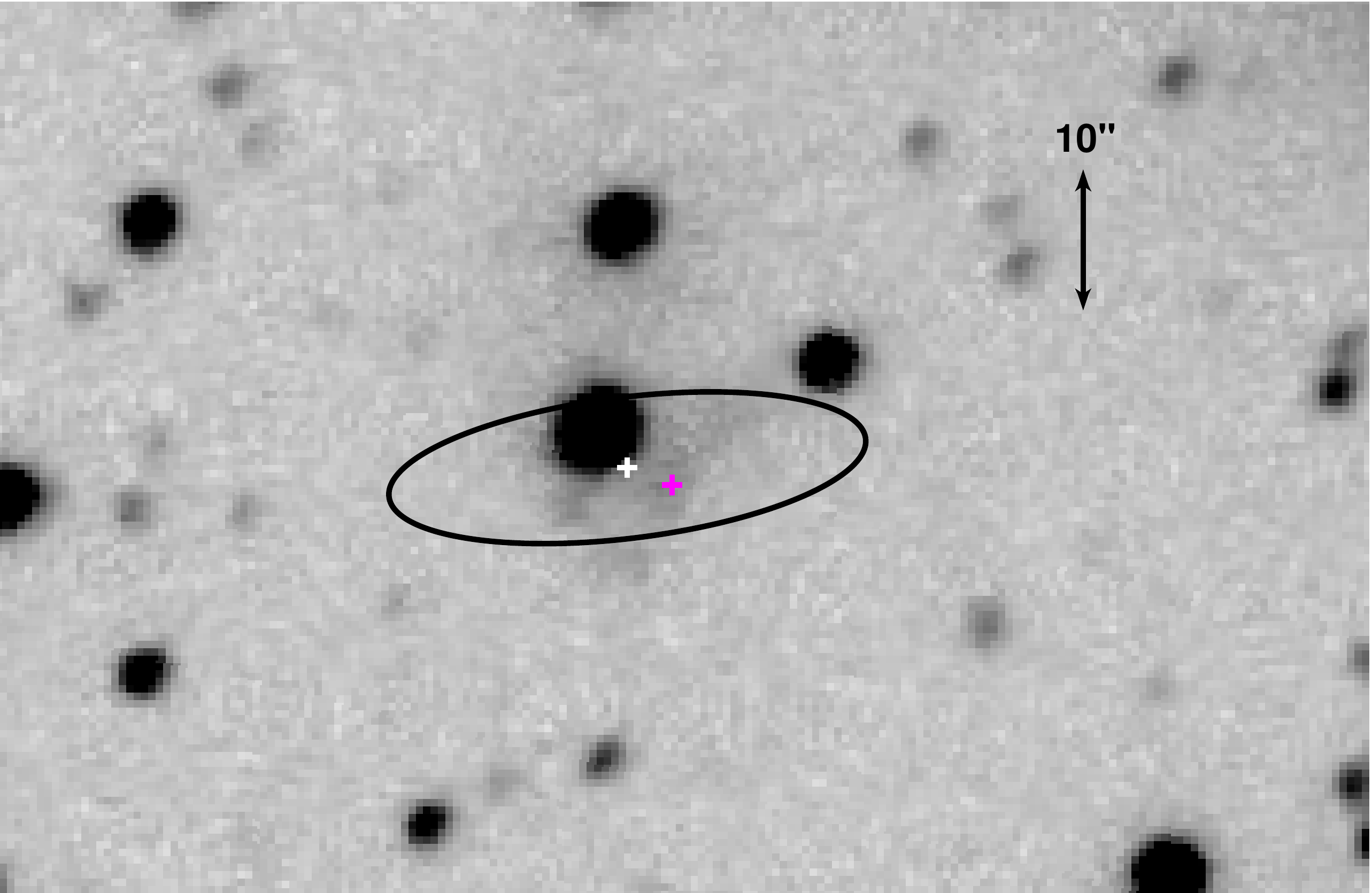}}
\caption{
IRAS 06471$-$0329:
Close-up of the CAHA image through the H$\alpha$ \emph{(top)} and \sii{} \emph{(bottom)} filters.
The positions from the Vizier Catalogue of the IRAS source (white)
and of the 2MASS source J06494021-0332523 (magenta)
are marked.
\label{156_lines}}
\end{figure}

\subsubsection{\object{IRAS 06471$-$0329}}

IRAS 06471$-$0329 is located near the CO boundary of \object{G216$-$2.5}, at a distance of $2110\pm 21$ pc \citep{Zuc20}.
In the following we present a short description of the source.
\begin{description}

    \item[\textit{First proposed counterpart of IRAS:}]
        A red star detected in the $I$-band \citep{Cam89}.
    \item[\textit{Later proposed counterpart of IRAS:}]
        One of the two bright stars in the $K$-band, close to the IRAS position \citep{Lee96}. 
        \begin{itemize}
            \item Very high reddening, with near-IR colors consistent with those of embedded YSOs.
            \item Associated with an IR nebulosity \citep{Ish02}.
            \item Absorption feature at 3.1 $\mu$m from H$_2$O ice, usually present in high-density interstellar clouds protected from UV radiation, characteristic of embedded objects.
        \end{itemize}

\end{description}

Figure \ref{156_cont} shows the image of the IRAS 06471$-$0329 field through a continuum filter. 
Close-ups of the narrow-band images in the H$\alpha$ and \sii{} line filters are shown in  Fig.{} \ref{156_lines}. 
We detected in all the images a weak nebular emission surrounding three point-like sources southwest of the bright star at the center of the field. 
This weak reflection nebula should correspond to the near-IR emission detected by \citet{Lee96} and \citet{Ish02}. 
 In the narrow-band line images (Fig.~\ref{156_lines}) we show the catalogue positions of the IRAS source (white) and the 2MASSX source \object{J06494021$-$0332523} (red).
As can be seen in Fig.~\ref{156_lines}, 
 the position of the 2MASSX source coincides with a weak star surrounded by nebular emission. 
Thus, we concluded that the optical counterpart of IRAS 06471$-$0329, instead of being the bright star at the center of the field, is most probably the more embedded near-IR source J06494021$-$0332523.

\section{Conclusions}

We obtained narrow-band images covering a wide FOV around a sample of IRAS sources that were proposed to be associated with YSOs, based mainly on their location in CC diagrams. 
The association of the IRAS source with young stellar objects has been confirmed from the images obtained.
Although these fields were observed in the past, our survey provides new data both for the IRAS counterparts themselves and for their environment. 
In particular:
\begin{itemize}

\item
In this survey, the first-ever images through the narrow-band, H$\alpha$ and \sii{}, line filters were obtained for nine IRAS sources (all the targets of groups II and III). 
The  IRAS sources of group II were known to have extended emission. 
However, previous imaging was only  made through broad-band filters, which do not allow to distinguish with confidence between reflected and shocked emission. 
The narrow-band images obtained confirm the nature of the  extended component of the emission associated with the targets.
In IRAS 02181+6817, a point-like target of group III, an elongated emission in the H$\alpha$ and \sii{} lines was detected, southeast of the source, which could be tracing a micro-jet powered by the source.

 \item
New images in the two emission lines, H$\alpha$ and \sii{}, were obtained for three sources of group I (IRAS 03220+3035, 04073+3800, and 04239+2436)
These sources were previously mapped only in one of the lines, or through a broad-band, red continuum filter. 
\item
In six of the mapped fields (group I), extended emission in the H$\alpha$ and \sii{} lines was detected, with no continuum  counterpart, tracing HH jets.  
In some targets (IRAS 00087+5833, 02236+7224, and 03220+3034), the jet emission was not associated with the IRAS target, but with another YSO in the mapped field. 
\item
The astrometric positions of most of the jet knots were not reported in the literature. 
Astrometry of the jet knots mapped in our images was performed, and is given in the section corresponding to each source.
In addition, new substructures were resolved within the knots of the jets of HH 196A and HH 196B, imaged in the field of IRAS 03220+3035, and HH 300, powered by IRAS 04239+2436.
Two new knotty emissions, closer to the IRAS source, labeled HH 300D0 and D1, were identified in the images.
 \item
For three of the observed fields (IRAS 00087+5833, 02236+7224, and 04073+3800) we were able to identify jet knots in previous optical and near-IR images in public archives (\emph{HST}, \emph{Spitzer}). Their positions were compared with our results from the CAHA observations, and near-IR counterparts of some knots were identified.
\item
We confirmed or proposed a different IRAS counterparts for seven targets of the sample (IRAS 00044+6521, 00087+5833, 04073+3800, 05302$-$0537, 05393+2235, 06249$-$1007, and 06471$-$0329).
To sustain these associations, the astrometric positions of the YSOs in the neighborhood of the target source were obtained.

\end{itemize}

\begin{acknowledgements}
This work has been partially supported by the Spanish MINECO grants 
AYA2014-57369-C3 and 
AYA2017-84390-C2 (cofunded with FEDER funds), and
MDM-2014-0369 of ICCUB (Unidad de Excelencia `Mar\'{\i}a de Maeztu').
This research has made use of the SIMBAD database, operated at CDS, Strasbourg, France.
\end{acknowledgements}

\end{document}